\documentclass[amsmath,amssymb,amsfonts,prd,twocolumn,superscriptaddress,nofootinbib,longbibliography]{revtex4-1}

\usepackage{graphicx}
\usepackage[usenames]{color}
\usepackage{epsfig}
\usepackage{multirow}
\usepackage{longtable}

\newcommand{\mb}{\mathbf}

\newcommand{\diff}{ \, \mathrm{d} }

\newcommand{\Ddel}{\delta_{\rm D}   }
\newcommand{ \e}{\mathrm{e}^ }

\def\Mpc{\, h^{-1} \, {\rm Mpc}}

\def\kvecMpc{\, h \, {\rm Mpc}^{-1}}
\def\kMpc{\, h \, {\rm Mpc}^{-1}}

\begin{document}

\newcommand{\beq}{\begin{equation}}
\newcommand{\eeq}{\end{equation}}
\newcommand{\beqa}{\begin{eqnarray}}
\newcommand{\eeqa}{\end{eqnarray}}

\newcommand{\lexp}{\mathop{\langle}}
\newcommand{\rexp}{\mathop{\rangle}}
\newcommand{\rexpc}{\mathop{\rangle_c}}

\def\d{\delta}
\def\te{\theta}
\def\ds{\delta_s}
\def\dt{\tilde \delta}
\def\dD{\delta_{\rm D}}
\def\del{\nabla}
\def\knl{k_{n\ell}}
\def\pl{L}
\def\tf{\tilde \phi}

\font\BF=cmmib10
\font\BFs=cmmib10 scaled 833
\def\k{{\hbox{\BF k}}}
\def\x{{\hbox{\BF x}}}
\def\r{{\hbox{\BF r}}}
\def\s{{\hbox{\BF s}}}
\def\ks{{\hbox{\BFs k}}}
\def\xs{{\hbox{\BFs x}}}
\def\p{{\hbox{\BF p}}}
\def\q{{\hbox{\BF q}}}
\def\v{{\hbox{\BF v}}}
\def\u{{\hbox{$u_z$}}}
\def\ub{{\hbox{\BF u}}}
\def\tk{\hat k}
\def\tvk{{\hat{\k}}}
\def\tvq{{\hat{\q}}}
\def\ttheta{\hat \theta}
\def\tphi{\hat \varphi}
\def\C{{\cal C}}
\def\M{{\cal M}}
\def\su{\sigma_u}

\def\mnras{MNRAS}
\def\aj{AJ}
\def\apj{ApJ}
\def\apjs{ApJS}
\def\aap{A \& A}
\def\apjl{ApJ. Lett.}
\def\araa{Ann. Rev. Astron. Astrophys.}
\def\jcap{JCAP}
\def\nat{Nature}
\def\pasj{Pub. Astron. Soc. Japan}
\def\pasp{Pub. Astron. Soc. Pacific}
\def\prd{Phys. Rev. D}
\def\physrep{Phys. Rep.}
\def\qjras{QJRAS}

\def\Gd{{\mathcal G}_2}
\def\Gt{{\mathcal G}_3}

\preprint{Version 1.10 }

\title{Gravity and Large-Scale Non-local Bias} 
\author{Kwan Chuen Chan} \email{kcc274@nyu.edu}
\affiliation{Center for Cosmology and Particle Physics, Department of Physics,  New York University, NY 10003, New York, USA} 
\author{Rom\'an Scoccimarro}
\affiliation{Center for Cosmology and Particle Physics, Department of Physics,  New York University, NY 10003, New York, USA} 
\author{Ravi K. Sheth} 
\affiliation{Center for Particle Cosmology, Department of Physics and Astronomy, University of Pennsylvania, 209 S 33rd Street,  Philadelphia, PA 19104, USA}
\affiliation{Abdus Salam International Centre for Theoretical Physics, Strada Costiera 11, 34151, Trieste, Italy }

\date{\today}

\begin{abstract}

For Gaussian primordial fluctuations the relationship between galaxy and matter overdensities, bias,  is most often assumed to be local at the time of observation in the large-scale limit. This hypothesis is however unstable under time evolution, we provide proofs under several (increasingly more realistic) sets of assumptions. In the simplest toy model galaxies are created locally and linearly biased at a single formation time, and subsequently move with the dark matter (no velocity bias) conserving their comoving number density (no merging). We show that, after this formation time, the bias becomes unavoidably non-local and non-linear at large scales. We identify the non-local gravitationally induced fields in which the galaxy overdensity can be expanded, showing that they can be constructed out of the invariants of the deformation tensor (Galileons), the main signature of which is a quadrupole field in second-order perturbation theory. In addition, we show that this result persists if we include an arbitrary evolution of the comoving number density of tracers. We then include velocity bias, and show that new contributions appear; these are related to the breaking of Galilean invariance of the bias relation, a dipole field being the signature at second order. 

We test these predictions by studying the dependence of halo overdensities in cells of fixed dark matter density:  measurements in simulations show that departures from the mean bias relation are strongly correlated with the non-local gravitationally induced fields identified by our formalism, suggesting that the halo distribution at the present time is indeed more closely related to the mass distribution at an earlier rather than present time.  However, the non-locality seen in the simulations is not fully  captured by assuming local bias in Lagrangian space.  The effects on non-local bias seen in the simulations are most important for the most biased halos, as expected from our predictions. 

Accounting for these effects when modeling galaxy bias is essential for correctly describing the  dependence on triangle shape of the galaxy bispectrum, and hence constraining cosmological parameters and primordial non-Gaussianity. We show that using our formalism we remove an important systematic in the determination of bias parameters from the galaxy bispectrum, particularly for luminous galaxies.

\end{abstract}

\maketitle

\section{Introduction}
Galaxies are one of the main probes of modern cosmology.  
However, the galaxy clustering amplitude depends on galaxy type, so not all types can be unbiased tracers of the dark matter~\cite{Kai8409}.  Therefore, understanding and accounting for this bias is important.  It is common to assume that this bias is a local and deterministic function of the dark matter density field (e.g. other properties of the field than the local overdensity, such as the tidal field, are assumed to produce negligible effects on the galaxy distribution), so the galaxy density contrast at any given time can be written as a Taylor series in the dark matter density at that time \cite{FryGaz9308}.  One of the main goals of this paper is to show that, if there is any time at which this is a good approximation, then it is not good at any other time.  A related goal is to argue is that this should be a better model at early than at later times, in a sense that will be made more precise later in this paper.  

In the galaxy distribution, one expects departures from the local deterministic bias model on scales where nonlinear baryon physics matters.  Nonetheless, on the scales larger than those associated with galaxy formation processes, the deterministic local bias is expected to be accurate, except for a possible constant shot-noise-type contribution~\cite{SchWei98,Mat99}. 
 This has motivated the use of the deterministic local biasing prescription for interpreting clustering measurements in galaxy surveys, in particular this model has been heavily used in interpreting measurements of three-point functions and other measures of non-Gaussianity~\cite{FryGaz9308,FriGaz94,Fry94,GazFri94,ScoFelFry0101,FelFriFry0102,VerHeaPer02,BauCroGaz04,CroGazBau04,JinBor0405,KaySutNic0406,PanSza0510,GazNorBau0512,NisKayHik0702,MarWecFri0801,McBConGar1012,McBConGar1007,Mar1108}.
 However, it is also common to assume that galaxies are closely associated with dark matter halos~\cite{CooShe02}.  So it is natural to ask if halo bias is a deterministic function of the local dark matter overdensity.  Numerical simulations indicate that, on scales of order 20 Mpc and less, halo abundance is not a deterministic function of the dark mass~\cite{MoWhi9609,1999MNRAS.304..767S,CasMoShe0207}.  This manifests as stochasticity in the relation between the galaxy and dark matter density fields at the present time \cite{DekLah9907,SomLemSig0101}.  If we distinguish between the stochasticity associated with some initial or formation time, and that due to evolution from this time to the time of observation -- then the question arises as to which matters more on the large scales which the next generation of galaxy surveys will probe.  

In the Excursion set model of halo formation, abundance and clustering, it is the initial fluctuation field which is fundamental~\cite{1991ApJ...379..440B,1993MNRAS.262..627L,She9811}.  In these models, the origin of the first source of stochasticity is relatively straightforward to understand:  the initial random fluctuation field is expected to have structure on arbitrary small scales, so the substructure within large patches of the same large scale overdensity may differ from one patch to another.  Whether or not a small patch forms a halo is known to be closely related to the initial overdensity of the patch.  If the initial overdensity is the only parameter which matters (e.g., in spherical evolution models, where tidal fields etc. play no role) then the fact that the small scale density is correlated with the density on larger scales produces stochasticity in halo abundances within large spheres of fixed initial overdensity.  Much of this scatter is just a sort of shot-noise which decreases as the cell size is increased \cite{1999MNRAS.304..767S}.  So, on large scales, a deterministic model for the bias can be quite accurate.  If halo formation depends on quantities other than local density~\cite{BonMye9603,SheMoTor01}, then this may contribute to the stochasticity seen in the initial conditions.  But if these other quantities are correlated over shorter scales than is the density, then their effects will be subdominant on large scales, and so they may be neglected in studies of sufficiently large scale bias.  In what follows, we will assume this is the case. 

That is to say, the main goal of this paper is to study departures from the local deterministic bias model which may appear on scales larger than those associated with galaxy or halo formation (i.e., above a few tens of Mpc).  We will show that, even if the bias is local and deterministic at some given time (which we will usually call the formation time), then subsequent nonlinear gravitational evolution will generate non-local bias.  In this respect, our results serve as a well-motivated model for non-local bias.  Other works on non-local bias have provided models~\cite{Mat9511,Mat99,McDRoy0908} based on statistical (as opposed to dynamical) considerations. The virtue of our approach is that it gives a concrete form of non-local bias that must be present even if formation bias is truly local, and we demonstrate for the first time their presence in numerical simulations. In addition, we show that our non-local bias model solves a systematic effect in the determination of the linear bias from bispectrum measurements for biased tracers.

Since evolution plays an important role in the discussion, we devote a substantial part of this paper to the study of the evolution of bias and how it generates non-local bias.  The evolution of halo bias, under the assumption that the number of halos was conserved and their motions were not biased relative to the mass, was first studied by \cite{MoWhi9609}.  They showed that the predictions for this evolution, based on a spherical collapse model for the dynamics, provided a good description of how halo bias evolves.  At linear order (linear theory evolution of the linear bias factor), this calculation agrees with that from combining the continuity equation with perturbation theory, again assuming no velocity bias \cite{Fry9604}.  At linear order, the perturbation theory approach can be generalized to include stochasticity and galaxy formation as a source \cite{Pen98,TegPee9806}.  However, going beyond linear order, either in evolution or in bias, is less straightforward.  

Evolution of the higher order bias factors was investigated in \cite{MoJinWhi97,ScoSheHui01}, but these works approximated the nonlinear gravitational evolution using the spherical collapse model.  This simplification  leads to the inaccurate conclusion that a local bias at formation stays local.  That gravitational evolution generates non-local bias can be seen from the results of~\cite{Fry9604} in second-order in perturbation theory, although this particular point was not noted in that work. The best known example is the limit of this result when the ``formation time" is taken to be at the far past, the so-called local Lagrangian bias, and was first emphasized in~\cite{CatLucMat9807} and further explored in~\cite{CatPorKam0011}. In this paper we develop a formalism that contains all these results in particular limits.  Moreover, it extends them 
  i) to higher-order in perturbation theory,
 ii) to include non-conservation of tracers 
     (arbitrary formation rate and merging), 
iii) to consider biased tracers that do not flow with the dark matter 
     (velocity bias).  
Non-local bias is particularly interesting in view of the fact that the local biasing prescription does not seem to agree well with simulations~\cite{RotPor1105,ManGaz1107}.  Our model of the non-locality generated by evolution gives a well-motivated model for non-local bias.  
  
This paper is organized as follows. 
In Section \ref{sec:EulerianBiasEvolution}, we develop a formalism to generalize previous work on bias evolution to include velocity bias.  We show that gravitational evolution induces a quadrupole, and hence non-locality of bias, on large scales.  If velocity bias is present, then a dipole is also induced.  We illustrate these effects for the case of the evolution of initially scale independent local bias.  

Section \ref{sec:LagrangianBiasEvolution} shows that, when there is no velocity bias, then the same results can be obtained from a Lagrangian formalism, provided the initial conditions are treated self-consistently.  In so doing, we demonstrate that Eulerian and Lagrangian treatments do, in fact, yield the same bispectrum; we discuss this in the context of what appear to be contradictory statements in the literature.  Section \ref{sec:GalaxyFormation} studies bias evolution when comoving number densities are not conserved, either because of merging, or because of the formation of new objects.  In this case also, no dipole contribution is generated if there is no velocity bias.  In Section \ref{sec:BiasGalileon}, we extend our calculation to third order (for the case of no velocity bias), and show that the structure of the non-local bias generated is most easily described by Galileon fields, with a dipole arising from breaking of Galilean invariance of the bias relation when there is velocity bias.  Appendix~\ref{appA} makes the connection between the conserved and non-conserved non-local bias in the most general terms. 

A comparison with simulations is done in Section~\ref{NumSim}, where we use the results of previous sections to motivate a search for correlations between the halo overdensity at fixed matter overdensity with the different non-local fields that our calculations singled out, finding signatures of non-local bias and its dependence on halo mass. In Section~\ref{BispNonLoc} we discuss the impact of non-local bias on the bispectrum, and quantify the magnitude of non-local bias in simulations.  A final section summarizes our conclusions.

Where necessary, we assume a flat $ \Lambda $CDM cosmology with $\Omega_{\rm m} =0.25$ and $\Omega_{\Lambda} = 0.75$. In this paper we use galaxies, halos, and biased tracers interchangeably.  Those readers interested in skipping the technical details and focussing on the main results, the detection of non-local bias in the simulations and their implications, can jump  directly to Section~\ref{NumSim}, where the main results derived previously are summarized.

\section{Non-local bias generation with conserved tracers}
\label{sec:EulerianBiasEvolution}

\subsection{Conserved Tracers with Velocity Bias}

We start by generalizing previous results~\cite{Fry9604,TegPee9806} on the evolution of a tracer density perturbation (galaxies or halos), under the assumption that they form at a single instant in time with local bias, and thereafter evolve conserving their comoving number density (we relax this assumption in Section~\ref{sec:GalaxyFormation}). In particular, we include the possibility that these tracers do not flow with the dark matter, and therefore have their own velocity field. To fully specify the evolution of their velocity field however one needs to make some assumptions, here we will assume that the tracers are massless so we can ignore their contribution  to the gravitational potential which is only sourced by the dark matter. This is a reasonable approximation for galaxies, since only about 20\% of the matter density is in baryons and an even smaller fraction of baryons is in galaxies~\cite{FukPee0412}.  At the large-scales of most interest, we can neglect dynamical friction and any pressure contribution, so we effectively treat the tracers as a pressure-less ideal fluid moving under the gravitational force generated by matter perturbations. In many respects, our approach is very similar to the perturbation theory treatment of two-fluids in~\cite{SomSmi1001}, a connection we will make more explicitly below (see also~\cite{EliKulPor1012,BerVanVer1109}).  In section~\ref{NumSim} we will apply our results to dark matter halos in simulations. In this case we are effectively assuming that halos may be treated as test particles (represented by their center of mass) whic move in the gravitational field due to the full matter distribution (i.e. all other halos). 

In what follows, we will make heavy use of results from perturbation theory (PT, see~\cite{BerColGaz02} for a review).  See section~\ref{sec:BiasGalileon} for a simpler approach (in real instead of Fourier space) that neglects velocity bias, but which goes to third-order in PT instead of the second-order calculations we do here.  We assume that our tracers (which we will henceforth denote as galaxies) are formed at a single instant, with a spatial distribution that is a local function of overdensity $\delta$, and a velocity bias that is linear. We thus have two density and two velocity fields, one each for matter and tracers, and equations of motion that follow from imposing conservation of mass and tracers (we go beyond conserved tracers in section~\ref{sec:GalaxyFormation}) and momentum conservation describing motion under the gravitational potential that is sourced by matter perturbations.

For a single-component fluid,  mass and momentum conservation can be combined into a single equation for a two-component ``vector'' which simplifies obtaining the evolution of density and velocity fields at once \cite{Sco98, Sco01}.  In what follows, we generalize this to a four-component vector equation for our two-component model.  That is, we consider  
 \begin{widetext}
 \begin{eqnarray}  
 \label{eq:DMcontinuity}
 \frac{\partial \delta }{ \partial \tau } + \theta & = & - \int \diff^3 k_1\diff^3 k_2\, \delta_{\rm D } (\mb{k}- \mb{k}_{12} ) \,\alpha (\mb{k}_1 , \mb{k}_2 )\, \theta ( \mb{k}_1 )\, \delta ( \mb{k}_2  ) ,   \\ 
 \label{eq:DMEuler}
 \frac{ \partial \theta }{ \partial \tau } + \mathcal{H}\,\theta + \frac{3}{2}  \mathcal{H}^2 \,\Omega_{\rm m}\,  \delta   &  =  & - \int \diff^3 k_1 \diff^3  k_2 \, \Ddel ( \mb{k} - \mb{k}_{12}  ) \,\beta(\mb{k}_1, \mb{k}_2 ) \,\theta ( \mb{k}_1 )\,  \theta ( \mb{k}_2  )  , \\ 
 \label{eq:Galcontinuity}
 \frac{\partial \delta_{ \rm g}  }{ \partial \tau } + \theta_{\rm g } & = & - \int \diff^3 k_1\diff^3 k_2\,  \Ddel (\mb{k}- \mb{k}_{12} )\, \alpha (\mb{k}_1 , \mb{k}_2 )\, \theta_{\rm g }( \mb{k}_1 )\, \delta_{\rm g } ( \mb{k}_2  )   , \\ 
 \label{eq:GalEuler}
 \frac{ \partial \theta_{ \rm g} }{ \partial \tau } + \mathcal{H}\,\theta_{\rm g} + \frac{3}{2}  \mathcal{H}^2\, \Omega_{\rm m} \, \delta   &  =  & - \int \diff^3 k_1 \diff^3  k_2 \, \Ddel ( \mb{k} - \mb{k}_{12}  )\, \beta(\mb{k}_1, \mb{k}_2 ) \,\theta_{\rm g }( \mb{k}_1 ) \, \theta_{\rm g}( \mb{k}_2  ) ,    
 \end{eqnarray}
  \end{widetext}
where $\delta$ and $\theta$ are the density contrast and velocity divergence of dark matter and $\delta_{\rm g}$ and $\theta_{\rm g}$ are the corresponding quantities for galaxies. $\tau $ is conformal time, $\mathcal{H} \equiv \diff \ln a / \diff \tau $,  and $ \mb{k}_{12} $ denotes $  \mb{k}_1 + \mb{k}_2 $.  The mode-coupling kernels $ \alpha $ and $\beta $ are defined as
 \beq 
 \alpha( \mb{k}_1, \mb{k}_2 )   =   \frac{ \mb{k}_{12} \cdot  \mb{k}_1 }{ k_1^2 }, \ \ \ \ \ 
 \beta(\mb{k}_1, \mb{k}_2  )  =  \frac{ k_{12}^2 ( \mb{k}_1 \cdot \mb{k}_2 ) }{    2 k_1^2  k_2^2  }
 \eeq
 We then introduce $ y = \ln D $ as the time variable, where $D$ is the linear growth factor for the matter perturbations satisfying 
 \begin{equation}
 \frac{\diff^2 D  }{ \diff \tau^2 } + \mathcal{H} \frac{ \diff  D }{ \diff \tau   } - \frac{ 3 }{ 2 }  \mathcal{H}^2 \Omega_{\rm m} D  = 0.   
 \end{equation} 
Since $f^2 = \Omega_{\rm m}$, with $ f = \diff y / \diff \ln a   $, is a very good approximation throughout the evolution~\cite{ScoColFry9803}, the equations of motion Eqs.~(\ref{eq:DMcontinuity}-\ref{eq:GalEuler}) can be written in compact form by defining a four-component  ``vector'' $\Psi $ as 
 \begin{eqnarray}
 \Psi  = 
 \begin{pmatrix}
 \delta   \\
 -  \theta  / f \mathcal{H}   \\ 
 \delta_{\rm g }  \\
 -  \theta_{\rm g}   / f \mathcal{H}   \\ 
 \end{pmatrix},
 \end{eqnarray}
which yields    
    \begin{equation}
 \label{eq:CompactEOM}
 \partial_{y} \Psi_a(\mb{k}) + \Omega_{ab} \Psi_{b}(\mb{k})  =  \gamma_{abc}\, \Psi_b(\mb{k}_1) \Psi_{c}(\mb{k}_2),  
 \end{equation}
 where integration over $\mb{k}_1$ and $\mb{k}_2  $ is implied and the entries of $\gamma_{abc}$ are zero except for  
 \begin{eqnarray}
 \gamma_{121} & = & \gamma_{343} = \Ddel( \mb{k} - \mb{k}_{12} )\, \alpha(  \mb{k}_1,  \mb{k}_2 ),  \\ 
 \gamma_{222} & = & \gamma_{444} =  \Ddel( \mb{k} - \mb{k}_{12} )\, \beta(  \mb{k}_1,  \mb{k}_2 ) ,   
 \end{eqnarray}
  and  the matrix $\Omega_{ab}$ reads 
  \begin{eqnarray}
 \Omega_{ab} = \left(
 \begin{array}{rrrr}
 0  &  -1   &  0   &   0      \\ 
 - \frac{ 3}{2 }  &  \frac{1}{2} & 0& 0  \\
 0 & 0 & 0  & -1 \\ 
 - \frac{ 3}{2 }  &  0  & 0 & \frac{1}{2}      \\ 
 \end{array} \right)
 \end{eqnarray}

 In this section we assume that galaxies are formed at a single epoch $y_*\equiv 0$ with linear density bias $b^*_1$ and linear velocity bias $b_{\rm v}^*$. Our choice of $y_*\equiv 0$ means that we can restore the more general time dependence by replacing $y \to \ln D/D^*$ in all formulas below.  The initial conditions can be handled conveniently by Laplace transforms. Taking the Laplace transform with respect to $ y $, Eq.~(\ref{eq:CompactEOM}) becomes 
 \begin{equation}
 \omega \tilde{\Psi}_a ( \omega ) - \phi_a + \Omega_{ab} \tilde{ \Psi}_b ( \omega ) = \gamma_{abc}\, [\tilde{ \Psi }_b * \tilde{\Psi}_c]( \omega ), 
 \end{equation}
 where  $ \tilde{\Psi}( \omega )$ represents the Laplace transform of $\Psi(y) $, $\phi_a=\Psi_a(y_*= 0 )$ is the initial condition   and 
 \begin{equation}
  [\tilde{ \Psi }_b * \tilde{\Psi}_c]( \omega) = \frac{1 }{2 \pi i } \int_{x -i \infty}^{x  + i \infty } \diff \omega'  \tilde{ \Psi }_b ( \omega' )  \tilde{ \Psi }_c ( \omega - \omega' ),    
 \end{equation}
for some $x$ in the region of convergence of  $  \tilde{ \Psi }  $.     Collecting the terms linear in $\tilde{ \Psi}$, we have 
 \begin{equation} 
 \tilde{ \Psi }_a( \omega) = \sigma_{ab}(\omega) ( \phi_b + \gamma_{bcd} \tilde{ \Psi}_c * \tilde{\Psi}_d ( \omega )  ),  
 \end{equation}
 with $ \sigma_{ab} = (\omega I + \Omega )^{-1}_{ab}$. Finally, we go back to the $y$-space by the taking the inverse Laplace transform 
 \begin{equation}
 \label{eq:EoMintegralform}
  \Psi_a(y) = g_{ab}(y) \phi_b + \int_{0}^{y} \diff y'  g_{ab}(y - y')  \gamma_{bcd}  \Psi_c(y')  \Psi_d( y' ), 
 \end{equation}
 where $g_{ab}(y)$, called the linear propagator \cite{Sco98,Sco01}, is given by 
 \begin{equation}
 g_{ab}(y) = \frac{ 1 }{ 2 \pi i } \int_{ \xi -i \infty}^{ \xi  + i \infty } \diff \omega\,  \sigma_{ab}(\omega)\, \mathrm{e}^{\omega y}, 
 \end{equation}
 where $\xi$ is a real number larger than the real parts of the poles of $\sigma $.   We then have,
 \begin{widetext}
 \begin{eqnarray}
 \label{gab} 
g_{ab} = 
 \left(
 \begin{array}{cccc}
  \frac{2}{5} \e{-3 y/2}+\frac{3 }{5}\e y  & -\frac{2}{5} \e{-3 y/2}+\frac{2 }{5} \e y & 0 & 0   \\
  -\frac{3}{5} \e {-3 y/2}+\frac{3 }{5} \e y   & \frac{3}{5} \e{-3 y/2}+\frac{2 }{5} \e y & 0 & 0   \\
  -1+\frac{2}{5} \e {-3 y/2}+\frac{3 \e y}{5} &   \left(-2-\frac{2}{5}
    \e {-3 y/2}+ 2 \e {-y/2} +\frac{2 \e y}{5}\right) & 1 & 2(1- \e {-y/2} )\\
  -\frac{3}{5} \e {-3 y/2}+\frac{3 \e y}{5} & \frac{3}{5} \e {-3
    y/2}-   \e {-y/2}  +\frac{2 \e y}{5}   & 0 & \e {-y/2}
 \end{array}
 \right).
 \end{eqnarray}
  \end{widetext}
We note that the $2\times 2$ block in the upper left corner is the same as the linear propagator for dark matter derived in \cite{Sco98,Sco01}. The linear propagator satisfies the relation 
\begin{equation}
 g_{ab}(y_1 +y_2) = g_{ac}(y_1) \ g_{cb}( y_2 ),
\end{equation}
which is the expected law for linear evolution generalized for arbitrary mixing of growing and decaying modes. 
The linear propagator has the usual (matter only) growing and decaying modes,
\beq
e_a^{(1)} = \left( \begin{array}{c} 1 \\ 1 \\ 1 \\ 1 \end{array} \right), \ \ \ \ \ 
e_a^{(2)} = \left( \begin{array}{c} 2/3 \\ -1 \\ 2/3 \\ -1 \end{array} \right)
\label{stdeigen}
\eeq 
i.e. $g_{ab}(y)\, e_b^{(1)} = \mathrm{e}^y \,e_a^{(1)}$ and $g_{ab}(y) \, e_b^{(2)} = \e{-3y/2} \,e_a^{(2)}$. In addition, there is an iso-density decaying mode $e_a^{(3)}$ and an  iso-density-velocity decaying mode $e_a^{(4)}$,
\beq
e_a^{(3)} = \left( \begin{array}{r} -\omega_2 \\ 0 \\ \omega_1 \\ 0 \end{array} \right), \ \ \ \ \ 
e_a^{(4)} = \left( \begin{array}{r} 2\omega_2 \\ -\omega_2 \\ -2\omega_1 \\ \omega_1 \end{array} \right)
\label{neweigen}
\eeq 
where here we restored (following~\cite{SomSmi1001}) temporarily a contribution to the overall mass density fraction of $\omega_1$ for matter and $\omega_2$ for galaxies ($\omega_1+\omega_2=1$). The first eigenmode here satisfies $g_{ab}(y)\, e_b^{(3)} = e_a^{(3)}$, corresponding to a constant mode with zero total density perturbation, while the second eigenmode satisfies $g_{ab}(y) \, e_b^{(4)} = \e{-y/2} \,e_a^{(4)}$ and corresponds to a vanishing total density and total velocity divergence perturbation. Our assumption of tracers as test particles (massless) means we have set $\omega_1=1$ and $\omega_2=0$ in our approximation. In the more general case, the same results we find here will apply with small corrections proportional to $\omega_2$ (see~\cite{SomSmi1001}).  Note that the standard eigenmodes $e_a^{(1,2)}$ are of course independent of $\omega_i$ as they correspond to in-phase motion of the two fluids as if they were one.

We have transformed the  equations of motion Eq.~(\ref{eq:CompactEOM}) into an integral equation Eq.~(\ref{eq:EoMintegralform}) with explicit dependence on initial conditions $\phi_a$ that can be solved perturbatively. To specify the initial conditions  we assume that they can be expanded as follows,
  \begin{widetext}      
 \begin{equation}
 \label{eq:psiexpansion}
 \phi_a ( \mb{k}) = \sum_{n} \int \diff^3 q_1 \dots  \diff^3 q_n\,  \Ddel ( \mb{k} - \mb{q}_{12 \dots n}  )\ \mathcal{I}_a^{(n)} ( \mb{q}_1 , \dots, \mb{q}_n  )\
 \delta_0 ( \mb{q}_1 )  \dots  \delta_0 ( \mb{q}_n ),  
 \end{equation}
where $\delta_0$ is the initial dark matter density contrast, and the vector $\Psi_a(\mb{k}, y) $ can be similarly expanded  as 
 \begin{equation}
 \label{eq:Psiexpansion}
 \Psi_a ( \mb{k}, y ) = \sum_{n} \int \diff^3 q_1 \dots  \diff^3 q_n \,  \Ddel ( \mb{k} - \mb{q}_{12 \dots n} ) \ \mathcal{K}_a^{(n)} ( \mb{q}_1 , \dots, \mb{q}_n , y ) \
  \delta_0 ( \mb{q}_1 )  \dots  \delta_0 ( \mb{q}_n ) .
 \end{equation}
 Putting Eqs.~(\ref{eq:psiexpansion})  and (\ref{eq:Psiexpansion}) into Eq.~(\ref{eq:EoMintegralform}), and collecting terms of the same order in $\delta_0$, we get a recursion relation for the $\mathcal{K}_a^{(n)}$ kernels, 
 \begin{eqnarray}
\label{eq:Kn_generalsol}
 \mathcal{ K }_a^{(n)} ( \mb{q}_1,  \dots  , \mb{q}_n , y  ) &  = &   g_{ab} ( y )\ \mathcal{ I }_b^{(n) } (  \mb{q}_1,  \dots  , \mb{q}_n   )  +  \sum_{ j = 1}^{n-1} \int_0^y \diff y' g_{ab} (y - y') \, \gamma_{bcd } ( \mb{k}_1 = \mb{q}_{1 \dots j  }  ,  \mb{k}_2 = \mb{q}_{j+1  \dots n  }  )     \\ 
 &  \times  &     \mathcal{ K }_c^{(j)} ( \mb{q}_1,  \dots  , \mb{q}_j , y'  ) \,     \mathcal{ K }_d^{(n-j)} ( \mb{q}_{ j+1} ,  \dots  , \mb{q}_n , y'  ) .  \nonumber    
 \end{eqnarray}
  \end{widetext}
 Note that the kernels $\mathcal{K}_a^{(n)}$ obtained from Eq.~(\ref{eq:Kn_generalsol}) are not symmetric in the arguments $\mb{q}_i$, but they can be symmetrized afterwards. Only the symmetric part contributes to $ \Psi_a$.

\subsection{Generation of Non-Local Bias}
\subsubsection{``Initial Conditions" at Formation}
We now explore the results of Eq.~(\ref{eq:Kn_generalsol}) to study the generation of non-local bias by gravitational evolution from local-bias initial conditions. That is, we assume that  biased tracers at formation time $t_*$ (corresponding to growth factor $D_*$ and $y=0$) can be written as a local function of matter density that is then Taylor expanded,
\begin{equation}
 \delta_g^* = \sum_k \frac{b_k^*}{k!}\,\delta^k_*,  \ \ \ \ \ \theta_g^* = b_{\rm v}^* \, \theta_*, 
\end{equation}
where we assumed that the tracers have velocities that are only linearly biased with respect to matter. In the examples below we assume $b_{\rm v}={\rm const.}$ but our results in this section also apply if the velocity bias is $k$-dependent. Small-scale velocity bias has been seen in simulations ~\cite{ColKlyKra0008,BerWeiBen0308,DieMooSta0408}, at the $10\%$ level. At large scales is predicted by peak theory~\cite{DesShe1001,DesCroSco1011} although in a statistical sense, i.e. peaks move locally with the dark matter but their statistics can be thought of as if there is a velocity bias that is $k$-dependent and goes to unity at very large scales as $k^2$. But the situation for baryons, and therefore galaxies (as opposed to halos), can be somewhat different, e.g. at early times $z \gtrsim 100$  the relative velocity between the dark matter and baryons is typically supersonic~\cite{TseHir1010,DalPenSel1011}, and there might be 
a non-trivial component to the relation between dark matter and baryons on large scales due to isocurvature modes, see e.g.~\cite{GriDorKam1112}.

We assume that the matter is in the growing mode, so, to linear order, the initial conditions kernel at formation time is given by,
\beqa
{\cal I}^{(1)}_a = \left( \begin{array}{c} 1 \\ 1 \\ b_1^* \\ b_{\rm v}^* \end{array} \right) &= &
e_a^{(1)} + [(b_1^*-1)+2(b_{\rm v}^*-1)] \, e_a^{(3)} \nonumber \\ & & +  \ (b_{\rm v}^*-1)\, e_a^{(4)} .
\label{linearICs}
\eeqa 
Here we have expanded the initial conditions in terms of the eigenmodes of the linear propagator (recall that $\omega_1=1$ and $\omega_2=0$ in Eq.~\ref{neweigen}). This makes clear that density bias excites the iso-density decaying mode and that velocity bias  excites, in addition, the iso-density-velocity decaying mode.  At second-order we have,
\beq
{\cal I}^{(2)}_a = 
\left( \begin{array}{c} 
F_2(\mb{k}_1, \mb{k}_2 ) \\ G_2(\mb{k}_1, \mb{k}_2 ) \\ b_1^* \, F_2(\mb{k}_1, \mb{k}_2  ) + b_2^*/2 \\ b_{\rm v}^* \, G_2(\mb{k}_1, \mb{k}_2    ) 
\end{array} \right),
\label{quadICs}
\eeq
where $F_2$ and $G_2$ are the second-order kernels that describe the density and velocity divergence to quadratic order in the linear matter fluctuations (see Eqs.~\ref{F20}-\ref{F22} below for the multipole expansion of $F_2$). They are generated by the second term in Eq.~(\ref{eq:Kn_generalsol}) during time evolution {\em up to} $y=0$ for matter fluctuations (which satisfies the same equations for $y<0$ as Eq.~(\ref{eq:Kn_generalsol}) restricted to $a=1,2$). They are functions of the wavectors through $\gamma_{bcd}$ because gravitational evolution is nonlocal. 
                  
At time $t>t_*$ (or redshift $z<z_*$) $\delta_g$ will be, precisely because of $\gamma_{bcd}$ in the second term in Eq.~(\ref{eq:Kn_generalsol}), a non-local function of $\delta$. We therefore are interested in separating out the non-local contribution to the galaxy density perturbations, 
\begin{equation}
\label{nonlocdg}
\delta_{\rm g}^{\rm Nloc}\equiv \delta_{\rm g} - \delta_{\rm g}^{\rm local} 
= \delta_{\rm g} - \sum_k \frac{b_k}{k!}\,\delta^k, 
\end{equation}
where $\delta_{\rm g}^{\rm local}$ is a local function of $\delta$ and thus can be expressed in terms of local bias parameters $b_k$ (the evolved version of the $b_k^*$'s). We will in fact construct $\delta_{\rm g}^{\rm Nloc}$ and $\delta_{\rm g}^{\rm local}$ order by order, e.g. by first substracting linear bias and analyzing local and non-local contributions at second order, then substracting local quadratic bias and analyzing what happens at third order, and so on. 

In this section we take the first step in this analysis:  to quantify the non-local contributions to second order.  To do so, it is convenient to define the field
\beq
\chi \equiv \delta_{\rm g} - b_1\, \delta
\label{chidef}
\eeq
which neglects all $k>1$ terms in $\delta_{\rm g}^{\rm Nloc}$ of Eq.~(\ref{nonlocdg}).  We can then study $\chi$ at second order and decompose it in Legendre polynomials ${\cal P}_\ell(\mu)$,
\beqa
\chi^{(2)}( \mb{x} ) &=& \int d^3 k_1 d^3 k_2\, \e{-i  \mb{k}_{12}  \cdot \mb{x} } \  \delta_0( \mb{k}_1)\,  \delta_0( \mb{k}_2) \nonumber \\ & & \times   \ 
 \sum_{\ell=0}^\infty \, {\cal P}_\ell(\mu)\  \chi_{\ell}^{(2)}(k_1,k_2),       
\label{LegDec}
\eeqa
where $\mu=\hat{k}_1 \cdot \hat{k}_2 $. A local contribution to $\chi^{(2)}$ should be proportional to $\delta_0^2$, and therefore corresponds to a monopole ($\ell=0$) contribution with $\chi_0$ independent of $k_i$.  Any $\ell>0$ piece cannot be written as local functions of $\delta$ and thus will be entirely the result of non-local contributions. As we shall see,  a quadrupole contribution ($\ell=2$) is inevitable for biased tracers, and velocity bias generates in addition a dipole ($\ell=1$). We shall not go beyond second order here, see Section~\ref{sec:BiasGalileon} for results to third order (see also Appendix~\ref{appA}), and the next step in the construction of Eq.~(\ref{nonlocdg}). 

 \begin{figure}[!t]
 \centering
 \includegraphics[ width=0.95\linewidth]{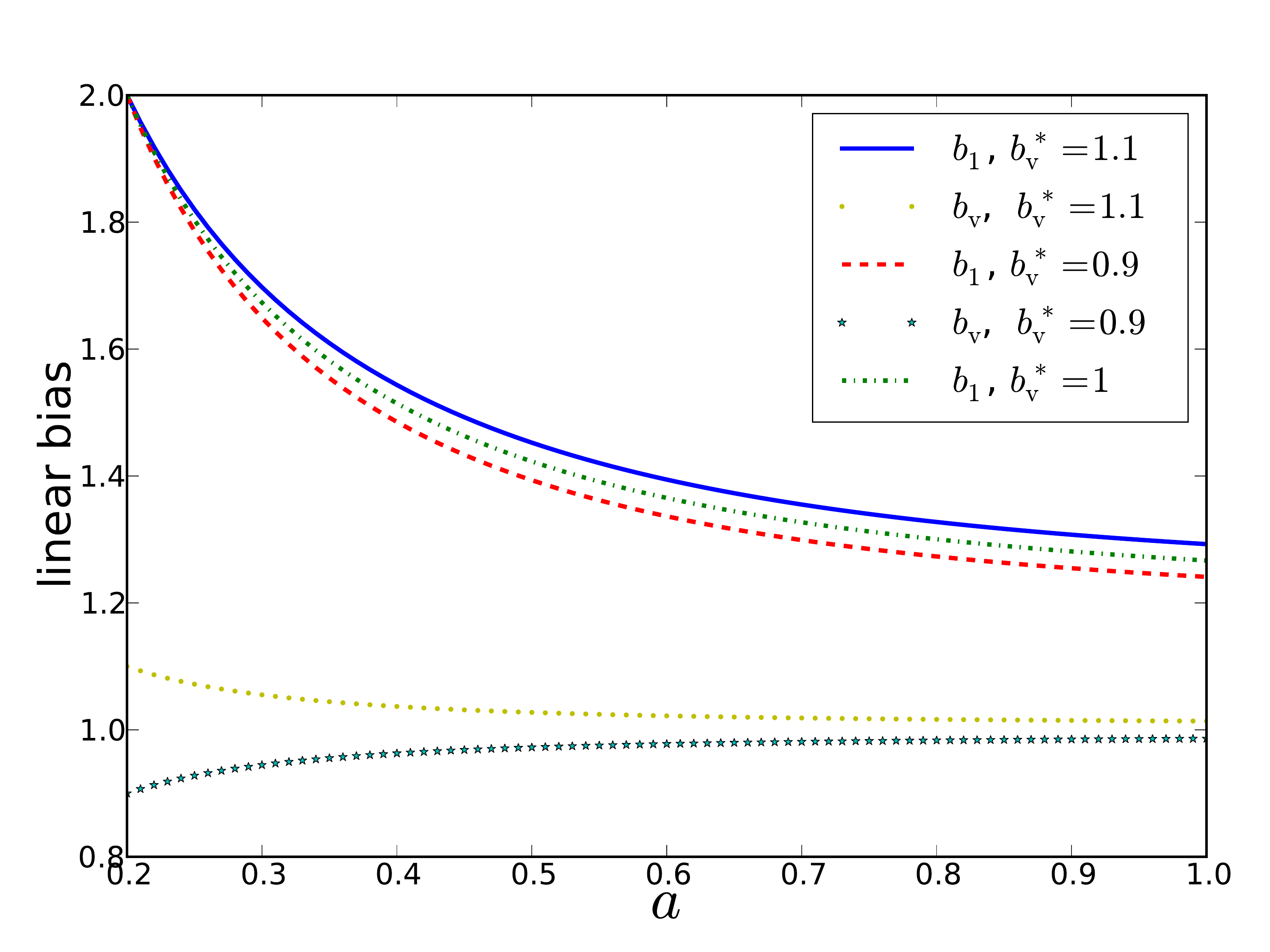}
 \caption{ The evolution of linear density bias $b_1$ (top three lines) and velocity bias $b_{\rm v}$ (bottom lines) as a function of the scale factor $a$ with the initial values $b^*_1=2$ and $b_{\rm v}^* =1.1 $, 1 and 0.9 respectively. A velocity bias larger than 1 slows down the decay of density bias slightly, while velocity bias less than 1 speeds it up. }
 \label{fig:linearbias}
 \end{figure} 

\subsubsection{Evolution of Linear Density and Velocity Bias}

We now turn to the solution of Eq.~(\ref{eq:Kn_generalsol}).  To linear order, only its first term contributes. From the decomposition into eigenmodes of the propagator, Eq.~(\ref{linearICs}), we can read off the evolution of each field, which is precisely of the local form given in the initial conditions but with a prescribed time dependence. For matter density and velocity fields we have linear growing-mode evolution,
\begin{equation}
 \mathcal{K}^{(1)}_1 (y) = \mathcal{K}^{(1)}_2 (y) = \e{y}, 
\end{equation}
while, for density and velocity bias we have, respectively, 
\begin{equation}
\label{eq:b1Evolution}
 b_1 \equiv \frac{\mathcal{K}^{(1)}_3 (y)}{\mathcal{K}^{(1)}_1 (y)} 
        =  1 + (b^*_1 - 1)\e {-y} + 2 (b_{\rm v}^* - 1) \e {-y} (1 - \e {-y/2}), 
\end{equation}
and 
\begin{equation} 
\label{eq:bvEvolution}
 b_{\rm v} \equiv \frac{\mathcal{K}^{(1)}_4 (y)}{\mathcal{K}^{(1)}_2 (y)} 
                 = 1 + (b_{\rm v}^* -1)\, \e{-3y/2}.  
\end{equation}
Note that when there is no velocity bias, $b_{\rm v}^* =1 $, we recover from Eq.~(\ref{eq:b1Evolution}) the  well-known result~\cite{MoWhi9609, Fry9604} 
\begin{equation}
\label{eq:FryBiasEvol}
  b_1 = 1 +  ( b^*_1  - 1  ) \, \e {-y},   
 \end{equation}
that density bias asymptotes to unity in the long-time limit if the comoving number density of tracers is conserved.  On the other hand, our generalization to $b_{\rm v}\ne 1$ does not agree with recent assumptions about the evolution of peaks in the initial density field~\cite{DesCroSco1011}, which do not show the presence of the iso-density-velocity mode contribution that gives the $\e{-3y/2}$ decay in Eqs.~(\ref{eq:b1Evolution}-\ref{eq:bvEvolution}). This disagreement results from different assumptions. 
Peaks move locally with the dark matter but their velocity statistics can be thought of as if they had a {\em statistical}  velocity bias that remains constant with evolution. Because of this difference in treatment, the peaks calculation cannot be directly compared to what we do here, although it is important to clarify which treatment is a more accurate description of velocity statistics of tracers. We hope to report on this in the near future.

Figure~\ref{fig:linearbias} shows $b_1$ and $b_{\rm v}$ as a function of the scale factor $a$. We have set $b^*_1=2$ and $b_{\rm v}^* = 1.1, 1$, and 0.9. 
Note that $b_{\rm v}^*>1$ slows down the relaxation of the density bias slightly while $b_{\rm v}^* <1$ speeds it up.   Velocity bias also relaxes to unity eventually.

\subsubsection{Quadratic Order: Emergence of Non-Local Bias}

Because the vertex $\gamma_{bcd}$ is quadratic in $\hat{\mb{k}}_1 \cdot  \hat{\mb{k}}_2  $, only $\ell\le 2$ multipole moments will be present.  For the matter density field,  \, 
the multipole expansion of $\mathcal{K}^{(2)}_1 (y)$ reads, 
\beqa
\label{F20}
       \mathcal{K}_{ 1, \ell=0}^{(2)} & =&  \frac{17}{21}\, \e {2y}, \\ 
\label{F21}
 \mathcal{K}_{1, \ell=1}^{(2)}  &=& \frac{1}{2} 
       \left( \frac{ k_1 }{k_2} + \frac{ k_2 }{k_1} \right)\, \e {2y} , \\
\label{F22}
 \mathcal{K}_{ 1, \ell=2}^{(2)} &=& \frac{4}{21}\, \e {2y}  .
\eeqa
These correspond to the multipole expansion of $\e{2y} F_2(\mb{k}_1, \mb{k}_2  )$. The monopole represents the second-order nonlinear growth in the spherical collapse dynamics,  the dipole the transport of matter by the velocity field, and the quadrupole describes tidal gravitational effects. 

The multipole moments for the galaxy density perturbation to second-order, $\chi^{(2)}_\ell$, are given by

\beqa
\label{chi20}
\chi^{(2)}_0 &=& \frac{b_2^*}{2} + {4\over 21}\epsilon_\delta + {2\over 21}\epsilon_{\rm v}\ 
\Big[3+14\e{y/2}-14\epsilon_{\rm v} + {21 \epsilon_\delta \over {\e{y}-1}}\Big]
 \nonumber \\ & & 
\\ 
\label{chi21}
\chi^{(2)}_1 &=&\Big(  \frac{k_1}{k_2} + \frac{k_2  }{k_1}   \Big) \ \epsilon_{\rm v}\ 
\Big[
-1+2\e{y/2}+{ \epsilon_\delta \over {\e{y}-1}}
\Big]
\\
\label{chi22}
\chi^{(2)}_2 &=& - {4\over 21}\epsilon_\delta +{4\over 21} \epsilon_{\rm v}\ 
\Big[
-12+14\e{y/2}+7 \epsilon_{\rm v}
\Big]
\eeqa
where $\epsilon_\delta$ and $\epsilon_{\rm v}$ are proportional to density and velocity bias, respectively:
\beq
\epsilon_\delta \equiv  (b_1-1)\ \e{y}\,(\e{y}-1), 
\label{epsilonD}
\eeq
\beq
\epsilon_{\rm v} \equiv  (b_{\rm v}-1)\ \e{y}\, (\e{y/2}-1), 
\label{epsilonV}
\eeq
and vanish for fully unbiased tracers. They are also defined to be zero at formation time ($y=0$), leaving only the prescribed monopole from local bias.  In the long-time limit ($y\to\infty$) they asymptote to 
\beq
 \epsilon_\delta \to \e{y}\, [(b_1^*-1)+2(b_{\rm v}^*-1)], \quad
 \epsilon_{\rm v} \to (b_{\rm v}^*-1).
\label{epsasymp}
\eeq
Thus, the effects proportional to $\epsilon_\delta$ dominate, but they are suppressed by another factor of $\e{y}$ compared to the usual second-order effects.  
Finally, note that when there is no velocity bias ($\epsilon_{\rm v}=0$) Eqs.~(\ref{chi20}-\ref{chi22}) reduce to,  
\beqa
\chi^{(2)}_0 = \frac{b_2^*}{2} + {4\over 21}\epsilon_\delta, \quad 
\chi^{(2)}_1 =0, \quad 
\chi^{(2)}_2 &=& - {4\over 21}\epsilon_\delta ,
\label{chinov}
\eeqa
with $\epsilon_\delta = (b_1^*-1)(\e{y}-1)$. In this case there is no dipole, and the induced quadrupole and monopole are opposite in sign. The induced structure when there is no velocity bias for conserved tracers is further explored in Section \ref{sec:BiasGalileon} to third-order in PT, and in Appendix~\ref{appA} for the non-conserved case. 

 \begin{figure}[!t]
 \centering
 \includegraphics[ width=0.95\linewidth]{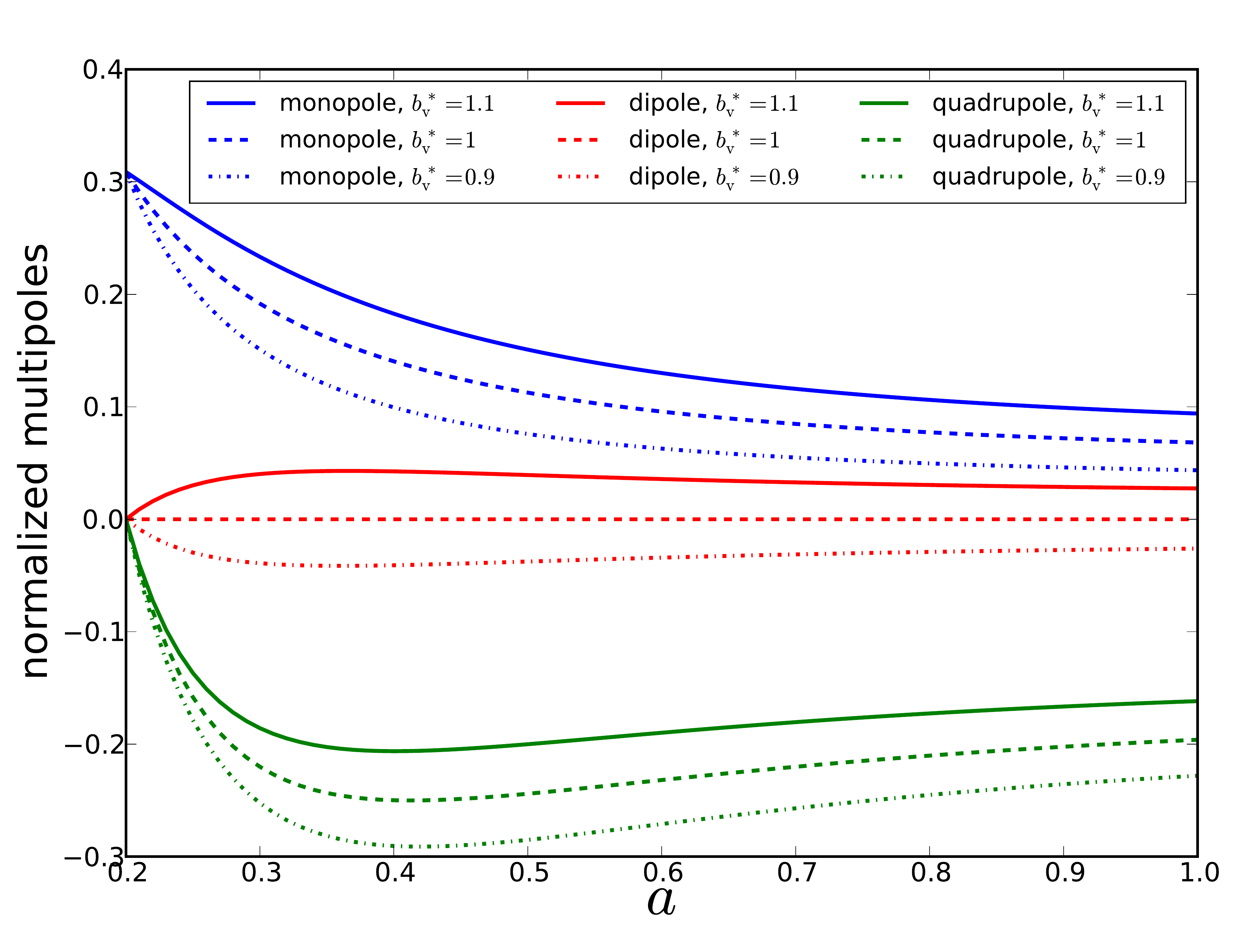}
 \caption{Emergence of non-local bias from local-bias initial conditions, as quantified by the evolution of the ratio of the galaxy multipoles $\chi^{(2)}_\ell$ (Eqs.~\ref{chi20}-\ref{chi22}) to the corresponding matter multipoles (Eqs.~\ref{F20}-\ref{F22}).  At formation ($a=0.2$), bias is local with $b_1^* = 2$ and $b_2^* =0.5$, i.e. there is only a monopole at second-order.  However, a quadrupole (three bottom lines) is generated at later times.  If there is velocity bias, then a dipole is also generated (three middle lines). The three lines for each multipole correspond to different choices for the initial velocity bias: $b_{\rm v}^*=1.1$ (solid), 1 (dashed) and 0.9 (dotted).}
 \label{fig:bias2vmultipoles} 
 \end{figure} 

Figure~\ref{fig:bias2vmultipoles} shows the evolution of the multipoles $\chi^{(2)}_\ell$ (normalized by their dark matter counterparts, Eqs.~\ref{F20}-\ref{F22}) as a function of $y$ for three different choices of velocity bias, $b_{\rm v}^*=1.1$ (solid), 1 (dashed) and 0.9 (dotted).  We see that even though the bias at formation ($a=0.2$) is local (only a monopole is present), higher-order multipoles get generated.  If there is no velocity bias then only a quadrupole gets generated; if $b_{\rm v}\ne 1$ then a dipole is also generated (with sign determined by $b_{\rm v}^*-1$).  All of these normalized multipoles eventually relax to zero because the galaxy multipoles grow more slowly than those of the dark matter. If $b_{\rm v}^* =1.1$, the relaxation of the monopole is slowed down,  whereas the quadrupole relaxes faster; the opposite holds for $ b_{\rm v}^* =0.9$. We see that even for a significant velocity bias of $10\%$, the generated dipole is only $10\%$ of that in the dark matter. A dipole contribution in galaxy bias can enhance the shift of the BAO peak in the correlation function~\cite{CroSco0801}, but since the dark matter dipole effect is at the percent level, velocity bias is unlikely to change this in any significant way except possibly for the very highly biased tracers. See Section~\ref{NumSim} for more discussion on the effects of such dipole term from numerical simulations.

 Thus, we see that non-local bias is inevitably induced by gravitational evolution, and that the locality assumption cannot be self-consistent.  In practice, because galaxy formation happens in a continuous fashion, we don't expect locality to be valid at any time, even if the formation bias were local.  We explore this in Section~\ref{sec:GalaxyFormation}.  In addition, there is no reason to expect the bias at formation to be purely local, even for dark matter halos, since the barrier for collapse is known to depend on quantities other than the overdensity (e.g.~\cite{SheMoTor01}).

\section{Comparison with Local Lagrangian Bias}
\label{sec:LagrangianBiasEvolution}

In this section, we would like to compare our results with those known  from the literature on local Lagrangian bias, which can be thought of as a particular limit of our results when formation time is at the far past ($z\to\infty$) and there is no velocity bias. While such calculations are usually done in Lagrangian PT (see e.g.~\cite{CatLucMat9807,CatPorKam0011}), clearly one should obtain the same results if done in Eulerian PT as we have used so far. It is however instructive to redo this calculation in a Lagrangian description and compare. 

Since there is no velocity bias and tracers are conserved, the continuity equations Eq.~(\ref{eq:DMcontinuity}) and Eq.~(\ref{eq:Galcontinuity}) can be used to relate $\delta_{\rm g}$ to $\delta$ through the matter velocity divergence field~\cite{CatLucMat9807},
\beq
 {d  \ln ( 1 +   \delta )  \over d\tau} = -\nabla\cdot \mb{u} = {d  \ln ( 1 +   \delta_{\rm g} )  \over d\tau}
 \label{contboth}
\eeq
where we used the Lagrangian or total derivative following the motion of a fluid element,
\beq
 {d\over d\tau} \equiv {\partial \over \partial \tau} + \mb{u}\cdot \nabla.
 \label{LagDer}
 \eeq
Upon integration of Eq.~(\ref{contboth}), we get 
\begin{equation} 
\ln [1 +   \delta_{\rm g}(\mb{x} )   ] =  \ln [ 1 +   \delta(\mb{x} )  ]  +  f( \mb{q} ), 
\end{equation} 
where $f(\mb{q}) $ is a function depending on initial fields at the Lagrangian coordinate $\mb{q} $ related to the Eulerian $\mb{x} $ through the displacement field 
\begin{equation}
\mb{x}(\mb{q},t) = \mb{q} + \Psi(\mb{q},t).  
\label{qmapx}   
\end{equation}
In terms of the initial condition $f(\mb{q} )  $ is clearly given by  $ \ln \{\, [1 +   \delta_{\rm g}(\mb{q}) ]\, /\, [ 1 +   \delta(\mb{q} )  ]\,\}  $ and thus we have,
\begin{equation} 
\label{eq:galaxymattereqLagrangian}
1 +   \delta_{\rm g}(\mb{x} )  =  \left( \frac{ 1 +   \delta_{\rm g}(\mb{q} )  } {  1 +   \delta(\mb{q} )   } \right) \  [ 1 +   \delta(\mb{x} )  ],
\end{equation} 
where the Lagrangian fields are evaluated at the initial time $t_*$. 
This is the same result as that given in~\cite{CatLucMat9807,CatPorKam0011,DesCroSco1011} except for the denominator $(1+ \delta (\mb{q})  )$, which was implicitly dropped in those works (it is however included in~\cite{BarMatRio1011}). However, to reproduce the decaying modes found in the previous section, this denominator is required.  

For comparison with the results in the previous section, we now assume local Lagrangian bias in Eq.~(\ref{eq:galaxymattereqLagrangian}) to quadratic order, 
\begin{equation}
1 + \delta_{\rm g} ( \mb{x} ) = [1 + \delta( \mb{ x}) ] 
 \frac{1 + b_1^{\rm L } \delta (\mb{q} ) + (b_2^{\rm L}/2)\,\delta^2(\mb{q}}
      {1 + \delta(\mb{q})}, 
\label{dgdLag}
\end{equation}
where the Lagrangian bias parameters $b_i^{\rm L}$ are the equivalent to the parameters $b_i^*$ in the previous section. To linear order, we can assume $\mb{q} \simeq \mb{x}$ in this equation, but to go to second order we need to include the displacement field to first order, i.e. in  the Zel'dovich approximation~(hereafter ZA, \cite{Zel7003}).  This is given by 
\begin{equation}
\Psi( \mb{q},t ) = \frac{D(t) - D(t_*)}{D(t_*)} \int \diff ^3 k \left(\frac{i  \mb{k}}{ k^2 }\right) \delta(\mb{k},t_*)\ \e { i \mb{k}\cdot \mb{q} }, 
\label{ZAdispl}
\end{equation}
where we have used the fact that, in the ZA, the decaying mode is constant~\cite{Sco98}.  Note that $ \Psi( \mb{q},t_* ) = 0$ at formation time $t_*$, as it should.  We emphasize again that this decaying mode, which is often neglected in the literature, must be included if one wishes to fully reproduce the results in the previous section to second order. 

To linear order Eq.~(\ref{dgdLag}) reads $\delta_{\rm g}^{(1)} \simeq \delta^{(1)}+b_1^{\rm L}\, \delta_*^{(1)}-\delta_*^{(1)}$, where all fields have the same argument $\mb{x}$. Therefore we deduce the Eulerian linear bias $b^{\rm E}_1 $
\begin{equation}
b^{\rm E}_1 = 1 + \frac{b_1^{\rm L} - 1   }{ {D / D_*}  } , 
\label{linLag}
\end{equation}
which is Eq.~(\ref{eq:FryBiasEvol}), with $y\equiv \ln  (D / D_*)$.  This seems different a priori from the often quoted relationship between linear Eulerian and Lagrangian bias $b^{\rm E}_1 = 1 + b_1^{\rm L}$. The reason is twofold: first, it is customary to define the Lagrangian bias with respect to the {\em extrapolated} linear density field  $\delta^{(1)}$ rather than the Lagrangian density field $\delta_*^{(1)}$ as we have done here, so the more standard definition is  instead
\beq
\tilde{b}^{ \rm L}_1 \equiv b_1^{\rm L} \, \left({D_*\over D}\right)
\label{b1Llit}
\eeq
and second, if we neglect the third term in Eq.~(\ref{linLag}) coming from the denominator in Eq.~(\ref{eq:galaxymattereqLagrangian}), then we recover the familiar $b^{\rm E}_1 = 1 + \tilde{b}_1^{\rm L}$. This second step is justified  for objects that are not arbitrarily close to unbiased in which case as $t_*\to 0$ and $D/D_*\to \infty $,  $b_1^{\rm L}$ increases without bound for objects with fixed  $b^{\rm E}_1$, so that $b_1^{\rm L}\gg 1$ in this limit. Although this step is unjustified for unbiased objects for which $b_1^{\rm L}= 1$, keeping only this term does no harm as its contribution to Eq.~(\ref{linLag}) vanishes as $t_*\to 0$. 

To second order in PT, Eq.~(\ref{dgdLag}) gives, after using Eq.~(\ref{qmapx}), 

\begin{widetext}

\begin{equation}
\delta_{\rm g}^{(2)} (\mb{x}) = 
\left[    1 + \frac{b_1^{\rm L} - 1   }{  \left( D / D_* \right)^2 }   \right]   \delta^{(2)}( \mb{x})   -   
\left(  \frac{b_1^{\rm L} - 1 }{ D / D_* }  \right)   \Psi \cdot \nabla \delta^{(1)}( \mb{x} ) +   
\left[  \frac{b_1^{\rm L} - 1 }{ D / D_*  }  + \frac{  b_2^{\rm L} / 2  -   b_1^{\rm L} + 1 }{  \left(D / D_* \right)^2 }      \right] ( \delta^{(1)} ( \mb{x} )  )^2  , 
\end{equation} 
which upon Fourier transform, after using Eq.~(\ref{ZAdispl}), we can write as the quadratic kernel for galaxies 
\begin{equation} 
\label{eq:K2Lagrangian}
\left({D_*\over D}\right)^2\, \mathcal{K}_3^{(2)}= 
\left[  \frac{5 }{ 7}  +  \frac{ b_1^{\rm L} - 1 } { \left( D / D_* \right) }  +   \frac{  \frac{b_2^{\rm L}}{ 2 } - \frac{2 }{ 7 }( b_1^{\rm L} - 1 )}{  \left( D / D_* \right)^2 }  \right] 
 +  
\left[ 1 +  \frac{b_1^{\rm L} - 1 }{ \left( D / D_* \right)  }   \right] \frac{\mu  }{2 } \left( \frac{k_1}{k_2} +   \frac{k_2}{k_1}  \right) 
 + 
 \frac{2}{7} \left[     1 + \frac{b_1^{\rm L} - 1   }{ \left( D / D_* \right)^2 }      \right] \mu^2 
\end{equation}
\end{widetext}
where we have used the second-order matter results  Eqs.~(\ref{F20}-\ref{F22}), and $\mu \equiv \hat{\mb{k}}_1 \cdot \hat{\mb{k}}_2$. It is easy to check that this equation agrees with Eq.~(\ref{chinov}) with $\epsilon_\delta = (b_1^{\rm L}-1)(\e{y}-1)$, after using that $\chi^{(2)} =  \mathcal{K}_3^{(2)} - b_1^{\rm E}\, \mathcal{K}_1^{(2)}$ with $b_1^{\rm E}$ given by Eq.~(\ref{linLag}) and $y =\ln(D/D_*)$.  

We can now take the limit $t_*\to 0$ in Eq.~(\ref{eq:K2Lagrangian}) as for the linear result above and compare with the results in the literature. Now we need to redefine the quadratic Lagrangian bias in terms of the present density fluctuations, as done for the linear bias in Eq.~(\ref{b1Llit}),
\beq
\tilde{b}^{ \rm L}_2 \equiv b_2^{\rm L} \, \left({D_*\over D}\right)^2
\label{b1Llit}
\eeq
and assuming $b_1^{\rm L}\gg 1$ as before we then get for Eq.~(\ref{eq:K2Lagrangian}), 
\begin{equation}
\label{eq:Catelan}
\left( \frac{5 }{ 7}  +  \tilde{ b}_1^{\rm L}      +   \frac{ \tilde{b}_2^{\rm L}  } { 2 }  \right)
 +  
 \left( 1 +  \tilde{b}_1^{\rm L}   \right) \frac{\mu  }{2 } \left( \frac{k_1}{k_2} +   \frac{k_2}{k_1}  \right) 
 + 
 \frac{2}{7}  \mu^2, 
\end{equation}
which agrees with Eq.~(8) in~\cite{CatPorKam0011}. Note however that in~\cite{CatPorKam0011}  it is argued that the dipole term (proportional to $\tilde{ b}_1^{\rm L}$ in Eq.~\ref{eq:Catelan}) is a new feature of local Lagrangian bias as opposed to local Eulerian bias. This interpretation is not correct:  in local Eulerian bias the second-order galaxy kernel is, apart from local contributions of quadratic bias, that of the matter multiplied by linear bias
 $b_1^{\rm E} = 1 +  \tilde{b}_1^{\rm L}$, 
so the precise amplitude of the dipole agrees with that in Eq.~(\ref{eq:Catelan}). Subtracting this local Eulerian piece to construct
 $\chi^{(2)} =  \mathcal{K}_3^{(2)} - b_1^{\rm E}\, \mathcal{K}_1^{(2)}$, 
shows that the new contributions are indeed of the form given by Eq.~(\ref{chinov}).  That is, the new qualitative contribution is a quadrupole term, not a dipole. As we showed in the previous section, an additional dipole will only appear if there is velocity bias.  The physical reason for this (breaking of the Galilean invariance of the bias relation) is discussed in Section~\ref{sec:BiasGalileon}.

Thus, we have shown that, in the appropriate limit, we reproduce known local Lagrangian bias results.  However, our approach in the previous section is more flexible as it does not require ``formation" to be in the distant past, and it also allows for velocity bias.  On the other hand, we have, so far, assumed that the comoving density of tracers is conserved.  We now discuss how to go beyond this assumption.

 \begin{figure*}[!t]
 \centering
 \includegraphics[ width=0.95\linewidth]{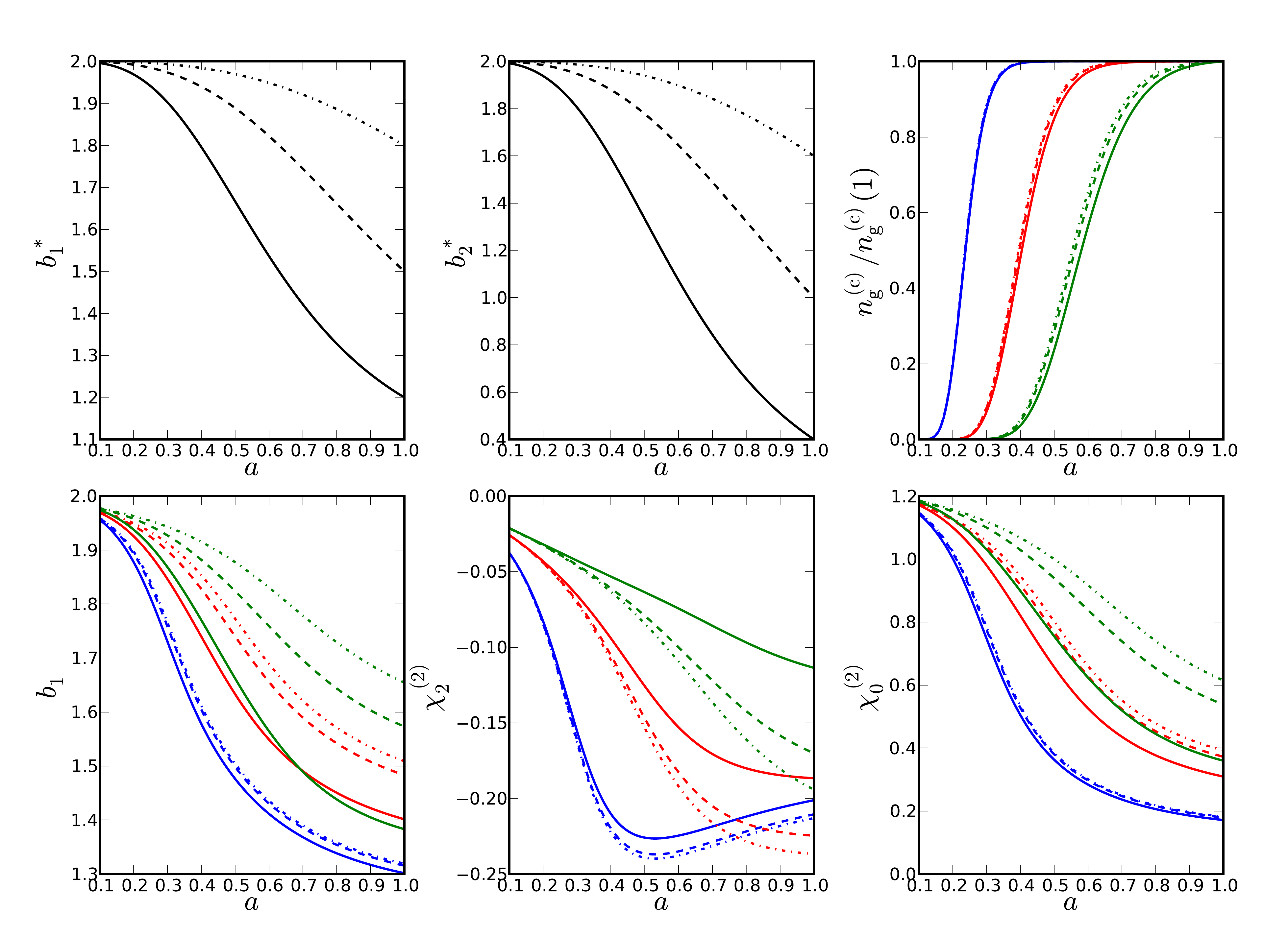}
 \caption{ Local bias parameters at formation $b_1^*$ (top-left panel) and $b_2^*$ (top-center), comoving number density (normalized by present value, top-right), 
linear bias (bottom-left), second-order galaxy bias quadrupole $\chi^{(2)}_2$  and monopole $\chi^{(2)}_0$ (normalized by dark matter values, bottom center and right panels) as a function of scale factor $a$. Each panel shows  three sets of values of $\alpha_1$ and $\alpha_2$, corresponding to $\{\alpha_1,\alpha_2\}=\{4, 1\}$ (solid), $\{1, 1\}$ (dashed), and $\{1, 4\}$ (dot-dashed).  To describe galaxy formation, we have used the toy model in Eq.~(\ref{profileA}) with  $\sigma_0=0.2$ and characteristic galaxy formation time $a_0$ equal to 0.3 (blue), 0.5 (red) and 0.7 (green) respectively.        }
 \label{fig:GFQsource}
 \end{figure*}

\section{Non-local bias generation with non-conserved tracers}
\label{sec:GalaxyFormation}

Galaxies form at a range of redshifts and merge.  So it is important to extend the previous results to the more realistic case when the comoving number density of galaxies changes with redshift due to some arbitrary source field $j$, which effectively includes the effects of galaxy formation and merging.  Our description here is similar to~\cite{TegPee9806} (see also~\cite{MatColLuc9703,MosColLuc9808}), but we shall extend the analysis to higher order in PT. For simplicity here we assume that the bias at formation is local (as we have done so far), Appendix~\ref{appA} discusses what happens in the most general case (see also Eq.~\ref{Nlocdgstar} below). 
The evolution equation for the physical galaxy number density $n_{\rm g} $ now becomes 
\begin{equation}
\frac{\partial n_{\rm g} }{ \partial \tau }  +  3 {\cal H} n_{\rm g} +   \nabla \cdot (  n_{\rm g} \mb{u} ) = A {\cal H}  j( \rho ).
\label{contWsource}
\end{equation}
Note that we factorized the source term into two components, $A$ and $j$, where $A$ roughly parametrizes the epoch of galaxy formation (e.g. following star formation history) and $j$  describes the effects of dark matter on the formation and merging of galaxies. Nonetheless, we stress that our main results are independent of the detailed functional form of $A$ and $j$, and we use the assumed functional forms only to make the plots shown in Fig.~\ref{fig:GFQsource}. For example, $A(t)$ can be a log-normal profile 
\begin{equation}  
A(t) = \frac{1}{a^3}  \e {- \frac{( \ln a - \ln a_0)^2 }{2 \sigma_0^2  } }, 
\label{profileA}
\end{equation} 
where $a_0 $ and $\sigma_0 $ are free parameters. For $j( \rho ) $  we take a simple quadratic form 
\begin{equation}
\label{eq:SourceJForm}
j( \rho ) = \alpha_1 \frac{\rho}{\rho_{ 0 } } +  \alpha_2 \left( \frac{\rho}{\rho_{ 0 } } \right)^2, 
\end{equation}
where $\alpha_1 $ and  $\alpha_2 $ are model parameters, and $\rho_{0}$ is the average matter density today. Appendix~\ref{appA} considers the implications of $j$ depending on non-local functions of $\delta$, or other fields. 

The second  term in Eq.~(\ref{contWsource}) can be eliminated if we use comoving rather than physical number densities, $n_{\rm g}^{\rm(c)} \equiv a^3  n_{\rm g} $, so we have 
\begin{equation} 
\label{eq:continuity_source}
\frac{\partial n_{\rm g}^{(\rm{c} )}  }{ \partial \ln a }  + \frac{1}{{\cal H}} \nabla \cdot (  n_{\rm g}^{(\rm{c} )}  \mb{u} ) = A^{\rm (c)}  j( \rho ),   
\end{equation}
where  $A^{\rm (c)} \equiv a^3 A $.  We then write,
\beq
n^{(\rm c)}_{\rm g}= \bar{n}_{\rm g}^{\rm(c)  }( 1 + \delta_{\rm g}^{(1)} + \delta_{\rm g}^{(2)})
\label{deltag}
\eeq
and solve Eq.~(\ref{eq:continuity_source}) by perturbation theory. We will also assume that there is no velocity bias, so that galaxies and matter share the same velocity field, which is known from solving the evolution of matter. We then expand the source term on the right hand side of Eq.~(\ref{eq:continuity_source}) to second order
\beq
A^{\rm (c) }  j({\rho}) \simeq A^{\rm (c) }  j(\bar{\rho})\Big[ 1  + \frac{j'( \bar{\rho} ) \bar{\rho} }{ j(\bar{\rho})}  \delta + \frac{1}{2}  \frac{ j''(\bar{\rho} ) \bar{\rho}^2  }{ j(\bar{\rho} ) }  \delta^2 \Big].
\label{jexpand}
\eeq
This invites us to interpret $ j'( \bar{\rho} ) \bar{\rho} /  j(\bar{\rho}) $ and  $ j''(\bar{\rho} ) \bar{\rho}^2 / j(\bar{\rho} )    $  as the instantaneous formation bias $b_1^*(t)$ and  $b_2^*(t)$ of the galaxies formed (or destroyed) at time $\ln a$
\begin{equation}
b_1^*(t) \equiv  \frac{j'( \bar{\rho} ) \bar{\rho} }{ j(\bar{\rho})} , \quad 
b_2^*(t) \equiv  \frac{ j''(\bar{\rho} ) \bar{\rho}^2  }{ j(\bar{\rho} ) }. 
\end{equation}
For example, the form of $j$ in Eq.~(\ref{eq:SourceJForm}) gives  
\begin{equation}
b_1^*(a) = 1 + \frac{\alpha_2 }{\alpha_1 a^3  + \alpha_2 }, \quad  b_2^*(a) = \frac{ 2 \alpha_2 }{\alpha_1 a^3  + \alpha_2 },  
\label{eq:b_a12}
\end{equation}
which imply a simple relation $b_1^* -1 = b_2^* / 2 $. This relation only holds at the formation time, as we will see evolution inevitably generates non-locality and breaks this. Note that in Eq.~(\ref{eq:b_a12})  the bias parameters have a pole when $\alpha_2$ is negative. This means that quadratic approximation is no longer valid and the higher order terms in the expansion are important. 

Figure~\ref{fig:GFQsource} shows in the top left and center panels $b_1^*$ and $b_2^* $ for three sets of values of $\alpha_1$ and $\alpha_2$, corresponding to $\{\alpha_1,\alpha_2\}=
\{4, 1\}$ (solid), $\{1, 1\}$ (dashed), and $\{1, 4\}$ (dot-dashed). As we mentioned before, these choices are just illustrative with no special physical significance, but serves to show a range of possibilities for the local biases at formation.

\subsection{Background Solution} 

We now look for the evolution of the galaxy comoving number density $\bar{n}_{\rm g}^{\rm (c)} $ by solving the background equation, 
\begin{equation}
\label{eq:background_eq}
\frac{\diff \bar{n}_{\rm g}^{\rm (c)  }   }{ \diff \ln a } = A^{\rm(c)}   j( \bar{\rho} ), 
\end{equation}
where $\bar{\rho } $ is the mean matter density. The solution is 
\begin{equation} 
\bar{n}_{\rm g}^{\rm(c)} = \int_{\ln a_{\rm ini} }^{\ln a  } \diff (\ln a )  \,  A^{(\rm c)} j( \bar{\rho} ),  
\end{equation} 
where we have assumed that there are no galaxies at $t=t_{\rm ini  }$.  The top right panel in Fig.~\ref{fig:GFQsource} shows the resulting $\bar{n}_{\rm g }^{\rm (c)}$ for three sets of parameters $\alpha_1 $ and $ \alpha_2$ in  three different backgound profiles $A$. As expected the background number density $\bar{n}_{\rm g }^{\rm (c)}$ is predominantly determined by the profile $A$.

\subsection{First-Order: The Evolution of Linear Bias}

To first order in PT we write $ n_{\rm g}^{\rm (c)}= \bar{n}_{\rm g}^{\rm (c)}( 1 + \delta_{\rm g}^{(1)}  )$ in Eq.~(\ref{eq:continuity_source}),  and using the background evolution in Eq.~(\ref{eq:background_eq}), we get 
\begin{equation} 
\label{eq:firstorder}
\frac{\partial \delta_{\rm g}^{(1)} }{ \partial  \ln a  }  + \frac{A^{\rm (c)}    }{\bar{n}_{\rm g}^{(\rm{c})}  }  j(\bar{\rho})  \delta_{\rm g}^{(1)}   =   -\frac{1}{ {\cal H} } \nabla \cdot \mb{u}^{(1)} +  \frac{ A^{\rm (c)}  }{\bar{n}_{\rm g}^{(\rm{c})} } j'(\bar{\rho}  ) \bar{\rho}  \delta^{(1)}, 
\end{equation}
which using linear theory evolution for matter with growth factor $D$ can be rewritten as 
\begin{equation} 
\frac{ \partial  \delta_{\rm g}^{(1)} }  {\partial \ln a }  + \frac{ A^{\rm (c)  } } { \bar{n}_{\rm g}^{( \rm{c} )}}  j(\bar{\rho} )   \delta_{\rm g}^{(1)}    =  \big(  f  +  \frac{ A^{\rm (c) } }{ \bar{n}_{\rm g}^{( \rm{c} )}} j'(\bar{\rho} ) \bar{\rho}   \big) D\, \delta_0 .   
\end{equation}
where $ f =  \diff \ln D / \diff \ln a $. Looking for a solution of the form $ \delta_{\rm g}^{ (1)} = D_{\rm g}(t)\, \delta_0  $, where $D_{\rm g}$ is the linear growth factor for galaxies, and after using Eq.~(\ref{eq:background_eq}), we arrive at      
\begin{equation}
\frac{ \diff  }{ \diff \ln a } (\bar{n}_{\rm g}^{\rm  (c )}  D_{\rm g}  ) = \big(  \bar{n}_{\rm g}^{\rm  (c )} f + A^{\rm (c)} j'( \bar{ \rho} ) \bar{ \rho}    \big) D. 
\label{eq:firstorderC1}
\end{equation}
And since $n_{\rm g}^{\rm  (c )}=0$ initially, we have 
\begin{equation}
\label{eq:C1sol} 
D_{\rm g}(t)  =  D(t) + \frac{1}{\bar{n}_{\rm g}^{\rm (c)} }  \int_{\ln a_{\rm ini}  }^{\ln a } \diff ( \ln a ) A^{\rm (c)}  D 
 \big(  -  j (\bar{ \rho })  + j'(\bar{ \rho} ) \bar{\rho} \big), 
\end{equation}
which agrees with~\cite{TegPee9806} after some simple redefinitions. We can express Eq.~(\ref{eq:C1sol}) in a more physical way that makes clear the independence of the detailed form of $A^{\rm (c)}$ and $j$, by using the comoving number density of galaxies as the integration variable,   
\begin{equation}
\label{eq:D1sol}
D_{\rm g}(t)  =  D(t) + \frac{1}{\bar{n}_{\rm g}^{\rm (c)} }    \int_{ 0 }^{  \bar{n}^{\rm(c)}_{\rm g} } \diff n_*  \,   \big(    b_1^*- 1   \big)  \ D_*, 
\end{equation}
which gives for the effective linear bias, $b_1\equiv D_{\rm g} / D $
\begin{equation}
\label{eq:b1evolvesource}
b_1(t)  =  1 + \frac{1}{\bar{n}_{\rm g}^{\rm (c)} D  }   \int_{ 0 }^{  \bar{n}^{\rm(c)}_{\rm g} } \diff  n_*\,     \big(  b_1^* -1 \big)  \ D_* , 
\end{equation}
where quantities inside the integral are evaluated at a time when the comoving number density of galaxies equals $n_*$. This equation is the generalization of Eq.~(\ref{eq:FryBiasEvol}) (with $y=\ln D/D_*$) for when galaxies form during a broad range of redshifts and are not necessarily conserved. We have assumed for simplicity that the relationship $\bar{n}^{\rm(c)}_{\rm g}$ to $\ln a$ is one-to-one, if not (because merging overcomes formation at some periods of time), one should just sum up over all contributions at a given value of $n$ in Eq.~(\ref{eq:b1evolvesource}). The key idea is that the simple result of the conserved-tracers case,  Eq.~(\ref{eq:FryBiasEvol}), gets generalized by simply weighting by instantaneous comoving number density. 

The form of Eq.~(\ref{eq:b1evolvesource}) is not practical for observational purposes since it is not feasible  to trace back all the way to when there were no galaxies.  A more useful  form is to use initial data at some (high) redshift when the growth factor was $D_i$ and there was some non-zero comoving number density $\bar{n}_{\rm gi}^{\rm (c)}$ with linear bias $b_{1i}$ .  One can then rewrite Eq.~(\ref{eq:b1evolvesource}) as,
\beq
\label{eq:b1evolvesource2}
b_1(t)  =  1 +  {\bar{n}_{\rm gi}^{\rm (c)} \over \bar{n}_{\rm g}^{\rm (c)} }\, (b_{1i}-1) 
{ D_i \over  D}+ 
\frac{1}{\bar{n}_{\rm g}^{\rm (c)} D  }   \int_{ \bar{n}_{\rm gi}^{\rm (c)} }^{  \bar{n}^{\rm(c)}_{\rm g} } \diff  n_*\,     \big(  b_1^* -1 \big)  \ D_* ,
\end{equation}
which follows from combining Eq.~(\ref{eq:b1evolvesource}) at times $t_i$ and $t$. This also makes rather clear that if the comoving number density does not change between $t_i$ and $t$ ($\bar{n}_{\rm gi}^{\rm (c)} = \bar{n}_{\rm g}^{\rm (c)} $), the last term vanishes and one recovers Eq.~(\ref{eq:FryBiasEvol}). When tracers are not conserved, we see that the second term in Eq.~(\ref{eq:b1evolvesource2}) is simply the usual (conserved case) decay of bias modulated by the evolution of $\bar{n}_{\rm g}^{\rm (c)}$ with  the result that this decay is slowed down by merging (decrease of $\bar{n}_{\rm g}^{\rm (c)}$). In addition, the third term provides an extra contribution that depends on the bias of galaxies that are formed or lost to merging, and the sign of this contribution depends on whether formation or merging dominates. 

Another way to see these effects is to go back to Eq.~(\ref{eq:firstorderC1}) and write an explicit differential equation for the linear bias,  by replacing $D_{\rm g}$ with $ b_1(t)\,  D $, which gives 
\begin{equation}
\label{eq:beff_evolvesource}
\frac{ \diff  } { \diff  \ln a }( b_1 - 1 )  = - ( b_1 - 1 )f   -  \frac{\diff \ln \bar{n}_{\rm g }^{\rm (c)} }{ \diff \ln a  }    \big(  b_1 - b_1^*  \big).    
\end{equation}
The first term on the right hand side is the usual term that drives the decay of bias due to the growth of the large scale structure. Indeed, neglecting the second source term one can solve Eq.~(\ref{eq:beff_evolvesource}) to recover Eq.~(\ref{eq:FryBiasEvol}). The second source term may speed up or slow down the decay of bias compared to the conserved model, depending on the signs of $ ({\rm d}\bar{n}_{\rm g }^{\rm (c)} /{\rm d} a ) $  (i.e. whether formation or merging dominates) and  $(b_1- b_1^*)$ specifying whether the galaxies created or destroyed are more or less biased than the overall bias.

In the bottom left panel of Fig.~\ref{fig:GFQsource}, we plot the the evolution of $b_1 $ for several choices of our parameters. It is interesting to note that as the characteristic galaxy formation time $a_0$ increases, the difference between the three models represented by $\{\alpha_1,\alpha_2\}$  (different line types in Fig~\ref{fig:GFQsource})  becomes more marked. From Eq.~(\ref{eq:b_a12}) we see that if $\alpha_2 $ is small compared to $\alpha_1$, then there is some epoch  where $(b_1 - b_1^* )$ is positive, and this can accelerate the decay of bias.  For the model $\{\alpha_1,\alpha_2\}= \{4 ,1\}$ (solid line in Fig.~\ref{fig:GFQsource}),  $(b_1- b_1^*  )$ becomes positive roughly when $a \gtrsim 0.5$  and the decay of bias is speeded up. This is particularly apparent for the model with $ a_0 = 0.7 $ (solid green) as galaxy formation occurs when  $(b_1- b_1^*)$ is positive.

Current galaxy surveys are deep enough to test the evolution of linear bias, something that will be done much more precisely in the near future. Comparison with theoretical predictions are often limited to the conserved-tracers case, Eq.~(\ref{eq:FryBiasEvol}), typically showing a decay of bias that is faster than predicted by this formula 
~\cite{ColKlyKra9909,MarLe-Men0511,WhiZheBro0702,BroZheWhi0808,WakCroSad0812,TojPer1007,KovPorLil1104}. These  deviations are interpreted in terms of merging and/or disruption of galaxies, and our more general formula, Eq.~(\ref{eq:b1evolvesource2}), can help understand the implications of such measurements for merging/formation rates.

\subsection{Second-Order: Generation of Non-Local Bias}

To find the second-order solution, we write $ n_{\rm g}^{(\rm{c} )} = \bar{n}_{\rm g}^{\rm(c)}( 1 + \delta_{\rm g}^{(1)}  +   \delta_{\rm g}^{(2)} ) $ in 
Eq.~(\ref{eq:continuity_source}), and using Eq.~(\ref{eq:background_eq}) and (\ref{eq:firstorder}), we obtain 
\begin{widetext}
\begin{equation}
\label{eq:secondorder}
 \frac{\partial \delta_{\rm g}^{(2)} }{ \partial \ln a }  +  \frac{ A^{\rm (c)} } {\bar{n}_{\rm g}^{(\rm{c})}  }  j(\bar{\rho}) \delta_{\rm g}^{(2)}   =   -\frac{1}{ {\cal H}} \nabla \cdot \mb{u}^{(2)}  -   \frac{1}{ {\cal H} } \nabla \cdot (  \delta_{\rm g}^{(1)}  \mb{u}^{(1)}   )   
 +   \frac{  A^{\rm (c)}  }{\bar{n}_{\rm g}^{(\rm{c})} } \Big[  j'(\bar{\rho}  ) \bar{\rho}  \delta^{(2)}  + \frac{1}{2} j''(\bar{\rho} ) \bar{\rho}^2 ( \delta^{(1)} )^2  \Big].    
\end{equation}
\end{widetext}

We then rewrite the source terms as before, in terms of $\bar{n}_{\rm g}^{\rm (c)}$, $b_1^*$ and $b_2^*$,  substract the linear bias contribution at second-order to define $\chi^{(2)}$ as in Eq.~(\ref{chidef}), and decompose it in terms of Legendre polynomials as in Eq.~(\ref{LegDec}).  This gives differential equations for the $\chi_\ell^{(2)}$. These can be integrated as before (see Eq.~\ref{eq:firstorderC1}) by using $\bar{n}_{\rm g}^{\rm (c)}$ as the integrating factor, to obtain (compare with Eq.~\ref{chinov})

\begin{eqnarray}   
\label{eq:D2source}
 \chi^{(2) }_{2} & = -& \frac{4}{21 \bar{n}_{\rm g}^{(c)} }   \int_{0}^{\bar{n}_{\rm g}^{\rm (c)} } \diff n_*   D_*(D-D_*)  ( b_1^* -1)    ,        \\ 
\label{eq:D1source}
\chi^{(2)}_{1} &=&  0 ,\\
\label{eq:D0source}
\chi^{(2)}_{0}& = &  -    \chi^{(2) }_{2}  + \frac{1}{\bar{n}_{\rm g}^{\rm (c)} }  \int_{0}^{\bar{n}_{\rm g}^{\rm (c)} } \diff  n_* D_*^2\,  \left({b_2^*\over 2}\right) . 
\end{eqnarray}

Note that the dipole moment is exactly zero,  as in the conserved case, irrespective of the functional form of the source term. In fact, we can recover the results of Eq.~(\ref{chinov}) by inserting $\Ddel ( n_*-   \bar{n}_{\rm gi}^{\rm (c)} )\, \bar{n}_{\rm gi} $ in the integral and noting that $n_{\rm g}^{\rm (c)} =  n_{\rm gi}^{\rm (c)}$ for the conserved-tracers case,  and $D_*\equiv 1$ in Section \ref{sec:EulerianBiasEvolution}.  Figure~\ref{fig:GFQsource}  shows the evolution of $ \chi^{(2) }_{2} $ (bottom center panel) and  $\chi^{(2)}_{0}$ (bottom right), normalized  by the corresponding dark matter multipole coefficients in Eq.~(\ref{F22}) and (\ref{F20}), respectively, as a function of scale factor $a$ for the different choices of the parameters that describe $A$ and $j$. 

It is interesting to ask if we can avoid inducing non-local bias  by a proper choice of sources.  Since locality is violated by the induced quadrupole term, we look for a solution of the equation $\chi^{(2)}_{2} = 0$.  That is,
\begin{equation}
  D \int_{0}^{\bar{n}_{\rm g}^{\rm (c)} } \diff n_* D_*  ( b_1^*-1 )          
  =  \int_{ 0 }^{ \bar{n}_{\rm g}^{\rm (c)} } \diff n_* D_*^2   (b_1^*-1 ).
\end{equation}
Differentiating this equation with respect to $\bar{n}_{\rm g}^{\rm (c)}  $ leads to
\begin{equation}
\int_{0 }^{\bar{n}_{ \rm g}^{\rm (c)} } \diff n_* D_* ( b_1^*-1) = 0 .
\end{equation}
Together with Eq.~(\ref{eq:D1sol}), this implies that $D_{\rm g} = D$, i.e. the galaxies must be unbiased. Hence biased tracers generically have non-local bias irrespective of their merging/formation history. 

In Appendix~\ref{appA} we study non-conserved tracers in more generality and show that the non-local bias can be related in a simple manner to that in the conserved case to arbitrary order.  Since the lack of conservation of tracers does not qualitatively change the structure of the non-local bias, we now go back to the conserved case and discuss the basic features that determine this structure.

\section{ Structure of the non-local bias to third order}
\label{sec:BiasGalileon}

In this section, we derive the form of the non-local bias to third order. In principle, the general solution Eq.~(\ref{eq:Kn_generalsol}) developed in Section~\ref{sec:EulerianBiasEvolution} (and generalized to non-conserved tracers in Appendix~\ref{appA}) would allow us to write down the solution immediately. However, as we see in Eq.~(\ref{chinov}),   $\chi_\ell^{(2)}$ exhibits an interesting structure when there is no velocity bias (preserved when going to non-conserved tracers, see Eqs.~\ref{eq:D2source}-\ref{eq:D0source}).  The full third-order Fourier-space solution is rather complicated, and naively applying Eq.~(\ref{eq:Kn_generalsol}) to  third order hardly gives us any insight. We find  it  more instructive to develop the solution in a different way.  

For simplicity we assume conserved tracers and no velocity bias, so the matter velocity field $\mb{u}$ controls the bias relation (see Appendix~\ref{appA} for generalization beyond these assumptions). We assume the velocity field is potential, and, as discussed in Section~\ref{sec:EulerianBiasEvolution}, it is convenient to work with a normalized velocity field
\beq
\mb{v}\equiv-{\mb{u}\over f{\cal H}}= \nabla \Phi_{\rm v}, \ \ \ \ \ \ \ \ \nabla\cdot \mb{v} =\nabla^2 \Phi_{\rm v} \equiv \theta_{\rm v},
\label{normvel}
\eeq
whose divergence agrees with density perturbations in linear theory,
 $ \theta_{\rm v}^{(1)} = \delta^{(1)}$.
We assume local bias at formation time $t_*$ (although see  Eq.~\ref{Nlocdgstar} below)
\beq
\delta_{\rm g}^*=b_1^*\, \delta_* + {b_2^*\over 2!}\, \delta_*^2 + {b_3^*\over 3!}\, \delta_*^3 + \ldots 
\label{ICsdeltag}
\eeq
 and look for the evolution of the bias relation, which can be divided into local bias evolution and induced non-local terms. Because the results to third order are complicated, we start by doing  the calculation in the Zel'dovich approximation (ZA), where it is easier because the dynamics of gravitational instability is local, and then extend it to the exact dynamics. This serves to highlight the similarities and differences from the exact dynamics, and how its non-locality affects the bias relation.

In the ZA, it is simple enough to see what to expect. The dynamics is given by the displacement field $\Psi$ that scales linearly with the growth factor (see Eqs.~\ref{qmapx} and~\ref{ZAdispl}). The displacement field is related directly to the velocity potential by $\Psi(\mb{q})= -\nabla \Phi_{\rm v}$, so the bias relation will be determined at large scales entirely by the second derivatives $\nabla_{ij} \Phi_{\rm v}$, which measure the variations in displacements that affect the clustering. The non-locality in the bias relation will then be determined by the scalar invariants of $\nabla_{ij} \Phi_{\rm v}$, since the galaxy density perturbation is a scalar under 3D rotations and translations. 

In three dimensions, there are only three principal invariants of $\nabla_{ij} \Phi_{\rm v}$.  These are the ``Galileons"~\cite{NicRatTri0903}
\begin{eqnarray} 
\label{eq:G1}
\mathcal{G}_1(\Phi_{\rm v})  &=& \nabla^2 \Phi_{\rm v} =\theta_{\rm v},   \\ 
\label{eq:G2}
\mathcal{G}_2(\Phi_{\rm v})  &=& ( \nabla_{ij} \Phi_{\rm v} )^2 - ( \nabla^2 \Phi_{\rm v} )^2 , \\ 
\label{eq:G3}
\mathcal{G}_3(\Phi_{\rm v})  &=& ( \nabla^2 \Phi_{\rm v} )^3 + 2 \nabla_{ij} \Phi_{\rm v} \nabla_{jk} \Phi_{\rm v} \nabla_{ki} \Phi_{\rm v} \nonumber \\
& & - 3 (\nabla_{ij} \Phi_{\rm v})^2  \nabla^2 \Phi_{\rm v} .
\end{eqnarray}
Note that $\lexp \mathcal{G}_2(\Phi_{\rm v}) \rexp = \lexp \mathcal{G}_3(\Phi_{\rm v}) \rexp = 0$. 
On the other hand, one can construct a similar description through the invariants of the deformation tensor. 
In terms of the eigenvalues $\lambda_i$ of the deformation tensor ${\cal D}_{ij}\equiv -\nabla_{\mb{q}_i}\Psi_j$, these are  
\beqa
\label{I1def}
I_1 &=& {\rm Tr}\, [ {\cal D}_{ij} ]= \lambda_1+\lambda_2+\lambda_3, \\
\label{I2def}
I_2 &=&  \lambda_1\lambda_2+\lambda_2\lambda_3+\lambda_3\lambda_1, \\
\label{I3def}
I_3 &=&  {\rm Det}\, [{\cal D}_{ij}] = \lambda_1\lambda_2\lambda_3,
\eeqa
making $\mathcal{G}_1=I_1$, $\mathcal{G}_2=-2\,I_2$ and $\mathcal{G}_3=6\,I_3$ {\em to leading order} (i.e. ignoring the distinction between $\mb{q}$ and $\mb{x}$). In linear theory $I_1= \delta^{(1)}=\theta_{\rm v}^{(1)}$, but in general there is no simple relation between $I_1$ and the Eulerian fields, unlike $\mathcal{G}_1$. Since we work in Eulerian coordinates we will use the Galileons rather than the invariants of the deformation tensor, although this distinction is not crucial. The invariants of the deformation tensor can also be related to other useful quantities in the theory of random Gaussian fields, such as the ellipticity and prolateness of the potential, see Appendix~\ref{eANDp}.

Galileons arise in theories with the linear shift symmetry $\Phi_{\rm v} \to \Phi_{\rm v} + \mb{V}\cdot \mb{x}+C$, which, in our 3D case corresponds precisely to Galilean invariance, since the equations of motion Eqs.~(\ref{eq:DMcontinuity}-\ref{eq:GalEuler}) are Galilean invariant~\cite{ScoFri9607}.  In theories of modified gravity, 4D Galileons arise as a remnant of a higher-dimensional Lorentz symmetry and Lagrangians containing Galileon interactions have equations of motion that are precisely second-order~\cite{NicRatTri0903}. In our case, the bias relation in the absence of velocity bias solely depends on $\Phi_{\rm v}$, which obeys second-order (in space) equations of motion. When there is velocity bias, however, there is a relative bulk flow between tracers and matter, which breaks the Galilean invariance of the bias relation, and this shows up as a dipole field at second-order in PT. This simple argument explains the basic structure of  non-local bias. 

Therefore, we expect that in the ZA dynamics the bias relation when there is no velocity bias will have the form up to third-order, apart from local terms in $\delta$
\beq
\delta_{\rm g}^{\rm Nloc}= \gamma_2\, \mathcal{G}_2\, (\Phi_{\rm v}) (1+\beta \, \delta) +  \gamma_3\,  \mathcal{G}_3(\Phi_{\rm v}) + \ldots
\label{deltaNlocZAform}
\eeq
with coefficients $\gamma_i\propto (b_1^*-1)$ and mixing $\beta\propto b_2^*$. Note that $\mathcal{G}_1$ does not appear as $\delta$ is already part of the local bias description, and the difference between $\mathcal{G}_1=\theta_{\rm v}$ and $\delta$ can be written in terms of Galileons (see e.g. Eq.~\ref{Delta2sol} below). When we extend these results to the exact dynamics we shall see that new contributions appear that are nonlocal in $\Phi_{\rm v}$. The second-order term in Eq.~(\ref{deltaNlocZAform}) was already shown in Section \ref{sec:EulerianBiasEvolution}, so the main task here is to show what happens to third order.

To proceed, we start by substracting the local linear term, and writing an equation of motion for $\chi  \equiv \delta_{\rm g} - b_1 \delta $, which is simply (after using Eq.~\ref{eq:FryBiasEvol})  
\begin{equation}
\label{eq:chi_evolve}
 \dot{\chi} = ( b_1 -1 )  \Delta   +   \nabla \cdot ( \chi \mb{v}   ), \ \ \ \ \ \ \ \Delta \equiv  \delta  -  \theta_{\rm v},
\end{equation}
where a dot denotes a derivative with respect to $y=\ln D$.  The equation of motion for $\Delta$ is, after using Eqs.~(\ref{eq:DMcontinuity}-\ref{eq:DMEuler}) 
\begin{equation}
\label{eq:DeltaEoMGalileon}
\dot{ \Delta }   + \frac{3}{2} \varepsilon  \Delta  =   \nabla \cdot \big( \Delta \mb{v} \big)  -  \mathcal{G}_2(\Phi_{\rm v}  ) . 
\end{equation}  
where $ \varepsilon=1,0$ for the exact dynamics (ED) or ZA, respectively, we have used the fact that 
\begin{equation}
\nabla \cdot \big[ ( \mb{v} \cdot \nabla ) \mb{v} \big] =  \mathcal{G}_2(\Phi_{\rm v}  )  + \nabla \cdot( \mb{v} \theta_{\rm v}),       
\end{equation}
and  that the  ZA corresponds to replacing the Poisson equation by  $\nabla^2 \Phi =  3 \mathcal{H}^2 \Omega_{\rm m} \theta / 2  $~\cite{MunSta9406,HuiBer9611}. 

Equations~(\ref{eq:chi_evolve}-\ref{eq:DeltaEoMGalileon}) make clear that the evolution of $\chi$ depends only on the velocity field through $\Delta$. Furthermore, by construction, $\chi$ and $\Delta$ vanish in linear theory, and since matter fluctuations are in the growing mode, $ \Delta^{(n)} \propto \e{n y} $, and one can solve Eq.~(\ref{eq:DeltaEoMGalileon}) immediately:  
\begin{equation}
\label{Deltasol}
\Delta^{(n)} = \frac{ 2 }{ 2n +3\varepsilon } \big[  \nabla \cdot (\Delta \mb{v})^{(n)} - \mathcal{G  }_2^{(n)}( \Phi_{\rm v}  )    \big].
\end{equation}
This gives a recursion relation starting from $ \Delta^{(1)}=0$. Note that $\mathcal{G  }_2^{(n)}$ denotes the $n^{\rm th}$-order contribution to $\mathcal{G  }_2$ because $\Phi_{\rm v}$ is a non-linear quantity. The second-order solution is straightforward,
\begin{equation}
\Delta^{(2)} = -\frac{ 2 }{ 4 + 3 \varepsilon}\ \mathcal{G}_2^{(2)}(\Phi_{\rm v}). 
\label{Delta2sol}
\end{equation}
When used in  Eq.~(\ref{eq:chi_evolve}), this gives (upon integration) the desired second-order solution:
\begin{equation}
\chi^{(2)}  =  \frac{b_2}{2}\, [\delta^{(1)}]^2    -    \frac{2 (b_1 -1 )}{ 4 + 3 \varepsilon } \,( 1 - \e{- y} ) \ \mathcal{G}_2^{(2)}(\Phi_{\rm v}),
\label{chi2solGal}
\end{equation}
after using the initial condition
 $\chi^{(2)}_* =(b_2^*/ 2) \, [ \delta^{(1)}_*]^2$ and 
\begin{eqnarray}
b_2  &=& b_2^* \, \e{-2y}. 
\label{b2evolved}
\end{eqnarray}
Comparing Eqs.~(\ref{deltaNlocZAform}) and~(\ref{chi2solGal}) we then identify
\beq
\gamma_2 =  -\frac{2 (b_1 -1 )}{ 4 + 3 \varepsilon } \,( 1 - \e{- y} ).
\label{gamma2EDZA}
\eeq
Equation~(\ref{chi2solGal}) agrees with Eq.~(\ref{chinov}) for the ED ($\varepsilon=1$), after recalling that the structure of $\mathcal{G  }_2$ yields monopole and quadrupole equal to each other but opposite in sign. We see from this that the change in dynamics (ED or ZA) only affects the amplitude of $\gamma_2$, not the structure of the second-order solution.


Let us now find out the third-order solution.  From Eq.~(\ref{eq:chi_evolve}) we have after using Eq.~(\ref{Deltasol}) 
\begin{equation}
\dot{\chi}^{(3)} =  \frac{2 (b_1  -1 ) }{ 6 + 3 \varepsilon} \big[ \nabla \cdot (\Delta^{(2)} \mb{v}^{(1)})  - \mathcal{G}_2^{(3)}(\Phi_{\rm v}  )  \big]  + \nabla \cdot \big( \chi^{(2)}  \mb{v}^{(1)} \big),
\label{chi3eom}
\end{equation}
which can be integrated to give, after using the  initial condition $\chi^{(3)}_* = b_2^* \,\delta^{(1)}_*  \delta^{(2)}_*   +  (b_3^* /6) \, [ \delta^{(1)}_* ]^3$, and Eqs.~(\ref{Delta2sol}) and~(\ref{chi2solGal})
\begin{widetext}
\begin{eqnarray} 
\label{eq:chi3long}
\chi^{(3)}& = &b_2\, \delta^{(1)} \delta^{(2)}  + \frac{b_3}{6}\, [ \delta^{(1)} ]^3 - \frac{2\,b_2}{ 4+3\varepsilon }\, ( 1-\e{-y} ) \, \delta^{(1)} \,  \mathcal{G}_2^{(2)}(\Phi_{\rm v}  )
\nonumber  \\
&- &\frac{2 (b_1 - 1 )}{ 4 + 3 \varepsilon} \Big[ \frac{8 + 3 \varepsilon }{6(2 + \varepsilon)} - \e{-y} + \frac{3 \varepsilon + 4 }{6(2 + \varepsilon )}  \e{-2 y}  \Big]  \nabla \cdot \Big(  \mathcal{G}_2^{(2)}(\Phi_{\rm v}) \mb{v}^{(1)}  \Big)  -   \frac{b_1 - 1}{6 + 3 \varepsilon }\, ( 1 - \e{-2y} )\, \mathcal{G}_2^{(3)}(\Phi_{\rm v} )    ,
\end{eqnarray}
\end{widetext}
where we  used $\delta^{(2)}=[\delta^{(1)}]^2+\mb{v}^{(1)}\cdot\nabla\delta^{(1)}+2 \mathcal{G}_2^{(2)}/(4+3\varepsilon)$ and 
\beq
b_3  =  b_3^* \, \e{-3y} -3 b_2 ( 1- \e{-y} ).
\label{b3evolved}
\eeq
The first two terms in the first line of Eq.~(\ref{eq:chi3long}) are precisely those expected of local bias; the last is the mixing term between local and non-local bias induced by a non-zero $b_2^*$, which identifies $\beta=b_2/(b_1-1)$  in Eq.~(\ref{deltaNlocZAform}) independent of the dynamics. 

To extract $\gamma_3$ from Eq.~(\ref{deltaNlocZAform}), and deviations from this equation in going from the ZA to the ED, we must first subtract from the second line of Eq.~(\ref{eq:chi3long}) the third-order contribution $\gamma_2\, \mathcal{G}_2^{(3)}$ implied in Eq.~(\ref{deltaNlocZAform}).  This gives, for the only remaining non-local contribution to third order, 
\begin{widetext}
\begin{equation} 
\delta_{\rm g}^{\rm Nloc} \supset
     \frac{ b_1 -1 }{ (4 + 3 \varepsilon )(6 + 3 \varepsilon ) }\Big[ (8 + 3 \varepsilon ) - 6(2 + \varepsilon )\e{-y}  
     + ( 4 + 3 \varepsilon ) \e{-2 y}  \Big] \Big[  \mathcal{G}_2^{(3)}(\Phi_{\rm v} )  -  \nabla \cdot (\mathcal{G}_2^{(2)}(\Phi_{\rm v}) \,    \mb{v}^{(1)}   )    \Big]. 
\label{dgNlocremain}
\end{equation}
\end{widetext}

To compute $\mathcal{G}_2^{(3)}(\Phi_{\rm v})$, we need the non-linear evolution of the velocity potential to second order, $ \Phi_{\rm v}= \Phi^{(1)}_{\rm v} + \Phi^{(2)}_{\rm v} $, which gives 
\begin{eqnarray} 
\label{eq:G2_2}
\mathcal{G}_2^{(2)}(\Phi_{\rm v}) &  = &  ( \nabla_{ij} \Phi^{(1)}_{\rm v} )^2 - ( \nabla^2 \Phi^{(1)}_{\rm v} )^2  \\
\label{eq:G2_3}
\mathcal{G}_2^{(3)}(\Phi_{\rm v}) &  = & 2(\nabla_{ij} \Phi^{(1)}_{ \rm v} \nabla_{ij} \Phi^{(2)}_{\rm v}  - \nabla^2 \Phi^{(1)}_{\rm v}   \nabla^2 \Phi^{(2)}_{\rm v}   ) .
\nonumber \\ &&
\end{eqnarray}

Our calculation so far holds for both the ZA and the ED. While $\mathcal{G}_2^{(2)}$ only depends on the linear potential (and thus it is independent of dynamics), $\mathcal{G}_2^{(3)}$ depends on the details of the dynamics through $\Phi^{(2)}_{\rm v}$. Therefore, we first compute Eq.~(\ref{dgNlocremain}) in the ZA, for which the second-order potential is straightforward: $\Phi^{(2)}_{\rm v, ZA} = [\nabla_i \Phi^{(1)}_{\rm v}]^2/2 $, and using Eqs.~(\ref{eq:G2_2}-\ref{eq:G2_3}) we see that, remarkably, 
\begin{equation}  
\mathcal{G}_{ 2 , \rm ZA}^{(3)}(\Phi_{\rm v} )  - \nabla \cdot (\mathcal{G}_2^{(2)}(\Phi_{\rm v} ) \, \mb{v}^{(1)}   )     =   \mathcal{G}_{ 3}(\Phi_{\rm v} ).
\label{G3miracle}
\end{equation}
This is the third Galileon operator to leading order.  Therefore, this completes the proof of Eq.~(\ref{deltaNlocZAform}), with a $\gamma_3$ coefficient that can be obtained from the amplitude in Eq.~(\ref{dgNlocremain}) by setting $\varepsilon=0$.

We now extend this result to the ED by using the fact that the ZA gives rise to the exact Galileon $\mathcal{G}_3$ in Eq.~(\ref{G3miracle}). We can obtain the ED second-order potential by noting that $\theta_{\rm v}^{(2)}=[\delta^{(1)}]^2+\mb{v}^{(1)}\cdot\nabla\delta^{(1)}+4 \mathcal{G}_2^{(2)}/(4+3\varepsilon)$, and thus 
\begin{eqnarray}
\nabla^2  \Phi_{\rm  v }^{  (2)} & = & \nabla^2  \Phi_{\rm v, ZA}^{ (2)} - {3\varepsilon \over (4+3\varepsilon)} \, \mathcal{G}_2^{(2)}(\Phi_{\rm v} ).
\label{Phi2ED}
\end{eqnarray} 
From this we deduce 
\begin{equation}
\Phi^{(2)}_{\rm v} = \Phi^{ (2)}_{\rm v, ZA}  + {3\varepsilon \over (4+3\varepsilon)} \, \Phi_{\rm 2LPT}, 
\end{equation}
where the potential $\Phi_{\rm 2LPT }  $ is precisely the 2LPT potential for the displacement field to second-order.  It obeys the Poisson equation~\cite{BouColHiv9504,Sco98}
\begin{equation}
\nabla^{2} \Phi_{\rm 2LPT} =  -   \mathcal{G}_2 (  \Phi^{(1)}_{\rm v}    ),   
\label{2LPTPoisson}
\end{equation}
which implies that Eq.~(\ref{G3miracle}) becomes 
\beqa
\mathcal{G}_{ 2}^{(3)}(\Phi_{\rm v} ) -\nabla \cdot (\mathcal{G}_2^{(2)}(\Phi_{\rm v} ) \, \mb{v}^{(1)}   ) &=        &       
\mathcal{G}_3(\Phi_{\rm v})  \\ & +&  \frac{6\, \varepsilon  }{ (4+3\varepsilon) } \mathcal{G}_2 ( \Phi^{(1)}_{ \rm v},  \Phi_{ \rm 2LPT } ) \nonumber 
\label{G3nomiracle}
\eeqa
where, with a slight abuse of notation, we have defined 
\begin{equation}
  \mathcal{G}_2 (\Phi^{(1)}_{\rm v}, \Phi_{\rm 2LPT }) \equiv   \nabla_{ij} \Phi^{(1)}_{\rm v} \nabla_{ij} \Phi_{\rm 2LPT } -  \nabla^2 \Phi^{(1)}_{\rm v} \nabla^2 \Phi_{\rm 2LPT  }  .     
\label{G2abuse}
\end{equation}
We thus obtain the main result of this section, that in the exact dynamics the non-local part of the bias relation is given by (setting $\varepsilon=1$)
\beqa
\label{deltaNlocform}
\delta_{\rm g}^{\rm Nloc}&=& \gamma_2\, \mathcal{G}_2\, (\Phi_{\rm v}) (1+\beta \, \delta)  \\& &+ \, \gamma_3  \left( \mathcal{G}_3(\Phi_{\rm v}) 
+{6\over 7}  \, \mathcal{G}_2 (\Phi^{(1)}_{\rm v}, \Phi_{\rm 2LPT }) \right) + \ldots \nonumber
\eeqa
with coefficients given in terms of the local bias parameters by,
\beqa
\gamma_2 &=& - \frac{2 }{ 7} ( b_1 -1) (1-\e{-y}),   \ \ \ \ \ \beta={b_2\over(b_1-1)}, \label{gamma2} \\
\gamma_3 &=&  \frac{1 }{ 63}( b_1 -1)(1-\e{-y}) (11-7\e{-y}).
\label{gamma3}
\eeqa

Equation~(\ref{deltaNlocform}) replaces Eq.~(\ref{deltaNlocZAform}) and includes the fact that the dynamics of gravitational instability is non-local; as a result of this, the bias relation is not only non-local in terms of the density, but also in terms of the velocity potential, through the non-local dependence of $\Phi_{\rm 2LPT }$ on $\Phi^{(1)}_{\rm v}$ from inverting the Poisson equation Eq.~(\ref{2LPTPoisson}).  This term can also be written using the (scaled) gravitational potential $\Phi$,

\beq
\mathcal{G}_2 (\Phi^{(1)}_{\rm v}, \Phi_{\rm 2LPT })= {7\over 4} \Big[ \mathcal{G}_2\, (\Phi)-\mathcal{G}_2\, (\Phi_{\rm v})
\Big],
\label{toPhi}
\eeq
where $\nabla^2\Phi\equiv\delta$. Therefore, one expects that the non-local bias relation to {\em any} order can be written in terms of $\mathcal{G}_2\, (\Phi_{\rm v})$, $\mathcal{G}_2\, (\Phi)$, $\mathcal{G}_3\, (\Phi_{\rm v})$, and $\mathcal{G}_3\, (\Phi)$ including nonlinear combinations of them and multiplications of them by powers of $\delta$. 

In the discussion so far we have assumed for simplicity that bias at formation is local. If we allow for non-local bias at formation, adding to Eq.~(\ref{ICsdeltag}) non-local terms given by

\beq
\delta_{\rm g}^{* \rm Nloc} = \gamma_2^*\, \mathcal{G}_2\, (\Phi_{\rm v}^*) + \gamma_3^*\, \mathcal{G}_3\, (\Phi_{\rm v}^*),
\label{Nlocdgstar}
\eeq
it is easy to check that Eq.~(\ref{deltaNlocform}) still holds, but with the coefficients in Eqs.~(\ref{gamma2}-\ref{gamma3}) changed to
\beqa
\label{newgamma2}
\gamma_2 &\to& \gamma_2+ \gamma_2^*\, \e{-2y}, \\
\gamma_3 &\to& \gamma_3 + \gamma_2^*\, \e{-2y}(1-\e{-y})+\gamma_3^*\, \e{-3y}.
\label{newgamma3}
\eeqa
In Appendix~\ref{appA} we generalize these results deriving how the $\delta_{\rm g}^{\rm Nloc}$ obtained here can be related to that in the case when tracers are not conserved.

Based on symmetry arguments~\cite{McDRoy0908} suggested a model of non-local bias with similar, but not identical, structure to what is derived here. Our results in the absence of velocity bias, Eq.~(\ref{deltaNlocform}), agree with them to second-order, while at third order our results differ somewhat: their $\delta s^2$ term corresponds to our term proportional to $\beta$, their $s^3$ term can account partially for $\mathcal{G}_3$ and their $st$ term basically corresponds to our non-local term given by Eq.~(\ref{toPhi}). However, they include an extra term to third order (which they call $\psi$)  which we find unnecessary as it can be written in terms of the other third-order contributions. Thus we find that the most general third-order non-local contributions in the absence of velocity bias contain only three, not four, free parameters. In our case, these parameters are not  free as they are connected by dynamical evolution to local bias (Eqs.~\ref{gamma2}-\ref{gamma3}), although if formation bias is non-local  (Eq.~\ref{Nlocdgstar}) there are extra free parameters (Eqs.~\ref{newgamma2}-\ref{newgamma3}).

\section{Large-scale non-local bias in Numerical simulations}
\label{NumSim}

\begin{table}
\caption{Halo Samples used in this paper}
\begin{ruledtabular}
\begin{tabular}{c c c  c}
Halo Sample &   $z$  &  $b_\times$ &  Mass bin [$10^{13}M_\odot/h$]\\
\hline
LMz0 & 0 & 1.43 & $4-7$  \\
MMz0 & 0 & 1.75 & $7-15$  \\
HMz0 & 0 & 2.66 & $>15$  \\ 
\hline
LMz0.5 & 0.5 & 1.88 & $3-5$  \\
MMz0.5 & 0.5 & 2.26 & $5-10$ \\
HMz0.5 & 0.5 & 3.29 & $>10$  \\ 
\hline
LMz1 & 1 & 2.43 & $2-3.1$  \\
MMz1 & 1 & 2.86 & $3.1-5.7$  \\
HMz1 & 1 & 3.99 & $>5.7$ 
\label{HaloSample}
\end{tabular} 
\end{ruledtabular} 
\end{table}

We now discuss how the ideas presented in the previous sections can be used to ascertain the extent to which the large scale bias of dark matter halos in numerical simulations is local. We use a set of $50$ simulations, each containing
$N_{par}=640^3$ particles within a comoving box-size of side $L_{\rm box}=1280 \Mpc$. The total comoving volume is thus approximately $105~(h^{-1} {\rm Gpc})^3$. Cosmological parameters were $\Omega_m=0.27$, $\Omega_{\Lambda}=0.73$,  $\Omega_b=0.046$ and $h=0.72$, together with scalar spectral index $n_s=1$ and normalization $\sigma_8=0.9$. The simulations were run using  Gadget2~\cite{2005MNRAS.364.1105S} with initial conditions set at $z_i=49$ using 2nd order-Lagrangian Perturbation Theory (2LPT)~\cite{Sco98,2006MNRAS.373..369C}. The halos are identified using the friends-of-friends algorithm with linking length equal to $0.2$ times the mean inter-particle separation. We divide our halo sample into three mass bins at each redshift $z=0, 0.5, 1$. Table~\ref{HaloSample} shows the main features of each of these, including the large-scale (linear) bias obtained from measuring the cross-power spectrum between halos and matter, i.e. $b_\times=P_{hm}/P_{mm}$, and averaging over scales $k \leq 0.05 \kvecMpc$.

\begin{figure}[!t]
 \centering
 \includegraphics[ width=0.95\linewidth]{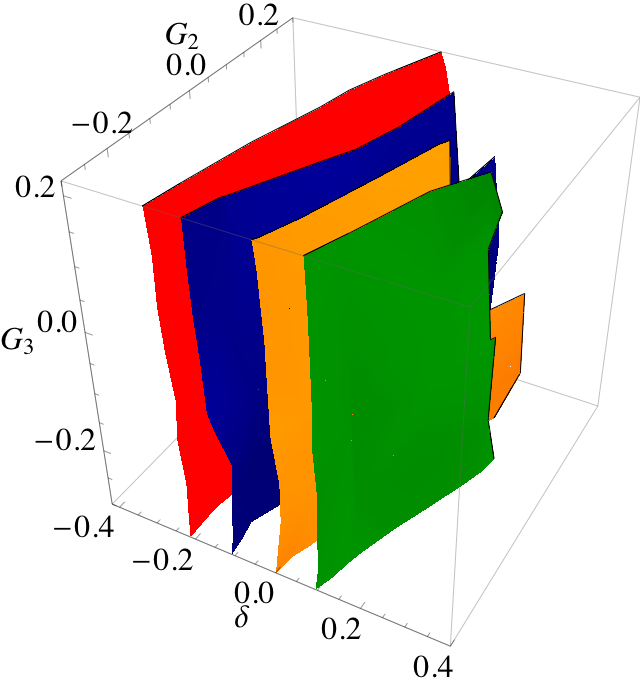}
 \caption{Illustration of non-local large-scale bias in numerical simulations for high-mass halos at $z=1$ (see HMz1 in Table~\protect\ref{HaloSample}). The  plot shows surfaces of constant $\delta_h=-0.3,0.1,0.5,0.9$ (from left to right, or red, blue, yellow, and green, respectively)  as a function $\delta$, $\Gd$ and $\Gt$. If large-scale bias were a local function of $\delta$, surfaces of constant $\delta_h$ would be $\delta={\rm const.}$~planes (see next figure).  Instead, there is significant tilt  ($\nabla \delta_h$ is not parallel to the $\delta$-axis) showing a non-negligible dependence on $\Gd$. All fields ($\delta$, $\Gd$, $\Gt$ and $\delta_h$) have been smoothed with a top-hat window of radius $R_s=40 \Mpc$.}
 \label{4DHMz1}
 \end{figure}

\begin{figure}[!t]
 \centering
 \includegraphics[ width=0.95\linewidth]{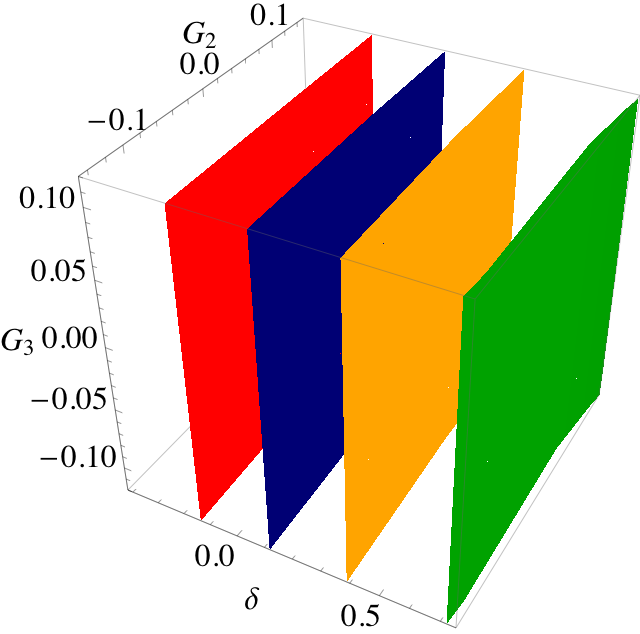}
 \caption{Same as Fig.~\ref{4DHMz1} but for low-mass halos at $z=0$ (see LMz0 in Table~\protect\ref{HaloSample}). For the least biased objects in our samples, bias becomes local.}
 \label{4DLMz0}
 \end{figure}

To assess the locality of large-scale halo bias in the simulations we proceed as follows.  We build the smoothed matter fluctuations by interpolating the dark matter particles in the simulation to a grid of size $N_{\rm grid}=180$ (corresponding to a grid separation of $ \simeq 7 \Mpc$), Fourier transforming using FFT's, multiplying by the Fourier transform of a real-space top-hat window function of radius $R_s=40\Mpc$, and Fourier transforming back to real space. We build the smoothed halo overdensity field similarly.  We build the smoothed Galileon fields ${\cal G}_2$ and ${\cal G}_3$ from the velocity field by first constructing the velocity potential $\Phi_{\rm v}$ (and velocity divergence $\theta_{\rm v}=\nabla^2 \Phi_{\rm v}$) by using a Delaunay tessellation to build the volume weighted velocity field on the grid (see~\cite{PueSco0908} for details), constructing the Galileon fields on the grid and then smoothing them as one does for any scalar field ($\delta$ or $\delta_h$) as explained above. That is,

\beq
\Gd(\x) = \int {\rm e}^{-i \k_{12}\cdot \x} \, (\mu_{12}^2-1)\, \theta_{\rm v}(\k_1) \theta_{\rm v}(\k_2)\, W_{12}\, d^3k_1 d^3k_2
\label{G2x}
\eeq
where $W_{12}\equiv W(k_{12}R_s)$, $\mu_{ij}\equiv \hat{k}_i\cdot \hat{k}_j$ and 
\beqa
\Gt(\x) &=& \int {\rm e}^{-i \k_{123}\cdot \x} \, (1+2 \mu_{12}\mu_{23}\mu_{31}-\mu_{12}^2-\mu_{23}^2-\mu_{31}^2) \nonumber \\
& & \times\, \theta_{\rm v}(\k_1) \theta_{\rm v}(\k_2)\theta_{\rm v}(\k_3)\, W_{123}\, d^3k_1 d^3k_2 d^3k_3.
\label{G3x}
\eeqa
We ignore, for simplicity, the extra non-local term depending on the 2LPT potential (see Eq.~\ref{deltaNlocform}). Including this term into the plots we present in this section does not change the results. 

Note that since the Galileon fields are non-linear combinations of (derivatives of) the velocity potential, this procedure {\em is not the same} as building the Galileon fields of the {\em smoothed} velocity potential, which would remove mode-couplings of the smoothing scale to smaller scales. This means that our smoothed Galileon fields depend to some extent on the choice of grid size (which effectively determines up to what scale we allow mode-couplings; in our case this is down to $ \simeq 7 \Mpc$). However, since the velocity power spectrum is suppressed compared to the density at small scales~\cite{2004PhRvD..70h3007S}, the dependence is not very strong, particularly because, in ${\cal G}_2$, the coupling to small-scale modes requires wave vectors to be anti-colinear in which case their  contribution to ${\cal G}_2$ vanishes. We have studied what happens if we increase $N_{\rm grid}$ and we see no significant change to the results presented below except for an increase in noise (from coupling to even smaller-scale modes). This is somewhat expected as one starts to probe couplings to scales comparable or smaller than the Lagrangian size halos. Ideally, one would use a grid size different for each halo sample so only scales larger than the respective Lagrangian radius are included in Eqs.~(\ref{G2x}-\ref{G3x}).

As a result of this procedure, at each grid point in the simulation box we have four fields smoothed at large-scales ($R_s=40\Mpc$): $\delta$,  ${\cal G}_2$, ${\cal G}_3$ and $\delta_h$.  If large-scale bias were local, $\delta_h$ would  depend only on $\delta$, and thus surfaces of constant $\delta_h$ should agree with those of constant $\delta$, in other words, $\nabla \delta_h$ in this three-dimensional space $(\delta,  {\cal G}_2, {\cal G}_3)$ should be parallel to the $\delta$-axis. Figure~\ref{4DHMz1} shows this construction for the highest mass bin at $z=1$, where the effects of non-local bias are the strongest: there is a clear tilt of the surfaces of constant $\delta_h$ in the ${\cal G}_2$ direction, but no discernible dependence on ${\cal G}_3$. Therefore, {\em in cells of fixed} $\delta$, where local bias would predict a constant $\delta_h$, {\em we see significant variations in $\delta_h$ that scale with the value of ${\cal G}_2$}. Note that at fixed $\delta$, $\delta_h$ is a decreasing function of ${\cal G}_2$, as predicted by our simple arguments in the previous sections. 

\begin{figure}[!t]
 \centering
 \includegraphics[ width=0.95\linewidth]{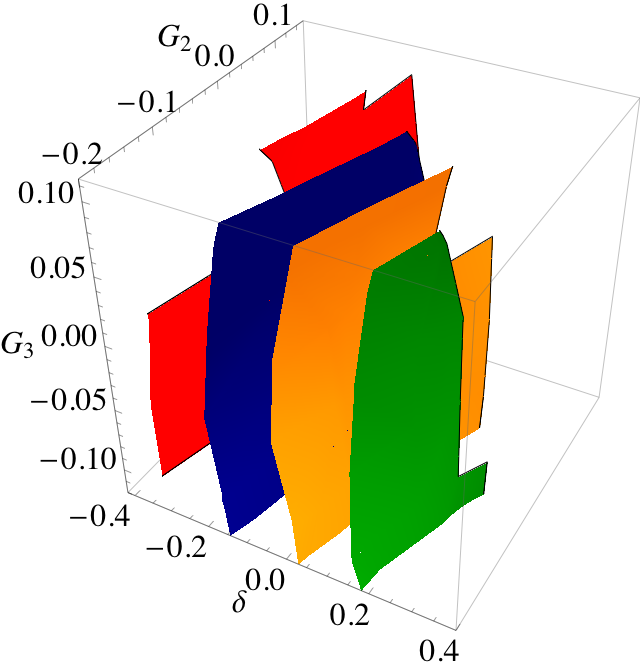}
 \caption{Same as Fig.~\ref{4DHMz1} but for high-mass halos at $z=0$ (see HMz0 in Table~\protect\ref{HaloSample}). }
 \label{4DHMz0}
  \end{figure}

Figure~\ref{4DLMz0} shows what happens in the lowest-mass bin at $z=0$, for which the large-scale linear bias is the smallest among our halo samples. We see now that bias does become local: planes of constant $\delta$ agree with constant $\delta_h$, and $\nabla \delta_h$ points along the $\delta$-axis. This is also in qualitative agreement with our simple  model developed in the previous sections. 

In Figure~\ref{4DHMz0} we show the same plot for high-mass halos at $z=0$, demonstrating that indeed  more biased objects at fixed $z$ do show more non-local large-scale bias. Again, $\nabla \delta_h$  has a significant component in the $\Gd$-direction, and little (if anything) in the $\Gt$-direction, showing that the results presented in Fig.~\ref{4DHMz1} are generic. Our model in the previous section does predict the dependence on $\Gd$ to be stronger than that on $\Gt$ but only by a factor of about two or so (the precise value depends on ``formation" time); the results from the simulations suggest that the suppression of the $\Gt$ amplitude is even greater. 

It is rather common (see e.g.~\cite{2007PhRvD..75f3512S,2009arXiv0912.0446M,RotPor1105,PolSmiPor1109} for recent examples) to present the bias relation from simulations in terms of a scatter plot of $\delta_h$ and $\delta$, which corresponds to projecting out the $\Gd$ and $\Gt$ directions in our Figs.~\ref{4DHMz1}-\ref{4DHMz0}. Because of the tilt in the $\Gd$ direction, a bias that is completely deterministic in $\delta$, $\Gd$ and $\Gt$ will lead, when projected into the $\delta$-axis, to a stochastic $\delta_h$ vs $\delta$ relation with the scatter simply coming from points with the same $\delta$ that have different $\Gd$. The question which arises is if the scatter seen in the $\delta_h$ vs $\delta$ relation can be explained by this projection effect, at least partially? One way to address this is to ask whether the scatter about the tilted planes with constant $\delta_h$ in the three-dimensional space $(\delta,  {\cal G}_2, {\cal G}_3)$ is significantly less than that seen in the 1D scatter plot of $\delta_h$ vs $\delta$. We find that indeed the multidimensional scatter is smaller than the 1D scatter, but only marginally so (with one exception, which we discuss in the next paragraph). This indicates that most of the scatter of the $\delta_h$ vs $\delta$ relation is not due to the dependence of $\delta_h$ on the ``hidden variables" $\Gd$ and $\Gt$. In fact, this scatter can be explained~\cite{1999MNRAS.304..767S,CasMoShe0207} in the context of the excursion-set model of halo formation by noting that the small-scale density field (whose excursions above the collapse threshold correspond to halo formation) has a stochastic relation to the large-scale density field $\delta$.  
 
\begin{figure}[!t]
 \centering
 \includegraphics[ width=0.95\linewidth]{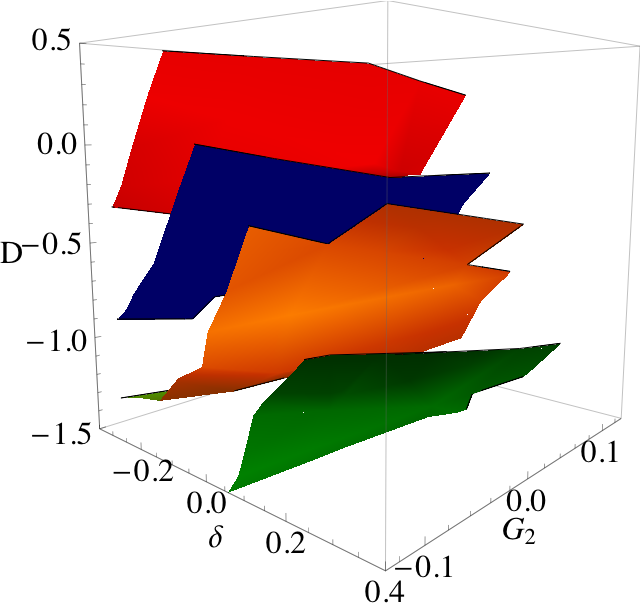}
 \caption{Same as Fig.~\ref{4DHMz1} (high-mass halos at $z=1$) but using the dipole field ${\cal D}$ rather than ${\cal G}_3$ as the third axis. There is significant dependence not only on ${\cal G}_2$ but also on ${\cal D}$, a signature of velocity bias, as expected for the most significant biased objects.}
 \label{4DHMz1Dip}
 \end{figure}

Having seen that there is little, if any, non-locality coming from $\Gt$ we look for the possible effects of velocity bias. From our model we expect that if there is velocity bias at the smoothing scale we consider ($R_s=40\Mpc$), then a dipole non-local term ${\cal D}$ will appear in the bias relation. As discussed before, a statistical velocity bias of halos is expected for the most  biased objects~\cite{DesShe1001,DesCroSco1011}, and while this statistical effect vanishes at large scales as $k^2$, the scale at which this is negligible can be very large for very massive halos. We thus repeat the process discussed above to construct ${\cal D}$ smoothed on $R_s=40 \Mpc$ at each grid point, 

\beq
{\cal D}(\x) = \int {\rm e}^{-i \k_{12}\cdot \x} \, {\mu_{12}} \Big({k_1\over k_2}\Big)\, \theta_{\rm v}(\k_1) \theta_{\rm v}(\k_2)\, W_{12}\, d^3k_1 d^3k_2
\label{Dx}
\eeq
and replace the $\Gt$-axis by a ${\cal D}$-axis for the halos in Fig.~\ref{4DHMz1} which are the most biased objects in our sample. These results are shown in Fig.~\ref{4DHMz1Dip}, and confirm there is a significant component of $\nabla \delta_h$ in the  ${\cal D}$-direction. This dipole dependence, unlike that on $\Gd$, quickly disappears as we consider less extreme objects, so it is relevant only for the most rare halos. It is however interesting to note that for this sample of highly biased objects, the multidimensional scatter of $\delta_h$ at fixed $\delta,\Gd,{\cal D}$, e.g. $\delta,\Gd,{\cal D}=0.15,0.08,0$ is 0.35, substantially smaller than the 1D scatter of 0.48 at fixed $\delta=0.15$. This indicates that a significant fraction of the scatter in the $\delta_h$-$\delta$ relation may be due to projection of non-local bias for highly biased halos.

\begin{figure}[!t]
 \centering
 \includegraphics[ width=0.95\linewidth]{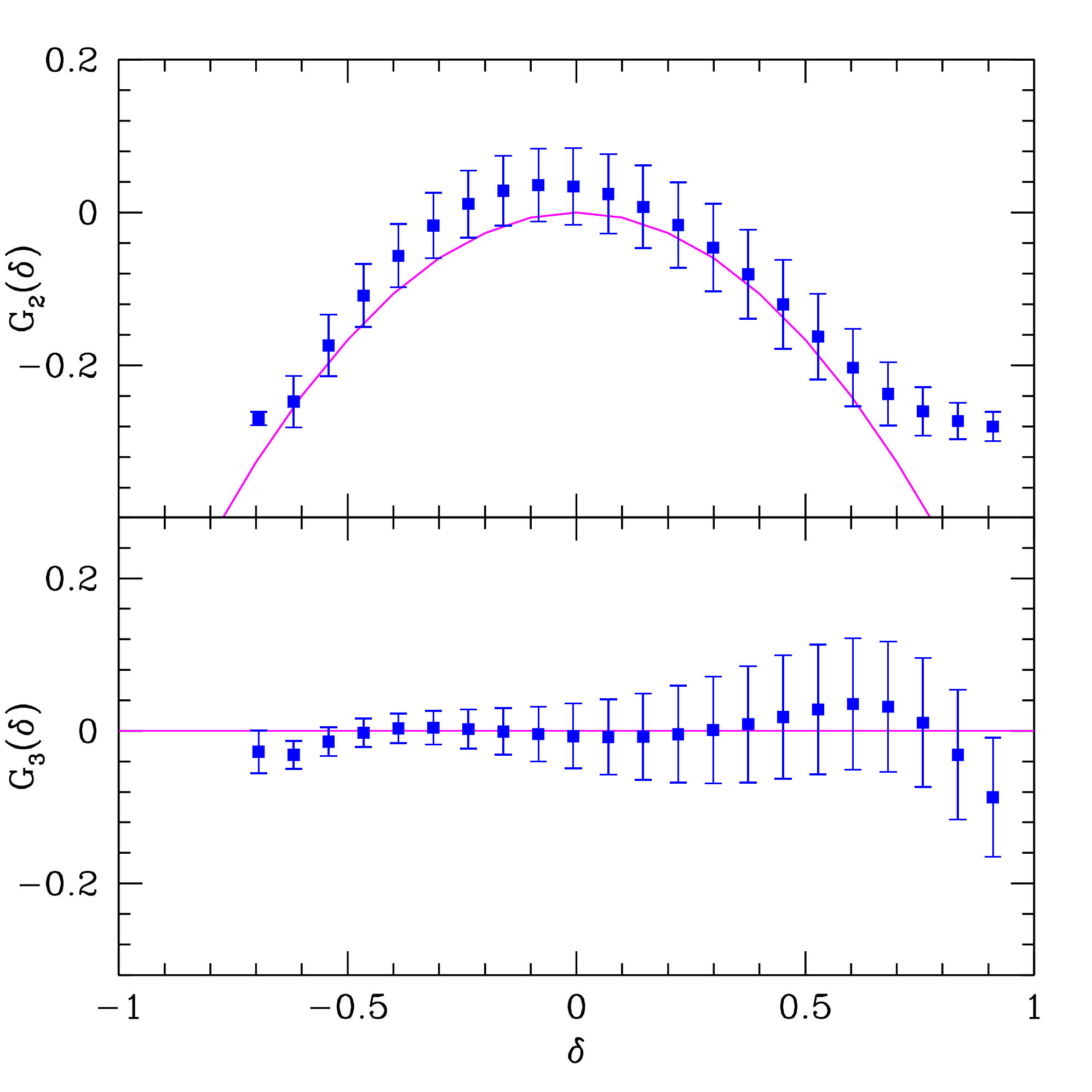}
 \caption{Mean and dispersion of ${\cal G}_2$ (top) and ${\cal G}_3$ (bottom) as a function of $\delta$ at $z=0$ for top-hat smoothing with $R_s=40 \Mpc$. Solid lines show the expected ${\cal G}_2 = -(2/3) \delta^2$ and ${\cal G}_3=0$.}
 \label{G2G3vsdelta}
 \end{figure} 

Finally, it is useful to ask how the mean and dispersion of $\Gd$ and $\Gt$ vary as a function of $\delta$. In Figure~\ref{G2G3vsdelta} we plot the mean (symbols) and dispersion (error bars) of $\Gd$ (top panel) and $\Gt$ (bottom panel) at fixed $\delta$ as a function of $\delta$ at $z=0$. The solid lines give the naive expectation corresponding to  averaging the angular dependence of the $\Gd$ and $\Gt$ kernels, i.e. $\Gd=-(2/3)\delta^2$ and $\Gt=0$, which are quite reasonable approximations. Therefore, a non-local bias relation of the form

\beq
\delta_h= b_1\, \delta + {b_2 \over 2}\, \delta^2 + \gamma_2 \, \Gd + \gamma_3 \, \Gt
\label{deltah-deltaG2G3}
\eeq
will look, when projected into the $\delta$-axis, as an effective local bias with

\beq
\delta_h^{\rm eff} \simeq b_1\, \delta + {(b_2-4 \gamma_2/3) \over 2}\, \delta^2. 
\label{deltah-deltaEFF}
\eeq
The relation $b_2^{\rm eff}=b_2-4 \gamma_2/3$ for the effective local quadratic bias will be useful to interpret the difference in quadratic bias parameters  between local vs non-local fits to the bispectrum obtained below. 

We now consider the effects of non-local bias from a different angle, in how it impacts the bispectrum. Along the way we quantify more precisely the amplitude of non-local bias terms at large scales.

\section{Non-local bias and the Bispectrum}
\label{BispNonLoc}

The bispectrum is sensitive to the actual couplings at second-order in perturbations in both the dark matter and the bias relation. Therefore, it can give a precise determination of how non-local bias is, in particular the amplitude of the $\Gd$ dependencies found from smoothed fields in the previous section. Here we concentrate on the bias relation up to second order, we thus include only local quadratic bias $b_2$ and the amplitude of the non-local effect through $\Gd$. As we found in the previous section, there is no significant detection of a $\Gt$ dependence, and the (quadratic) dipole dependence is only significant for the most biased samples, which correspond to extremely rare halos. Thus we study the bias relation given by

\beq
\delta_h= b_1\, \delta + {b_2 \over 2}\, \delta^2 + \gamma_2 \, \Gd,
\label{deltah-delta}
\eeq
where we recall that for $\gamma_2=-2(b_1-1)/7$, this corresponds to assuming local {\em Lagrangian} bias (ie. $y \to \infty$ in our simple model). In this way we can simultaneously test for local Eulerian ($\gamma=0$), local Lagrangian ($\gamma=-2(b_1-1)/7$) and more generic non-local bias. We note that a similar test (of local Eulerian vs Lagrangian bias) was performed in the PSCz galaxy survey bispectrum~\cite{FelFriFry0102} with the result that Eulerian local bias was a slightly better fit to the galaxy bispectrum. Our tests in this paper are in a very different regime, as PSCz galaxies are anti-biased while our halos are positively biased (see Table~\ref{HaloSample}). 

To avoid dealing with the complications of the inaccuracy of Poisson shot-noise subtraction for halos (see appendix A in~\cite{2007PhRvD..75f3512S}) that complicates interpreting deviations from the local bias description, here we just study the halo-matter-matter bispectrum $b_{hmm}$ rather than the the halo bispectrum~\cite{SefCroDes1111,PolSmiPor1109},

\beq
\langle \delta_h(\k_1) \delta(\k_2) \delta(\k_3) \rangle = \dD(\k_{123})\ b_{hmm}(k_1,k_2,k_3).
\label{bhmm}
\eeq
Note that this is not a symmetric function of the wavectors. We thus define a symmetrized quantity~\cite{PolSmiPor1109},

\beq
B_{hmm} \equiv {1\over 3} (b_{hmm} + b_{mhm} + b_{mmh}),
\label{Bhmm}
\eeq
which from  Eq.~(\ref{deltah-delta}) obeys

\beq
B_{hmm}= b_1 \, B_{123} + {b_2\over 3}\, \Sigma_{123} + {2\over 3} \gamma_2 \, {\cal K}_{123}
\label{Bbias}
\eeq
where $B_{123}$ is the matter bispectrum and 
\beq
\Sigma_{123}= P_1\, P_2 + {\rm cyc.}, \ \ \ \ \ {\cal K}_{123}= (\mu_{12}^2-1)\, P_1\, P_2 + {\rm cyc.},
\label{SigK}
\eeq
with $\mu_{12}$ the cosine of the angle between $\k_1$ and $\k_2$. The kernel ${\cal K}_{123}$ vanishes for colinear trianges where $\mu_{ij}= \pm 1$, thus the non-local correction is most important for isosceles triangles. 

We measured the halo-matter-matter bispectrum $B_{hmm}$ and matter bispectrum $B_{123}$ from the 50 realizations of the simulations at the three redshift outputs. The triangles included in the bispectrum analysis correspond to all triangles with sides from twice the fundamental mode ($2k_f\simeq 0.01 \kvecMpc$) up to $k\leq 0.1 \kvecMpc$, binned in units of  $2k_f$, yielding 150 binned triangles (corresponding to $\sim 1.2\times 10^8$ fundamental triangles of sides inside the prescribed bins and all possible orientations). This together with the measured matter power spectrum can be used in Eq.~(\ref{Bbias}) to fit for the parameters $b_1$, $b_2$ and $\gamma_2$. In what follows we discuss such constraints for all the halo samples.

\begin{table}[!t]
\caption{Local Eulerian bias parameters $b_1$ and $b_2$ obtained from halo-matter-matter bispectrum fits for all triangles with $k<0.1 \kMpc$. We also include the large-scale bias $b_\times$ obtained from the halo-matter power spectrum, to be compared with $b_1$. The last column indicates the goodness of the fit assuming a diagonal covariance matrix ($N_{\rm dof}=148$).}
\begin{ruledtabular}
\begin{tabular}{c c c c c}
 Sample &   $b_\times$ &  $b_1$ & $b_2$ & $\chi^2$/dof  \\
\hline
LMz0 & 1.43 & $1.42\pm 0.01$ & $-0.91 \pm 0.03$ & 1.86 \\
MMz0 & 1.75 & $1.71\pm 0.01$ & $-0.55 \pm 0.03$ & 1.29 \\
HMz0 & 2.66 & $2.37\pm 0.02$ & $2.98 \pm 0.07$ & 3.74 \\ 
\hline
LMz0.5 & 1.88 & $1.77\pm 0.01$ & $-0.15 \pm 0.03$ & 0.91 \\
MMz0.5 & 2.26 & $2.13\pm 0.01$ & $0.67 \pm 0.03$ & 0.87 \\
HMz0.5 & 3.29 & $2.84\pm 0.03$ & $5.89 \pm 0.10$ & 3.77 \\ 
\hline
LMz1 & 2.43 & $2.22\pm 0.01$ & $1.27 \pm 0.04$ & 0.89 \\
MMz1 & 2.86 & $2.62\pm 0.02$ & $2.77 \pm 0.06$ & 1.07 \\
HMz1 & 3.99 & $3.41\pm 0.05$ & $9.98 \pm 0.14$ & 3.42 
\label{BiasParamE}
\end{tabular} 
\end{ruledtabular} 
\end{table} 

Table~\ref{BiasParamE} shows the results from fitting Eulerian local bias ($\gamma_2=0$)  to the relation in Eq.~(\ref{Bbias}). For comparison, in this and other tables, we reproduce the value of the large-scale linear bias obtained from the halo-matter cross spectrum $b_\times$. Note that for the lowest biased objects in our sample, LMz0, the linear bias obtained from the bispectrum $b_1$ agrees with $b_\times$, but this agreement  disappears for all other samples, giving a significantly smaller $b_1$ than the  large-scale linear bias $b_\times$ shown by the power spectrum, increasingly so for more biased objects. Recently,~\cite{PolSmiPor1109} found a similar result for halos with more than 20 particles at $z=0$.  Here, we highlight the mass and redshift dependence of this issue in more detail. As shown in~\cite{PolSmiPor1109}, had we used the reduced bispectrum $Q=B/\Sigma$ rather the bispectrum itself to find the bias parameters, then we would have found the opposite result, i.e. a linear bias $b_1$ {\em smaller} than $b_\times$. To explain why, let us for definiteness define a reduced halo-matter-matter bispectrum by

\beq
Q_{hmm} \equiv {B_{hmm}\over (P_\times(k_1)P_\times(k_2)+ {\rm cyc.})} = {B_{hmm}\over b_\times^2\Sigma_{123}}.
\label{Qhmm}
\eeq
Thus, while the halo-matter-matter bispectrum fits yield $b_{1B}$ and $b_{2B}$ with $B_{hmm}= b_{1B} B + b_{2B} \Sigma/3$, the reduced bispectrum yield parameters $b_{1Q}$ and $b_{2Q}$ with $Q_{hmm}=Q/b_{1Q}+b_{2Q}/3b_{1Q}^2$.  These are related by
\beq
b_{1Q}=b_\times \, \Big({b_\times\over b_{1B}}\Big), \ \ \ \ \ b_{2Q} = b_{2B}\, \Big({b_\times\over b_{1B}}\Big)^2;
\eeq
therefore, if $b_{1B}>b_\times$, then $b_{1Q}<b_\times$. Similarly, for halo bispectra (rather than halo-matter-matter), the relationship between reduced and un-reduced bispectra linear bias is instead $b_{1Q}=b_\times \,({b_\times/ b_{1B}})^3$, an even bigger difference (i.e. the relative deviation of $b_{1Q}$ from $b_{\times}$ is three times larger than for $b_{1B}$). These disagreements will be resolved shortly by including non-local bias.

\begin{table}[!t]
\caption{Eulerian bias parameters $b_1$ and $b_2$ obtained from doing a {\em Lagrangian} local bias model fit to the bispectrum.}
\begin{ruledtabular}
\begin{tabular}{c c c c c}
 Sample &   $b_\times$ &  $b_1$ & $b_2$ & $\chi^2$/dof  \\
\hline
LMz0 & 1.43 & $1.48\pm 0.01$ & $-1.26 \pm 0.04$ & 2.12 \\
MMz0 & 1.75 & $1.81\pm 0.01$ & $-1.15 \pm 0.03$ & 1.36 \\
HMz0 & 2.66 & $2.59\pm 0.02$ & $1.78 \pm 0.07$ & 2.73 \\ 
\hline
LMz0.5 & 1.88 & $1.87\pm 0.01$ & $-0.79 \pm 0.04$ & 0.94 \\
MMz0.5 & 2.26 & $2.30\pm 0.01$ & $-0.26 \pm 0.04$ & 0.72 \\
HMz0.5 & 3.29 & $3.12\pm 0.03$ & $4.34 \pm 0.11$ & 2.91 \\ 
\hline
LMz1 & 2.43 & $2.40\pm 0.02$ & $0.27 \pm 0.05$ & 0.77 \\
MMz1 & 2.86 & $2.85\pm 0.02$ & $1.45 \pm 0.06$ & 0.82 \\
HMz1 & 3.99 & $3.77\pm 0.05$ & $7.97 \pm 0.16$ & 2.74 
\label{BiasParamL}
\end{tabular} 
\end{ruledtabular} 
\end{table}

Table~\ref{BiasParamL} shows the analogous results when the bias is assumed to be local in {\em Lagrangian} space, equivalent to assuming $\gamma_2=-2(b_1-1)/7$ in Eq.~(\ref{deltah-delta}). The results in this case are somewhat mixed. At $z=0$ the results are worse than for the Eulerian case, except at high mass. At higher redshifts, the Lagrangian results show improvement, particularly at $z=1$, but there are still some significant discrepancies between $b_1$ and $b_\times$, and in any case the $\chi^2/{\rm dof}$ are not very convincing. 

Finally, Table~\ref{BiasParamNL} shows the results for the non-local bias model with the amplitude of  $\Gd$ being fit for. The results show now a significant improvement, in particular $b_1$ is always within two-sigma of the $b_\times$ values, for all redshifts and halo masses considered. We note that the average (over all halo samples) deviations of $b_1$ from $b_\times$ are  $11\sigma, 4.5\sigma$ and $1.5\sigma$ for Eulerian, Lagrangian and non-local bias fits to the bispectrum, respectively. Thus we reject local Eulerian and Lagrangian bias models at high significance. The price to pay in fitting for $\gamma_2$ as well is an increase in the $b_1$ error bars, by a factor of almost two.  

\begin{figure}[!t]
 \centering
 \includegraphics[ width=0.95\linewidth]{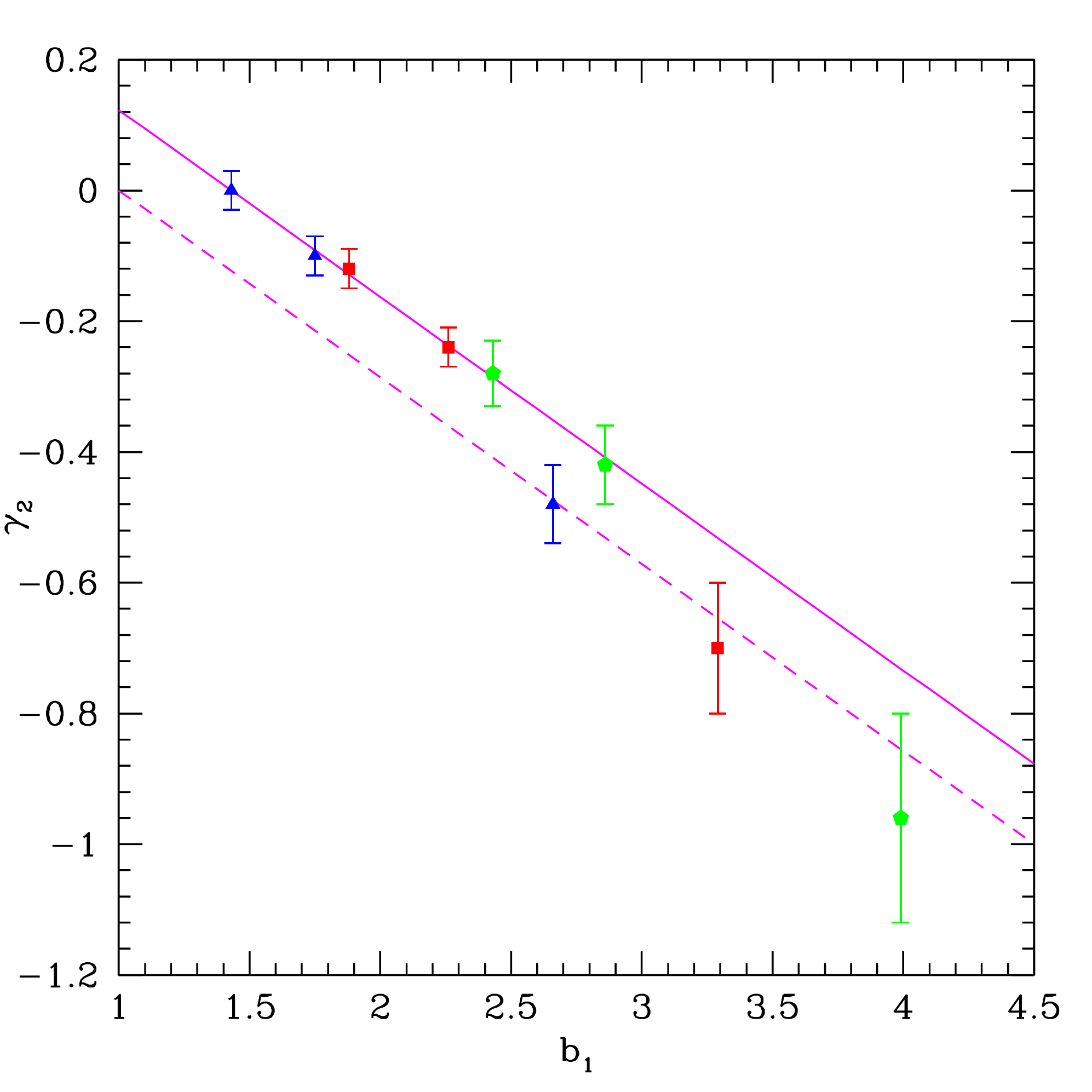}
 \caption{Dependence of the ${\cal G}_2$ non-local amplitude $\gamma_2$ (see Table~\protect\ref{BiasParamNL}) bispectrum fits on the linear bias ($b_1=b_\times$) for the three halo mass bins at three different redshifts, $z=0$ (blue triangles), $z=0.5$ (red squares) and $z=1$ (green pentagons). The dashed line shows the predictions of local Lagrangian bias, and the solid line corresponds to $\gamma_2=-2(b_1-1.43)/7$.}
 \label{gamma2res}
 \end{figure}

The values for $\gamma_2$ in Table~\ref{BiasParamNL} show a clear dependence with linear bias, which is plotted in Fig.~\ref{gamma2res} (using the more precise value of $b_\times$ as linear bias).  We see that the results fall mostly along along a ``universal" line given by $-2(b_1-1.43)/7$ (solid line), except for the most biased halos at each redshift which fall below this line (closer to the Lagrangian bias result, shown in dashed line). However it is precisely these highly biased objects that may have extra non-local contributions (such as a dipole, as discussed in the last section), so it is not clear at this point how reliable this behavior is. On the other hand, note that the solid line in Fig.~\ref{gamma2res} {\em is not} a fit to the data, but it serves to illustrate deviations from local Lagrangian bias for our least biased samples. More work is needed to see whether one could  understand these results from theoretical arguments. We note however that it is not surprising that bias is not local in Lagrangian space, even in simple extensions of the excursion set of halo formation the barrier for collapse is known to depend on other quantities than the overdensity $\delta$, mostly on the ellipticity parameter $e$~\cite{SheMoTor01}.  Appendix~\ref{eANDp} shows the relationship between ellipticity $e$, prolateness $p$ and the invariants of the deformation tensor or Galileons.

\begin{table}[!t]
\caption{Eulerian bias parameters $b_1$ and $b_2$ and non-local $\gamma_2$ parameter obtained from doing a quadratic non-local bias model fit to the bispectrum. For comparison purposes, note that a non-zero $\gamma_2$  gives an effective $-(4/3)\gamma_2$ contribution to $b_2$ (see top panel in Fig.~\ref{G2G3vsdelta}). Here $N_{\rm dof}=147$.}
\begin{ruledtabular}
\begin{tabular}{c c c c c c}
 Sample &   $b_\times$ &  $b_1$ & $b_2$ & $\gamma_2$ & $\chi^2$/dof  \\
\hline
LMz0 & 1.43 & $1.42\pm 0.02$ & $-0.92 \pm 0.08$ & $-0.01 \pm 0.03$ & 1.87 \\
MMz0 & 1.75 & $1.76\pm 0.02$ & $-0.81 \pm 0.08$ & $-0.10 \pm 0.03$& 1.19 \\
HMz0 & 2.66 & $2.61\pm 0.04$ & $1.71 \pm 0.18$ & $-0.48 \pm 0.06$& 2.74 \\ 
\hline
LMz0.5 & 1.88 & $1.83\pm 0.02$ & $-0.46 \pm 0.09$ & $-0.12 \pm 0.03$& 0.84 \\
MMz0.5 & 2.26 & $2.24\pm 0.02$ & $0.05 \pm 0.09$ & $-0.24 \pm 0.03$& 0.67 \\
HMz0.5 & 3.29 & $3.16\pm 0.06$ & $4.10 \pm 0.28$ & $-0.70 \pm 0.10$& 2.91 \\ 
\hline
LMz1 & 2.43 & $2.35\pm 0.03$ & $0.57 \pm 0.13$ & $-0.28 \pm 0.05$& 0.74 \\
MMz1 & 2.86 & $2.80\pm 0.03$ & $1.70 \pm 0.16$ & $-0.42 \pm 0.06$& 0.80 \\
HMz1 & 3.99 & $3.84\pm 0.08$ & $7.55 \pm 0.41$ & $-0.96 \pm 0.16$& 2.73 
\label{BiasParamNL}
\end{tabular} 
\end{ruledtabular} 
\end{table}

We see then that the presence of non-local bias ($\Gd$) required from the multi-dimensional plots in the previous section is confirmed by the bispectrum analysis, which shows that including such terms solves a systematic error in the determination of the linear bias, increasing for more biased objects. This is important because this systematic error would otherwise affect the determination of cosmological parameters from a bispectrum analysis (see~\cite{2006PhRvD..74b3522S}), particularly for luminous  galaxies (such as LRGs in SDSS) that populate high-mass halos.  The extra dependence on  $\Gd$ is also important in at least two more aspects: it introduces a dependence on triangle shape that is degenerate with brane-induced modifications of gravity~\cite{2009PhRvD..80j4005C,2009PhRvD..80j4006S}, and also mimics an equilateral-type primordial non-Gaussianity signature (see Fig.~1 in~\cite{ScoHuiMan1108}). Therefore, for all these reasons, it is important that such dependencies are taken into account when doing bispectrum analyses in galaxy surveys, extending what was done already in~\cite{FelFriFry0102} by considering both Eulerian and Lagrangian local bias models.

\section{Conclusions}

In this paper we studied the non-localities induced in the bias relation by gravitational evolution, providing results under a number of different scenarios. In the simplest case, galaxies form at a single time and evolve conserving their comoving number density (no merging) following the dark matter (no velocity bias). In this case, even if galaxies initially are locally and linearly biased, they develop non-local and non-linear bias that can be best described by the invariants of the deformation tensor (Galileons, see Eqs.~\ref{eq:G2}-\ref{eq:G3}). The main signature of this is a quadrupole field to second-order (already present in~\cite{Fry9604}, and best known from local Lagrangian bias~\cite{CatLucMat9807,CatPorKam0011}). We also derive for the first time the evolution of the bias to third order (Eq.~\ref{deltaNlocform}) and show that it is not only non-local in the density but also in the potential, as a result of the non-local nature of gravitational instability. 

We generalized the second-order result in several directions. First, we considered what happens when merging and formation of galaxies is taken into account, showing that the non-local bias to second-order can still be written due to symmetry reasons in the same Galileon form  but with an amplitude that depends on the evolution of the comoving number density that cannot be made to vanish (and thus keep bias local) unless galaxies are unbiased. Along the way, we generalized the standard formula of the time evolution of linear bias~\cite{MoWhi9609,Fry9604} to the case where merging is important (see Eq.~\ref{eq:b1evolvesource2}).  We also extended the results to the case where there is velocity bias $b_{\rm v}$, finding its time evolution in linear theory, Eq.~(\ref{eq:bvEvolution}). We showed that when there is velocity bias then an extra non-local contribution appears at second-order.  This is related to the breaking of Galilean invariance in the bias relation proportional to $(b_{\rm v}-1)$.  This relative motion between tracers and matter generates a dipole term (see Eqs.~\ref{chi20}-\ref{chi22}). Appendix~\ref{appA} generalizes these results to the case when tracers are not conserved, finding how the non-local bias relation in this case is related to conserved tracers, even if bias at formation and merging is non-local.

To test these results we proceeded in two different ways: by studying the bias relation in simulations at the level of the halo field compared to the matter field (Section~\ref{NumSim}), and statistically by measuring the halo bispectrum in simulations and comparing to the predictions of non-local bias (Section~\ref{BispNonLoc}). 

Regarding the halo bias relation, we found that halo overdensities in cells of {\em fixed} dark matter density indeed vary with the value of the quadratic Galileon field (describing the strength of the tidal field) as expected from our calculation (with stronger dependence for more biased objects, see Figs.~\ref{4DHMz1}-\ref{4DHMz0}). This is a {\em direct} demonstration that such non-local terms are present in the bias relation for dark matter halos. For highly biased halos at high redshift we also found some evidence for the effects of a dipole field (see Fig.~\ref{4DHMz1Dip}), which might be related to velocity bias. More work is needed to establish this precisely in the simulations.

The halo bispectrum measurements in the simulations show a clear detection of the effects of non-local bias as the linear bias increases (see Table~\ref{BiasParamNL} and Fig.~\ref{gamma2res}), fixing a systematic error in the determination of linear bias (also seen recently in~\cite{PolSmiPor1109}) when assuming local Eulerian bias (see Table~\ref{BiasParamE}).  We also performed local Lagrangian bias fits to the halo bispectrum, showing that this model also leads to systematic errors in the determination of the linear bias, although it becomes better for highly biased objects (but the resulting chi-square values are still large). We obtain an approximate relation valid for the least biased halos in our sample (solid line in Fig.~\ref{gamma2res}) between the amplitude of the non-local term $\gamma_2$ and the linear bias inspired by a simple modification of the local Lagrangian bias model that would be interesting to test in other cosmologies and in other linear bias regimes.  Our results indicate that viewed in Lagrangian space bias is also non-local, this can be checked simply by redoing the multidimensional scatter plots presented in Section~\ref{NumSim} in Lagrangian space. 

Accounting for these effects when modeling galaxy bias is important for correctly describing the  dependence on triangle shape of the galaxy bispectrum, and hence constraining cosmological parameters and primordial non-Gaussianity. The non-locality of bias also introduces new angular dependences in the redshift-space power spectrum and bispectrum (since it modifies the angular dependence of perturbation theory kernels) that will be important to include and test against simulations. We hope to report on this in the near future.

 \acknowledgements 
 We thank M.~Crocce, V.~Desjacques, E.~Gazta\~naga, E.~Sefusatti, R.~Smith, M.~Takada, R.~Tojeiro, and I.~Zehavi for useful discussions. 
We thank the participants of the PTChat workshop in Saclay, September 2011, where these results were presented, for comments and discussions. KCC acknowledges the support by James Arthur Graduate Assistantship and Mark Leslie Graduate Assistantship.  RS was partially supported by grants NSF AST-1109432 and NASA NNA10A171G, and thanks RKS and the Physics and Astronomy Department at the University of Pennsylvania  for hospitality during a sabbatical visit in Fall 2009, where this collaboration was started.  RKS was supported in part by NSF AST-0908241 and NASA NNX11A125G.

\appendix

\section{More on Non-Conserved Tracers }
\label{appA}

We have demonstrated in Section~\ref{sec:GalaxyFormation} that even when the tracers are not conserved, due to formation and merging, the structure of the non-local bias up to second order is the same as in the conserved case. Here we extend these results to third order and beyond.  Furthermore, here we also keep the source function  $j$ general, allowing a dependence on other fields in addition to $\delta$, allowing for non-locality in the galaxy formation/merging processes. 

We write $n_{\rm g}^{\rm (c)}  \equiv   \bar{ n}_{\rm g}^{\rm (c)}    ( 1 + \delta_{\rm g} ) $ and introduce a generic dependence of the source on fluctuations through $\psi $ by 
\begin{equation} 
\label{eq:ngand psi}
    j(\rho , \Theta ) \equiv  j( \bar{\rho} , \bar{\Theta} )\,  [ 1 + \psi( \delta ,  \zeta  )   ] ,
\end{equation}
where, in addition to $\delta $, the source function $j$ also depends on $\Theta$, which collectively denotes any extra fields (may even be non-local, e.g. $\mathcal{G}_2 $).  More specifically, the fluctuating part of $j$ is given by 
 \begin{equation}
 \label{fluctj}
\psi( \delta ,  \zeta  ) =    b_1^* \delta + \frac{1}{ 2 } b_2^* \delta^2 +    \frac{1}{ 6 } b_3^* \delta^3   + \dots  + c_1^* \zeta  + \dots        ,  
\end{equation}
where 
\begin{equation} 
b_i^*  \equiv   \bar{ \rho}^i   \frac{\partial^i j }{ \partial \rho^i } \Big|_{ \bar{\rho} , \bar{ \Theta}  }  , \quad 
c_i^*   \equiv   \bar{ \Theta}^i  \frac{\partial^i j }{ \partial \Theta^i } \Big|_{ \bar{\rho} , \bar{\Theta}  } , \quad
\zeta \equiv \frac{  \Theta - \bar{\Theta  } } {  \bar{\Theta  } } . 
\end{equation}
Note that if $\bar{\Theta  }=0$ (e.g. as in the case ${\Theta  }=\mathcal{G}_2$) one can still apply this, as $c_i^* \zeta^i$ is well-behaved in $\bar{\Theta  }$. Using Eq.~(\ref{eq:ngand psi}) in Eq.~(\ref{eq:continuity_source}), we get 
\begin{equation} 
\label{eq:sourceeqgeneral}
\frac{  \partial   } { \partial  y  } ( \bar{n}_{\rm g} ^{ \rm (c) }   \delta_{\rm g} )  -   \bar{n}_{\rm g} ^{ \rm (c)}  \nabla \cdot [ (1+ \delta_{\rm g}  ) \,\mb{v} ]      = \dot{\bar{n}}_{\rm g} ^{ \rm (c)} \,   \psi( \delta, \zeta   ),   
\end{equation}
where a dot denotes a derivative with respect to  $y$  and we have used Eq.~(\ref{eq:background_eq}) to eliminate the background contributions.  We now construct the equation of motion for $ \chi=\delta_{\rm g}-b_1\delta  $ (see Eq.~\ref{chidef}) for the non-conserved case. It obeys,
\begin{eqnarray}
\label{eq:chieq_source}
\frac{  \partial }{   \partial y}  (   \bar{n}_{\rm g} ^{ \rm (c) }    \chi ) & =  &  [ ( b_1 - 1 ) \Delta + \nabla \cdot ( \chi \mb{v}  ) ]\bar{n}_{\rm g} ^{ \rm (c) }     
+ \dot{\bar{n}}_{\rm g} ^{ \rm (c)} (\psi - b_1^* \delta  )         \nonumber 
   \\
     &= &   \bar{n}_{\rm g} ^{ \rm (c)}  \dot{\chi}_{\rm cons}+ \dot{\bar{n}}_{\rm g} ^{ \rm (c)}\, (\psi - b_1^* \delta  )    ,
\end{eqnarray} 
where ${\chi}_{\rm cons}$ denotes the solution when the tracers are conserved (see Eq.~\ref{eq:chi_evolve}). 
In deriving Eq.~(\ref{eq:chieq_source}), we have also used Eq.~(\ref{eq:sourceeqgeneral}), Eq.~(\ref{eq:beff_evolvesource}) and the continuity equation for dark matter.  When  tracers are conserved,  $\dot{\bar{n}}_{\rm g} ^{ \rm (c) }=0 $ and thus Eq.~(\ref{eq:chieq_source}) reduces to Eq.~(\ref{eq:chi_evolve}). The general solution for $\chi$ can then be written in terms of that in the conserved case, namely

\begin{eqnarray}
\label{eq:chisolNC}
 \bar{n}_{\rm g} ^{ \rm (c) }\chi & =&\int_0^y \diff y_*  \,  n_* \, \dot{\chi}_{\rm cons} + \int_0^{  \bar{n}_{\rm g} ^{ \rm (c) }  } \diff  n_*   \,   (  \psi - b_1^* \delta    ), \ \ \   
\end{eqnarray}
which says that the solution for non-conserved tracers is essentially that for conserved tracers weighted by the evolution of the number density (first term) plus a term that depend on the sources that describe galaxy formation and merging ($\psi$). If these sources are local functions of $\delta$, then this extra term does not lead to any non-local contributions, and the non-locality of bias is precisely of the same form as in the conserved case (with slightly different coefficients that depend on the evolution of the comoving number density).  If, on the other hand, galaxy formation or merging depends on non-local functions of $\delta$ (e.g. the velocity divergence, the tidal field) then an extra non-local contribution to galaxy bias gets generated by the second term in Eq.~(\ref{eq:chisolNC}). 

All this implies that the non-local part of the galaxy bias in the general case is related to that for conserved tracers by,

\begin{eqnarray}
\label{deltaGNlocGEN}
 \bar{n}_{\rm g} ^{ \rm (c) }\delta_{\rm g}^{\rm Nloc} & =&\int_0^y \diff y_*  \,  n_* \, \dot{\delta}_{\rm g, cons}^{\rm Nloc}
 + \int_0^{  \bar{n}_{\rm g} ^{ \rm (c) }  } \diff  n_*   \,   \psi^{\rm Nloc},  \nonumber \\ & & 
\end{eqnarray}
or, integrating by parts,
\begin{eqnarray}
\label{deltaGNlocGEN2}
\delta_{\rm g}^{\rm Nloc} & =& \delta_{\rm g, cons}^{\rm Nloc} + {1 \over  \bar{n}_{\rm g} ^{ \rm (c) } }\int_0^{  \bar{n}_{\rm g} ^{ \rm (c) }  } {\diff  n_*}
  \,  ( \psi^{\rm Nloc}-\delta_{\rm g, cons}^{\rm Nloc} ). \nonumber \\ & & 
\end{eqnarray}

This gives our most general expression for the non-local part of galaxy bias when  formation and/or merging cannot be neglected. Although we have implicitly assumed there is no velocity bias, it's easy to check that Eq.~(\ref{deltaGNlocGEN2}) is also valid when there is velocity bias, which only changes $\delta_{\rm g, cons}^{\rm Nloc}$ through dipole terms but not the relationship given by Eq.~(\ref{deltaGNlocGEN2}) itself.

\section{relation of ${\cal G}_2$ and ${\cal G}_3$ to $e$ and $p$}
\label{eANDp}

 The terms induced by gravity are present due to tidal fields, described by derivatives of the  gravitational potential and velocity potential (proportional to each other in linear theory).  To describe the shape of the gravitational potential it is common to introduce the ellipticity $e$ and prolateness $p$, defined from the eigenvalues $\lambda_i$ ($i=1,2,3$) of $\nabla_{ij}\Phi$:
\beqa
 \delta &\equiv& \lambda_1 + \lambda_2 + \lambda_3, \\
 e &\equiv& \frac{\lambda_1 - \lambda_3}{2 \delta}, \\
 p &\equiv& \frac{\lambda_1 + \lambda_3 - 2\lambda_2}{2\delta}.
\label{ep}
\eeqa
This set of parameters is used in triaxial evolution models of nonlinear structure formation~\cite{BonMye9603,SheMoTor01}.  However, because $e$ and $p$ are ratios of the eigenvalues, it is not obvious that they are the best choice of parameters in a perturbative analysis.  In particular, one might have wondered if the rotationally invariariant quantities, 
\beqa
 I_1 &=& {\rm Tr}(D) = \sum_i \lambda_i = \delta, \\
 I_2 &=& \lambda_1\lambda_2 + \lambda_1\lambda_3 + \lambda_2\lambda_3, \\
 I_3 &=& {\rm Det}(D) = \prod_i \lambda_i 
\eeqa
are more relevant.  When expressed in terms of $(\delta,e,p)$ these are 
\beq
 I_2 = \frac{\delta^2}{3}[1 - (3e^2 + p^2)],\qquad
 I_3 = \frac{\delta^3}{27}(1-2p)[(1 + p)^2 - 9e^2]
\eeq
Since the $I_j$ do not depend on taking ratios of the eigenvalues, they, or other quantities built from them, have considerable appeal.  One such combination is 
\beq
 \delta = I_1, \qquad
    r^2 = I_1^2 - 3I_2, \qquad 
    u^3 = \frac{2I_1^3 - 9I_1I_2 + 27 I_3}{9}. \nonumber\\
\eeq
Despite the appearance of $I_1$ in their definition, $r$ and $u$ are actually independent of $I_1$.  Moreover, they are precisely the quantities which arise in a perturbative analysis of the ellipsoidal collapse model:  $J_1$ and $J_2$ of~\cite{OhtKayTar0406} are our $r^2$ and $9u^3$ respectively.  Notice that $e$, $p$, $r^2$ and $u^3$ all vanish for a spherically symmetric perturbation ($\lambda_1 = \lambda_2 = \lambda_3 = \delta/3$).

\bibliography{masterbiblio}

\begin{thebibliography}{89}%
\makeatletter
\providecommand \@ifxundefined [1]{%
 \@ifx{#1\undefined}
}%
\providecommand \@ifnum [1]{%
 \ifnum #1\expandafter \@firstoftwo
 \else \expandafter \@secondoftwo
 \fi
}%
\providecommand \@ifx [1]{%
 \ifx #1\expandafter \@firstoftwo
 \else \expandafter \@secondoftwo
 \fi
}%
\providecommand \natexlab [1]{#1}%
\providecommand \enquote  [1]{``#1''}%
\providecommand \bibnamefont  [1]{#1}%
\providecommand \bibfnamefont [1]{#1}%
\providecommand \citenamefont [1]{#1}%
\providecommand \href@noop [0]{\@secondoftwo}%
\providecommand \href [0]{\begingroup \@sanitize@url \@href}%
\providecommand \@href[1]{\@@startlink{#1}\@@href}%
\providecommand \@@href[1]{\endgroup#1\@@endlink}%
\providecommand \@sanitize@url [0]{\catcode `\\12\catcode `\$12\catcode
  `\&12\catcode `\#12\catcode `\^12\catcode `\_12\catcode `\%12\relax}%
\providecommand \@@startlink[1]{}%
\providecommand \@@endlink[0]{}%
\providecommand \url  [0]{\begingroup\@sanitize@url \@url }%
\providecommand \@url [1]{\endgroup\@href {#1}{\urlprefix }}%
\providecommand \urlprefix  [0]{URL }%
\providecommand \Eprint [0]{\href }%
\providecommand \doibase [0]{http://dx.doi.org/}%
\providecommand \selectlanguage [0]{\@gobble}%
\providecommand \bibinfo  [0]{\@secondoftwo}%
\providecommand \bibfield  [0]{\@secondoftwo}%
\providecommand \translation [1]{[#1]}%
\providecommand \BibitemOpen [0]{}%
\providecommand \bibitemStop [0]{}%
\providecommand \bibitemNoStop [0]{.\EOS\space}%
\providecommand \EOS [0]{\spacefactor3000\relax}%
\providecommand \BibitemShut  [1]{\csname bibitem#1\endcsname}%
\let\auto@bib@innerbib\@empty
\bibitem [{\citenamefont {{Kaiser}}(1984)}]{Kai8409}%
  \BibitemOpen
  \bibfield  {author} {\bibinfo {author} {\bibfnamefont {N.}~\bibnamefont
  {{Kaiser}}},\ }\bibfield  {title} {\enquote {\bibinfo {title} {{On the
  spatial correlations of Abell clusters}},}\ }\href {\doibase 10.1086/184341}
  {\bibfield  {journal} {\bibinfo  {journal} {\apjl}\ }\textbf {\bibinfo
  {volume} {284}},\ \bibinfo {pages} {L9--L12} (\bibinfo {year}
  {1984})}\BibitemShut {NoStop}%
\bibitem [{\citenamefont {{Fry}}\ and\ \citenamefont
  {{Gaztanaga}}(1993)}]{FryGaz9308}%
  \BibitemOpen
  \bibfield  {author} {\bibinfo {author} {\bibfnamefont {J.~N.}\ \bibnamefont
  {{Fry}}}\ and\ \bibinfo {author} {\bibfnamefont {E.}~\bibnamefont
  {{Gaztanaga}}},\ }\bibfield  {title} {\enquote {\bibinfo {title} {{Biasing
  and hierarchical statistics in large-scale structure}},}\ }\href {\doibase
  10.1086/173015} {\bibfield  {journal} {\bibinfo  {journal} {\apj}\ }\textbf
  {\bibinfo {volume} {413}},\ \bibinfo {pages} {447--452} (\bibinfo {year}
  {1993})},\ \Eprint {http://arxiv.org/abs/arXiv:astro-ph/9302009}
  {arXiv:astro-ph/9302009} \BibitemShut {NoStop}%
\bibitem [{\citenamefont {{Scherrer}}\ and\ \citenamefont
  {{Weinberg}}(1998)}]{SchWei98}%
  \BibitemOpen
  \bibfield  {author} {\bibinfo {author} {\bibfnamefont {R.~J.}\ \bibnamefont
  {{Scherrer}}}\ and\ \bibinfo {author} {\bibfnamefont {D.~H.}\ \bibnamefont
  {{Weinberg}}},\ }\bibfield  {title} {\enquote {\bibinfo {title} {{Constraints
  on the Effects of Locally Biased Galaxy Formation}},}\ }\href {\doibase
  10.1086/306113} {\bibfield  {journal} {\bibinfo  {journal} {\apj}\ }\textbf
  {\bibinfo {volume} {504}},\ \bibinfo {pages} {607--+} (\bibinfo {year}
  {1998})}\BibitemShut {NoStop}%
\bibitem [{\citenamefont {{Matsubara}}(1999)}]{Mat99}%
  \BibitemOpen
  \bibfield  {author} {\bibinfo {author} {\bibfnamefont {T.}~\bibnamefont
  {{Matsubara}}},\ }\bibfield  {title} {\enquote {\bibinfo {title}
  {{Stochasticity of Bias and Nonlocality of Galaxy Formation: Linear
  Scales}},}\ }\href@noop {} {\bibfield  {journal} {\bibinfo  {journal} {\apj}\
  }\textbf {\bibinfo {volume} {525}},\ \bibinfo {pages} {543--553} (\bibinfo
  {year} {1999})}\BibitemShut {NoStop}%
\bibitem [{\citenamefont {{Frieman}}\ and\ \citenamefont
  {{Gaztanaga}}(1994)}]{FriGaz94}%
  \BibitemOpen
  \bibfield  {author} {\bibinfo {author} {\bibfnamefont {J.{}A.}\ \bibnamefont
  {{Frieman}}}\ and\ \bibinfo {author} {\bibfnamefont {E.}~\bibnamefont
  {{Gaztanaga}}},\ }\bibfield  {title} {\enquote {\bibinfo {title} {{The
  three-point function as a probe of models for large-scale structure}},}\
  }\href@noop {} {\bibfield  {journal} {\bibinfo  {journal} {\apj}\ }\textbf
  {\bibinfo {volume} {425}},\ \bibinfo {pages} {392--402} (\bibinfo {year}
  {1994})}\BibitemShut {NoStop}%
\bibitem [{\citenamefont {{Fry}}(1994)}]{Fry94}%
  \BibitemOpen
  \bibfield  {author} {\bibinfo {author} {\bibfnamefont {J.{}N.}\ \bibnamefont
  {{Fry}}},\ }\bibfield  {title} {\enquote {\bibinfo {title} {{Gravity, bias,
  and the galaxy three-point correlation function}},}\ }\href@noop {}
  {\bibfield  {journal} {\bibinfo  {journal} {Physical Review Letters}\
  }\textbf {\bibinfo {volume} {73}},\ \bibinfo {pages} {215--219} (\bibinfo
  {year} {1994})}\BibitemShut {NoStop}%
\bibitem [{\citenamefont {{Gaztanaga}}\ and\ \citenamefont
  {{Frieman}}(1994)}]{GazFri94}%
  \BibitemOpen
  \bibfield  {author} {\bibinfo {author} {\bibfnamefont {E.}~\bibnamefont
  {{Gaztanaga}}}\ and\ \bibinfo {author} {\bibfnamefont {J.{}A.}\ \bibnamefont
  {{Frieman}}},\ }\bibfield  {title} {\enquote {\bibinfo {title} {{Bias and
  high-order galaxy correlation functions in the APM galaxy survey}},}\
  }\href@noop {} {\bibfield  {journal} {\bibinfo  {journal} {\apjl}\ }\textbf
  {\bibinfo {volume} {437}},\ \bibinfo {pages} {L13--L16} (\bibinfo {year}
  {1994})}\BibitemShut {NoStop}%
\bibitem [{\citenamefont {{Scoccimarro}}\ \emph
  {et~al.}(2001{\natexlab{a}})\citenamefont {{Scoccimarro}}, \citenamefont
  {{Feldman}}, \citenamefont {{Fry}},\ and\ \citenamefont
  {{Frieman}}}]{ScoFelFry0101}%
  \BibitemOpen
  \bibfield  {author} {\bibinfo {author} {\bibfnamefont {R.}~\bibnamefont
  {{Scoccimarro}}}, \bibinfo {author} {\bibfnamefont {H.~A.}\ \bibnamefont
  {{Feldman}}}, \bibinfo {author} {\bibfnamefont {J.~N.}\ \bibnamefont
  {{Fry}}}, \ and\ \bibinfo {author} {\bibfnamefont {J.~A.}\ \bibnamefont
  {{Frieman}}},\ }\bibfield  {title} {\enquote {\bibinfo {title} {{The
  Bispectrum of IRAS Redshift Catalogs}},}\ }\href {\doibase 10.1086/318284}
  {\bibfield  {journal} {\bibinfo  {journal} {\apj}\ }\textbf {\bibinfo
  {volume} {546}},\ \bibinfo {pages} {652--664} (\bibinfo {year}
  {2001}{\natexlab{a}})},\ \Eprint
  {http://arxiv.org/abs/arXiv:astro-ph/0004087} {arXiv:astro-ph/0004087}
  \BibitemShut {NoStop}%
\bibitem [{\citenamefont {{Feldman}}\ \emph {et~al.}(2001)\citenamefont
  {{Feldman}}, \citenamefont {{Frieman}}, \citenamefont {{Fry}},\ and\
  \citenamefont {{Scoccimarro}}}]{FelFriFry0102}%
  \BibitemOpen
  \bibfield  {author} {\bibinfo {author} {\bibfnamefont {H.~A.}\ \bibnamefont
  {{Feldman}}}, \bibinfo {author} {\bibfnamefont {J.~A.}\ \bibnamefont
  {{Frieman}}}, \bibinfo {author} {\bibfnamefont {J.~N.}\ \bibnamefont
  {{Fry}}}, \ and\ \bibinfo {author} {\bibfnamefont {R.}~\bibnamefont
  {{Scoccimarro}}},\ }\bibfield  {title} {\enquote {\bibinfo {title}
  {{Constraints on Galaxy Bias, Matter Density, and Primordial Non-Gaussianity
  from the PSCz Galaxy Redshift Survey}},}\ }\href {\doibase
  10.1103/PhysRevLett.86.1434} {\bibfield  {journal} {\bibinfo  {journal}
  {Physical Review Letters}\ }\textbf {\bibinfo {volume} {86}},\ \bibinfo
  {pages} {1434--1437} (\bibinfo {year} {2001})},\ \Eprint
  {http://arxiv.org/abs/arXiv:astro-ph/0010205} {arXiv:astro-ph/0010205}
  \BibitemShut {NoStop}%
\bibitem [{\citenamefont {{Verde}}\ \emph {et~al.}(2002)\citenamefont
  {{Verde}}, \citenamefont {{Heavens}}, \citenamefont {{Percival}},
  \citenamefont {{Matarrese}}, \citenamefont {{Baugh}}, \citenamefont
  {{Bland-Hawthorn}}, \citenamefont {{Bridges}}, \citenamefont {{Cannon}},
  \citenamefont {{Cole}}, \citenamefont {{Colless}}, \citenamefont {{Collins}},
  \citenamefont {{Couch}}, \citenamefont {{Dalton}}, \citenamefont {{De
  Propris}}, \citenamefont {{Driver}}, \citenamefont {{Efstathiou}},
  \citenamefont {{Ellis}}, \citenamefont {{Frenk}}, \citenamefont
  {{Glazebrook}}, \citenamefont {{Jackson}}, \citenamefont {{Lahav}},
  \citenamefont {{Lewis}}, \citenamefont {{Lumsden}}, \citenamefont {{Maddox}},
  \citenamefont {{Madgwick}}, \citenamefont {{Norberg}}, \citenamefont
  {{Peacock}}, \citenamefont {{Peterson}}, \citenamefont {{Sutherland}},\ and\
  \citenamefont {{Taylor}}}]{VerHeaPer02}%
  \BibitemOpen
  \bibfield  {author} {\bibinfo {author} {\bibfnamefont {L.}~\bibnamefont
  {{Verde}}}, \bibinfo {author} {\bibfnamefont {A.{}F.}\ \bibnamefont
  {{Heavens}}}, \bibinfo {author} {\bibfnamefont {W.{}J.}\ \bibnamefont
  {{Percival}}}, \bibinfo {author} {\bibfnamefont {S.}~\bibnamefont
  {{Matarrese}}}, \bibinfo {author} {\bibfnamefont {C.{}M.}\ \bibnamefont
  {{Baugh}}}, \bibinfo {author} {\bibfnamefont {J.}~\bibnamefont
  {{Bland-Hawthorn}}}, \bibinfo {author} {\bibfnamefont {T.}~\bibnamefont
  {{Bridges}}}, \bibinfo {author} {\bibfnamefont {R.}~\bibnamefont {{Cannon}}},
  \bibinfo {author} {\bibfnamefont {S.}~\bibnamefont {{Cole}}}, \bibinfo
  {author} {\bibfnamefont {M.}~\bibnamefont {{Colless}}}, \bibinfo {author}
  {\bibfnamefont {C.}~\bibnamefont {{Collins}}}, \bibinfo {author}
  {\bibfnamefont {W.}~\bibnamefont {{Couch}}}, \bibinfo {author} {\bibfnamefont
  {G.}~\bibnamefont {{Dalton}}}, \bibinfo {author} {\bibfnamefont
  {R.}~\bibnamefont {{De Propris}}}, \bibinfo {author} {\bibfnamefont {S.{}P.}\
  \bibnamefont {{Driver}}}, \bibinfo {author} {\bibfnamefont {G.}~\bibnamefont
  {{Efstathiou}}}, \bibinfo {author} {\bibfnamefont {R.{}S.}\ \bibnamefont
  {{Ellis}}}, \bibinfo {author} {\bibfnamefont {C.{}S.}\ \bibnamefont
  {{Frenk}}}, \bibinfo {author} {\bibfnamefont {K.}~\bibnamefont
  {{Glazebrook}}}, \bibinfo {author} {\bibfnamefont {C.}~\bibnamefont
  {{Jackson}}}, \bibinfo {author} {\bibfnamefont {O.}~\bibnamefont {{Lahav}}},
  \bibinfo {author} {\bibfnamefont {I.}~\bibnamefont {{Lewis}}}, \bibinfo
  {author} {\bibfnamefont {S.}~\bibnamefont {{Lumsden}}}, \bibinfo {author}
  {\bibfnamefont {S.}~\bibnamefont {{Maddox}}}, \bibinfo {author}
  {\bibfnamefont {D.}~\bibnamefont {{Madgwick}}}, \bibinfo {author}
  {\bibfnamefont {P.}~\bibnamefont {{Norberg}}}, \bibinfo {author}
  {\bibfnamefont {J.{}A.}\ \bibnamefont {{Peacock}}}, \bibinfo {author}
  {\bibfnamefont {B.{}A.}\ \bibnamefont {{Peterson}}}, \bibinfo {author}
  {\bibfnamefont {W.}~\bibnamefont {{Sutherland}}}, \ and\ \bibinfo {author}
  {\bibfnamefont {K.}~\bibnamefont {{Taylor}}},\ }\bibfield  {title} {\enquote
  {\bibinfo {title} {{The 2dF Galaxy Redshift Survey: the bias of galaxies and
  the density of the Universe}},}\ }\href@noop {} {\bibfield  {journal}
  {\bibinfo  {journal} {\mnras}\ }\textbf {\bibinfo {volume} {335}},\ \bibinfo
  {pages} {432--440} (\bibinfo {year} {2002})}\BibitemShut {NoStop}%
\bibitem [{\citenamefont {{Baugh}}\ \emph {et~al.}(2004)\citenamefont
  {{Baugh}}, \citenamefont {{Croton}}, \citenamefont {{Gazta{\~n}aga}},
  \citenamefont {{Norberg}}, \citenamefont {{Colless}}, \citenamefont
  {{Baldry}}, \citenamefont {{Bland-Hawthorn}}, \citenamefont {{Bridges}},
  \citenamefont {{Cannon}}, \citenamefont {{Cole}}, \citenamefont {{Collins}},
  \citenamefont {{Couch}}, \citenamefont {{Dalton}}, \citenamefont {{De
  Propris}}, \citenamefont {{Driver}}, \citenamefont {{Efstathiou}},
  \citenamefont {{Ellis}}, \citenamefont {{Frenk}}, \citenamefont
  {{Glazebrook}}, \citenamefont {{Jackson}}, \citenamefont {{Lahav}},
  \citenamefont {{Lewis}}, \citenamefont {{Lumsden}}, \citenamefont {{Maddox}},
  \citenamefont {{Madgwick}}, \citenamefont {{Peacock}}, \citenamefont
  {{Peterson}}, \citenamefont {{Sutherland}},\ and\ \citenamefont
  {{Taylor}}}]{BauCroGaz04}%
  \BibitemOpen
  \bibfield  {author} {\bibinfo {author} {\bibfnamefont {C.~M.}\ \bibnamefont
  {{Baugh}}}, \bibinfo {author} {\bibfnamefont {D.~J.}\ \bibnamefont
  {{Croton}}}, \bibinfo {author} {\bibfnamefont {E.}~\bibnamefont
  {{Gazta{\~n}aga}}}, \bibinfo {author} {\bibfnamefont {P.}~\bibnamefont
  {{Norberg}}}, \bibinfo {author} {\bibfnamefont {M.}~\bibnamefont
  {{Colless}}}, \bibinfo {author} {\bibfnamefont {I.~K.}\ \bibnamefont
  {{Baldry}}}, \bibinfo {author} {\bibfnamefont {J.}~\bibnamefont
  {{Bland-Hawthorn}}}, \bibinfo {author} {\bibfnamefont {T.}~\bibnamefont
  {{Bridges}}}, \bibinfo {author} {\bibfnamefont {R.}~\bibnamefont {{Cannon}}},
  \bibinfo {author} {\bibfnamefont {S.}~\bibnamefont {{Cole}}}, \bibinfo
  {author} {\bibfnamefont {C.}~\bibnamefont {{Collins}}}, \bibinfo {author}
  {\bibfnamefont {W.}~\bibnamefont {{Couch}}}, \bibinfo {author} {\bibfnamefont
  {G.}~\bibnamefont {{Dalton}}}, \bibinfo {author} {\bibfnamefont
  {R.}~\bibnamefont {{De Propris}}}, \bibinfo {author} {\bibfnamefont {S.~P.}\
  \bibnamefont {{Driver}}}, \bibinfo {author} {\bibfnamefont {G.}~\bibnamefont
  {{Efstathiou}}}, \bibinfo {author} {\bibfnamefont {R.~S.}\ \bibnamefont
  {{Ellis}}}, \bibinfo {author} {\bibfnamefont {C.~S.}\ \bibnamefont
  {{Frenk}}}, \bibinfo {author} {\bibfnamefont {K.}~\bibnamefont
  {{Glazebrook}}}, \bibinfo {author} {\bibfnamefont {C.}~\bibnamefont
  {{Jackson}}}, \bibinfo {author} {\bibfnamefont {O.}~\bibnamefont {{Lahav}}},
  \bibinfo {author} {\bibfnamefont {I.}~\bibnamefont {{Lewis}}}, \bibinfo
  {author} {\bibfnamefont {S.}~\bibnamefont {{Lumsden}}}, \bibinfo {author}
  {\bibfnamefont {S.}~\bibnamefont {{Maddox}}}, \bibinfo {author}
  {\bibfnamefont {D.}~\bibnamefont {{Madgwick}}}, \bibinfo {author}
  {\bibfnamefont {J.~A.}\ \bibnamefont {{Peacock}}}, \bibinfo {author}
  {\bibfnamefont {B.~A.}\ \bibnamefont {{Peterson}}}, \bibinfo {author}
  {\bibfnamefont {W.}~\bibnamefont {{Sutherland}}}, \ and\ \bibinfo {author}
  {\bibfnamefont {K.}~\bibnamefont {{Taylor}}},\ }\bibfield  {title} {\enquote
  {\bibinfo {title} {{The 2dF Galaxy Redshift Survey: hierarchical galaxy
  clustering}},}\ }\href {\doibase 10.1111/j.1365-2966.2004.07962.x} {\bibfield
   {journal} {\bibinfo  {journal} {\mnras}\ }\textbf {\bibinfo {volume}
  {351}},\ \bibinfo {pages} {L44--L48} (\bibinfo {year} {2004})}\BibitemShut
  {NoStop}%
\bibitem [{\citenamefont {{Croton}}\ \emph {et~al.}(2004)\citenamefont
  {{Croton}}, \citenamefont {{Gazta{\~n}aga}}, \citenamefont {{Baugh}},
  \citenamefont {{Norberg}}, \citenamefont {{Colless}}, \citenamefont
  {{Baldry}}, \citenamefont {{Bland-Hawthorn}}, \citenamefont {{Bridges}},
  \citenamefont {{Cannon}}, \citenamefont {{Cole}}, \citenamefont {{Collins}},
  \citenamefont {{Couch}}, \citenamefont {{Dalton}}, \citenamefont {{De
  Propris}}, \citenamefont {{Driver}}, \citenamefont {{Efstathiou}},
  \citenamefont {{Ellis}}, \citenamefont {{Frenk}}, \citenamefont
  {{Glazebrook}}, \citenamefont {{Jackson}}, \citenamefont {{Lahav}},
  \citenamefont {{Lewis}}, \citenamefont {{Lumsden}}, \citenamefont {{Maddox}},
  \citenamefont {{Madgwick}}, \citenamefont {{Peacock}}, \citenamefont
  {{Peterson}}, \citenamefont {{Sutherland}},\ and\ \citenamefont
  {{Taylor}}}]{CroGazBau04}%
  \BibitemOpen
  \bibfield  {author} {\bibinfo {author} {\bibfnamefont {D.~J.}\ \bibnamefont
  {{Croton}}}, \bibinfo {author} {\bibfnamefont {E.}~\bibnamefont
  {{Gazta{\~n}aga}}}, \bibinfo {author} {\bibfnamefont {C.~M.}\ \bibnamefont
  {{Baugh}}}, \bibinfo {author} {\bibfnamefont {P.}~\bibnamefont {{Norberg}}},
  \bibinfo {author} {\bibfnamefont {M.}~\bibnamefont {{Colless}}}, \bibinfo
  {author} {\bibfnamefont {I.~K.}\ \bibnamefont {{Baldry}}}, \bibinfo {author}
  {\bibfnamefont {J.}~\bibnamefont {{Bland-Hawthorn}}}, \bibinfo {author}
  {\bibfnamefont {T.}~\bibnamefont {{Bridges}}}, \bibinfo {author}
  {\bibfnamefont {R.}~\bibnamefont {{Cannon}}}, \bibinfo {author}
  {\bibfnamefont {S.}~\bibnamefont {{Cole}}}, \bibinfo {author} {\bibfnamefont
  {C.}~\bibnamefont {{Collins}}}, \bibinfo {author} {\bibfnamefont
  {W.}~\bibnamefont {{Couch}}}, \bibinfo {author} {\bibfnamefont
  {G.}~\bibnamefont {{Dalton}}}, \bibinfo {author} {\bibfnamefont
  {R.}~\bibnamefont {{De Propris}}}, \bibinfo {author} {\bibfnamefont {S.~P.}\
  \bibnamefont {{Driver}}}, \bibinfo {author} {\bibfnamefont {G.}~\bibnamefont
  {{Efstathiou}}}, \bibinfo {author} {\bibfnamefont {R.~S.}\ \bibnamefont
  {{Ellis}}}, \bibinfo {author} {\bibfnamefont {C.~S.}\ \bibnamefont
  {{Frenk}}}, \bibinfo {author} {\bibfnamefont {K.}~\bibnamefont
  {{Glazebrook}}}, \bibinfo {author} {\bibfnamefont {C.}~\bibnamefont
  {{Jackson}}}, \bibinfo {author} {\bibfnamefont {O.}~\bibnamefont {{Lahav}}},
  \bibinfo {author} {\bibfnamefont {I.}~\bibnamefont {{Lewis}}}, \bibinfo
  {author} {\bibfnamefont {S.}~\bibnamefont {{Lumsden}}}, \bibinfo {author}
  {\bibfnamefont {S.}~\bibnamefont {{Maddox}}}, \bibinfo {author}
  {\bibfnamefont {D.}~\bibnamefont {{Madgwick}}}, \bibinfo {author}
  {\bibfnamefont {J.~A.}\ \bibnamefont {{Peacock}}}, \bibinfo {author}
  {\bibfnamefont {B.~A.}\ \bibnamefont {{Peterson}}}, \bibinfo {author}
  {\bibfnamefont {W.}~\bibnamefont {{Sutherland}}}, \ and\ \bibinfo {author}
  {\bibfnamefont {K.}~\bibnamefont {{Taylor}}},\ }\bibfield  {title} {\enquote
  {\bibinfo {title} {{The 2dF Galaxy Redshift Survey: higher-order galaxy
  correlation functions}},}\ }\href {\doibase 10.1111/j.1365-2966.2004.08017.x}
  {\bibfield  {journal} {\bibinfo  {journal} {\mnras}\ }\textbf {\bibinfo
  {volume} {352}},\ \bibinfo {pages} {1232--1244} (\bibinfo {year}
  {2004})}\BibitemShut {NoStop}%
\bibitem [{\citenamefont {{Jing}}\ and\ \citenamefont
  {{B{\"o}rner}}(2004)}]{JinBor0405}%
  \BibitemOpen
  \bibfield  {author} {\bibinfo {author} {\bibfnamefont {Y.~P.}\ \bibnamefont
  {{Jing}}}\ and\ \bibinfo {author} {\bibfnamefont {G.}~\bibnamefont
  {{B{\"o}rner}}},\ }\bibfield  {title} {\enquote {\bibinfo {title} {{The
  Three-Point Correlation Function of Galaxies Determined from the Two-Degree
  Field Galaxy Redshift Survey}},}\ }\href {\doibase 10.1086/383343} {\bibfield
   {journal} {\bibinfo  {journal} {\apj}\ }\textbf {\bibinfo {volume} {607}},\
  \bibinfo {pages} {140--163} (\bibinfo {year} {2004})},\ \Eprint
  {http://arxiv.org/abs/arXiv:astro-ph/0311585} {arXiv:astro-ph/0311585}
  \BibitemShut {NoStop}%
\bibitem [{\citenamefont {{Kayo}}\ \emph {et~al.}(2004)\citenamefont {{Kayo}},
  \citenamefont {{Suto}}, \citenamefont {{Nichol}}, \citenamefont {{Pan}},
  \citenamefont {{Szapudi}}, \citenamefont {{Connolly}}, \citenamefont
  {{Gardner}}, \citenamefont {{Jain}}, \citenamefont {{Kulkarni}},
  \citenamefont {{Matsubara}}, \citenamefont {{Sheth}}, \citenamefont
  {{Szalay}},\ and\ \citenamefont {{Brinkmann}}}]{KaySutNic0406}%
  \BibitemOpen
  \bibfield  {author} {\bibinfo {author} {\bibfnamefont {I.}~\bibnamefont
  {{Kayo}}}, \bibinfo {author} {\bibfnamefont {Y.}~\bibnamefont {{Suto}}},
  \bibinfo {author} {\bibfnamefont {R.~C.}\ \bibnamefont {{Nichol}}}, \bibinfo
  {author} {\bibfnamefont {J.}~\bibnamefont {{Pan}}}, \bibinfo {author}
  {\bibfnamefont {I.}~\bibnamefont {{Szapudi}}}, \bibinfo {author}
  {\bibfnamefont {A.~J.}\ \bibnamefont {{Connolly}}}, \bibinfo {author}
  {\bibfnamefont {J.}~\bibnamefont {{Gardner}}}, \bibinfo {author}
  {\bibfnamefont {B.}~\bibnamefont {{Jain}}}, \bibinfo {author} {\bibfnamefont
  {G.}~\bibnamefont {{Kulkarni}}}, \bibinfo {author} {\bibfnamefont
  {T.}~\bibnamefont {{Matsubara}}}, \bibinfo {author} {\bibfnamefont
  {R.}~\bibnamefont {{Sheth}}}, \bibinfo {author} {\bibfnamefont {A.~S.}\
  \bibnamefont {{Szalay}}}, \ and\ \bibinfo {author} {\bibfnamefont
  {J.}~\bibnamefont {{Brinkmann}}},\ }\bibfield  {title} {\enquote {\bibinfo
  {title} {{Three-Point Correlation Functions of SDSS Galaxies in Redshift
  Space: Morphology, Color, and Luminosity Dependence}},}\ }\href@noop {}
  {\bibfield  {journal} {\bibinfo  {journal} {\pasj}\ }\textbf {\bibinfo
  {volume} {56}},\ \bibinfo {pages} {415--423} (\bibinfo {year} {2004})},\
  \Eprint {http://arxiv.org/abs/arXiv:astro-ph/0403638}
  {arXiv:astro-ph/0403638} \BibitemShut {NoStop}%
\bibitem [{\citenamefont {{Pan}}\ and\ \citenamefont
  {{Szapudi}}(2005)}]{PanSza0510}%
  \BibitemOpen
  \bibfield  {author} {\bibinfo {author} {\bibfnamefont {J.}~\bibnamefont
  {{Pan}}}\ and\ \bibinfo {author} {\bibfnamefont {I.}~\bibnamefont
  {{Szapudi}}},\ }\bibfield  {title} {\enquote {\bibinfo {title} {{The monopole
  moment of the three-point correlation function of the two-degree Field Galaxy
  Redshift Survey}},}\ }\href {\doibase 10.1111/j.1365-2966.2005.09407.x}
  {\bibfield  {journal} {\bibinfo  {journal} {\mnras}\ }\textbf {\bibinfo
  {volume} {362}},\ \bibinfo {pages} {1363--1370} (\bibinfo {year} {2005})},\
  \Eprint {http://arxiv.org/abs/arXiv:astro-ph/0505422}
  {arXiv:astro-ph/0505422} \BibitemShut {NoStop}%
\bibitem [{\citenamefont {{Gazta{\~n}aga}}\ \emph {et~al.}(2005)\citenamefont
  {{Gazta{\~n}aga}}, \citenamefont {{Norberg}}, \citenamefont {{Baugh}},\ and\
  \citenamefont {{Croton}}}]{GazNorBau0512}%
  \BibitemOpen
  \bibfield  {author} {\bibinfo {author} {\bibfnamefont {E.}~\bibnamefont
  {{Gazta{\~n}aga}}}, \bibinfo {author} {\bibfnamefont {P.}~\bibnamefont
  {{Norberg}}}, \bibinfo {author} {\bibfnamefont {C.~M.}\ \bibnamefont
  {{Baugh}}}, \ and\ \bibinfo {author} {\bibfnamefont {D.~J.}\ \bibnamefont
  {{Croton}}},\ }\bibfield  {title} {\enquote {\bibinfo {title} {{Statistical
  analysis of galaxy surveys - II. The three-point galaxy correlation function
  measured from the 2dFGRS}},}\ }\href {\doibase
  10.1111/j.1365-2966.2005.09583.x} {\bibfield  {journal} {\bibinfo  {journal}
  {\mnras}\ }\textbf {\bibinfo {volume} {364}},\ \bibinfo {pages} {620--634}
  (\bibinfo {year} {2005})},\ \Eprint
  {http://arxiv.org/abs/arXiv:astro-ph/0506249} {arXiv:astro-ph/0506249}
  \BibitemShut {NoStop}%
\bibitem [{\citenamefont {{Nishimichi}}\ \emph {et~al.}(2007)\citenamefont
  {{Nishimichi}}, \citenamefont {{Kayo}}, \citenamefont {{Hikage}},
  \citenamefont {{Yahata}}, \citenamefont {{Taruya}}, \citenamefont {{Jing}},
  \citenamefont {{Sheth}},\ and\ \citenamefont {{Suto}}}]{NisKayHik0702}%
  \BibitemOpen
  \bibfield  {author} {\bibinfo {author} {\bibfnamefont {T.}~\bibnamefont
  {{Nishimichi}}}, \bibinfo {author} {\bibfnamefont {I.}~\bibnamefont
  {{Kayo}}}, \bibinfo {author} {\bibfnamefont {C.}~\bibnamefont {{Hikage}}},
  \bibinfo {author} {\bibfnamefont {K.}~\bibnamefont {{Yahata}}}, \bibinfo
  {author} {\bibfnamefont {A.}~\bibnamefont {{Taruya}}}, \bibinfo {author}
  {\bibfnamefont {Y.~P.}\ \bibnamefont {{Jing}}}, \bibinfo {author}
  {\bibfnamefont {R.~K.}\ \bibnamefont {{Sheth}}}, \ and\ \bibinfo {author}
  {\bibfnamefont {Y.}~\bibnamefont {{Suto}}},\ }\bibfield  {title} {\enquote
  {\bibinfo {title} {{Bispectrum and Nonlinear Biasing of Galaxies:
  Perturbation Analysis, Numerical Simulation, and SDSS Galaxy Clustering}},}\
  }\href@noop {} {\bibfield  {journal} {\bibinfo  {journal} {\pasj}\ }\textbf
  {\bibinfo {volume} {59}},\ \bibinfo {pages} {93--106} (\bibinfo {year}
  {2007})},\ \Eprint {http://arxiv.org/abs/arXiv:astro-ph/0609740}
  {arXiv:astro-ph/0609740} \BibitemShut {NoStop}%
\bibitem [{\citenamefont {{Mar{\'{\i}}n}}\ \emph {et~al.}(2008)\citenamefont
  {{Mar{\'{\i}}n}}, \citenamefont {{Wechsler}}, \citenamefont {{Frieman}},\
  and\ \citenamefont {{Nichol}}}]{MarWecFri0801}%
  \BibitemOpen
  \bibfield  {author} {\bibinfo {author} {\bibfnamefont {F.~A.}\ \bibnamefont
  {{Mar{\'{\i}}n}}}, \bibinfo {author} {\bibfnamefont {R.~H.}\ \bibnamefont
  {{Wechsler}}}, \bibinfo {author} {\bibfnamefont {J.~A.}\ \bibnamefont
  {{Frieman}}}, \ and\ \bibinfo {author} {\bibfnamefont {R.~C.}\ \bibnamefont
  {{Nichol}}},\ }\bibfield  {title} {\enquote {\bibinfo {title} {{Modeling the
  Galaxy Three-Point Correlation Function}},}\ }\href {\doibase 10.1086/523628}
  {\bibfield  {journal} {\bibinfo  {journal} {\apj}\ }\textbf {\bibinfo
  {volume} {672}},\ \bibinfo {pages} {849--860} (\bibinfo {year} {2008})},\
  \Eprint {http://arxiv.org/abs/0704.0255} {arXiv:0704.0255} \BibitemShut
  {NoStop}%
\bibitem [{\citenamefont {{McBride}}\ \emph
  {et~al.}(2010{\natexlab{a}})\citenamefont {{McBride}}, \citenamefont
  {{Connolly}}, \citenamefont {{Gardner}}, \citenamefont {{Scranton}},
  \citenamefont {{Scoccimarro}}, \citenamefont {{Berlind}}, \citenamefont
  {{Marin}},\ and\ \citenamefont {{Schneider}}}]{McBConGar1012}%
  \BibitemOpen
  \bibfield  {author} {\bibinfo {author} {\bibfnamefont {C.~K.}\ \bibnamefont
  {{McBride}}}, \bibinfo {author} {\bibfnamefont {A.~J.}\ \bibnamefont
  {{Connolly}}}, \bibinfo {author} {\bibfnamefont {J.~P.}\ \bibnamefont
  {{Gardner}}}, \bibinfo {author} {\bibfnamefont {R.}~\bibnamefont
  {{Scranton}}}, \bibinfo {author} {\bibfnamefont {R.}~\bibnamefont
  {{Scoccimarro}}}, \bibinfo {author} {\bibfnamefont {A.~A.}\ \bibnamefont
  {{Berlind}}}, \bibinfo {author} {\bibfnamefont {F.}~\bibnamefont {{Marin}}},
  \ and\ \bibinfo {author} {\bibfnamefont {D.~P.}\ \bibnamefont
  {{Schneider}}},\ }\bibfield  {title} {\enquote {\bibinfo {title}
  {{Three-Point Correlation Functions of SDSS Galaxies: Constraining
  Galaxy-Mass Bias}},}\ }\href@noop {} {\bibfield  {journal} {\bibinfo
  {journal} {ArXiv e-prints}\ } (\bibinfo {year} {2010}{\natexlab{a}})},\
  \Eprint {http://arxiv.org/abs/1012.3462} {arXiv:1012.3462 [astro-ph.CO]}
  \BibitemShut {NoStop}%
\bibitem [{\citenamefont {{McBride}}\ \emph
  {et~al.}(2010{\natexlab{b}})\citenamefont {{McBride}}, \citenamefont
  {{Connolly}}, \citenamefont {{Gardner}}, \citenamefont {{Scranton}},
  \citenamefont {{Newman}}, \citenamefont {{Scoccimarro}}, \citenamefont
  {{Zehavi}},\ and\ \citenamefont {{Schneider}}}]{McBConGar1007}%
  \BibitemOpen
  \bibfield  {author} {\bibinfo {author} {\bibfnamefont {C.~K.}\ \bibnamefont
  {{McBride}}}, \bibinfo {author} {\bibfnamefont {A.~J.}\ \bibnamefont
  {{Connolly}}}, \bibinfo {author} {\bibfnamefont {J.~P.}\ \bibnamefont
  {{Gardner}}}, \bibinfo {author} {\bibfnamefont {R.}~\bibnamefont
  {{Scranton}}}, \bibinfo {author} {\bibfnamefont {J.~A.}\ \bibnamefont
  {{Newman}}}, \bibinfo {author} {\bibfnamefont {R.}~\bibnamefont
  {{Scoccimarro}}}, \bibinfo {author} {\bibfnamefont {I.}~\bibnamefont
  {{Zehavi}}}, \ and\ \bibinfo {author} {\bibfnamefont {D.~P.}\ \bibnamefont
  {{Schneider}}},\ }\bibfield  {title} {\enquote {\bibinfo {title}
  {{Three-Point Correlation Functions of SDSS Galaxies: Luminosity and Color
  Dependence in Redshift and Projected Space}},}\ }\href@noop {} {\bibfield
  {journal} {\bibinfo  {journal} {ArXiv e-prints}\ } (\bibinfo {year}
  {2010}{\natexlab{b}})},\ \Eprint {http://arxiv.org/abs/1007.2414}
  {arXiv:1007.2414 [astro-ph.CO]} \BibitemShut {NoStop}%
\bibitem [{\citenamefont {{Mar{\'{\i}}n}}(2011)}]{Mar1108}%
  \BibitemOpen
  \bibfield  {author} {\bibinfo {author} {\bibfnamefont {F.}~\bibnamefont
  {{Mar{\'{\i}}n}}},\ }\bibfield  {title} {\enquote {\bibinfo {title} {{The
  Large-scale Three-point Correlation Function of Sloan Digital Sky Survey
  Luminous Red Galaxies}},}\ }\href {\doibase 10.1088/0004-637X/737/2/97}
  {\bibfield  {journal} {\bibinfo  {journal} {\apj}\ }\textbf {\bibinfo
  {volume} {737}},\ \bibinfo {eid} {97} (\bibinfo {year} {2011})},\ \Eprint
  {http://arxiv.org/abs/1011.4530} {arXiv:1011.4530 [astro-ph.CO]} \BibitemShut
  {NoStop}%
\bibitem [{\citenamefont {{Cooray}}\ and\ \citenamefont
  {{Sheth}}(2002)}]{CooShe02}%
  \BibitemOpen
  \bibfield  {author} {\bibinfo {author} {\bibfnamefont {A.}~\bibnamefont
  {{Cooray}}}\ and\ \bibinfo {author} {\bibfnamefont {R.}~\bibnamefont
  {{Sheth}}},\ }\bibfield  {title} {\enquote {\bibinfo {title} {{Halo models of
  large scale structure}},}\ }\href@noop {} {\bibfield  {journal} {\bibinfo
  {journal} {\physrep}\ }\textbf {\bibinfo {volume} {372}},\ \bibinfo {pages}
  {1--129} (\bibinfo {year} {2002})}\BibitemShut {NoStop}%
\bibitem [{\citenamefont {{Mo}}\ and\ \citenamefont
  {{White}}(1996)}]{MoWhi9609}%
  \BibitemOpen
  \bibfield  {author} {\bibinfo {author} {\bibfnamefont {H.~J.}\ \bibnamefont
  {{Mo}}}\ and\ \bibinfo {author} {\bibfnamefont {S.~D.~M.}\ \bibnamefont
  {{White}}},\ }\bibfield  {title} {\enquote {\bibinfo {title} {{An analytic
  model for the spatial clustering of dark matter haloes}},}\ }\href@noop {}
  {\bibfield  {journal} {\bibinfo  {journal} {\mnras}\ }\textbf {\bibinfo
  {volume} {282}},\ \bibinfo {pages} {347--361} (\bibinfo {year} {1996})},\
  \Eprint {http://arxiv.org/abs/arXiv:astro-ph/9512127}
  {arXiv:astro-ph/9512127} \BibitemShut {NoStop}%
\bibitem [{\citenamefont {{Sheth}}\ and\ \citenamefont
  {{Lemson}}(1999)}]{1999MNRAS.304..767S}%
  \BibitemOpen
  \bibfield  {author} {\bibinfo {author} {\bibfnamefont {R.{}K.}\ \bibnamefont
  {{Sheth}}}\ and\ \bibinfo {author} {\bibfnamefont {G.}~\bibnamefont
  {{Lemson}}},\ }\bibfield  {title} {\enquote {\bibinfo {title} {{Biasing and
  the distribution of dark matter haloes}},}\ }\href@noop {} {\bibfield
  {journal} {\bibinfo  {journal} {\mnras}\ }\textbf {\bibinfo {volume} {304}},\
  \bibinfo {pages} {767--792} (\bibinfo {year} {1999})}\BibitemShut {NoStop}%
\bibitem [{\citenamefont {{Casas-Miranda}}\ \emph {et~al.}(2002)\citenamefont
  {{Casas-Miranda}}, \citenamefont {{Mo}}, \citenamefont {{Sheth}},\ and\
  \citenamefont {{Boerner}}}]{CasMoShe0207}%
  \BibitemOpen
  \bibfield  {author} {\bibinfo {author} {\bibfnamefont {R.}~\bibnamefont
  {{Casas-Miranda}}}, \bibinfo {author} {\bibfnamefont {H.~J.}\ \bibnamefont
  {{Mo}}}, \bibinfo {author} {\bibfnamefont {R.~K.}\ \bibnamefont {{Sheth}}}, \
  and\ \bibinfo {author} {\bibfnamefont {G.}~\bibnamefont {{Boerner}}},\
  }\bibfield  {title} {\enquote {\bibinfo {title} {{On the distribution of
  haloes, galaxies and mass}},}\ }\href {\doibase
  10.1046/j.1365-8711.2002.05378.x} {\bibfield  {journal} {\bibinfo  {journal}
  {\mnras}\ }\textbf {\bibinfo {volume} {333}},\ \bibinfo {pages} {730--738}
  (\bibinfo {year} {2002})},\ \Eprint
  {http://arxiv.org/abs/arXiv:astro-ph/0105008} {arXiv:astro-ph/0105008}
  \BibitemShut {NoStop}%
\bibitem [{\citenamefont {{Dekel}}\ and\ \citenamefont
  {{Lahav}}(1999)}]{DekLah9907}%
  \BibitemOpen
  \bibfield  {author} {\bibinfo {author} {\bibfnamefont {A.}~\bibnamefont
  {{Dekel}}}\ and\ \bibinfo {author} {\bibfnamefont {O.}~\bibnamefont
  {{Lahav}}},\ }\bibfield  {title} {\enquote {\bibinfo {title} {{Stochastic
  Nonlinear Galaxy Biasing}},}\ }\href {\doibase 10.1086/307428} {\bibfield
  {journal} {\bibinfo  {journal} {\apj}\ }\textbf {\bibinfo {volume} {520}},\
  \bibinfo {pages} {24--34} (\bibinfo {year} {1999})},\ \Eprint
  {http://arxiv.org/abs/arXiv:astro-ph/9806193} {arXiv:astro-ph/9806193}
  \BibitemShut {NoStop}%
\bibitem [{\citenamefont {{Somerville}}\ \emph {et~al.}(2001)\citenamefont
  {{Somerville}}, \citenamefont {{Lemson}}, \citenamefont {{Sigad}},
  \citenamefont {{Dekel}}, \citenamefont {{Kauffmann}},\ and\ \citenamefont
  {{White}}}]{SomLemSig0101}%
  \BibitemOpen
  \bibfield  {author} {\bibinfo {author} {\bibfnamefont {R.~S.}\ \bibnamefont
  {{Somerville}}}, \bibinfo {author} {\bibfnamefont {G.}~\bibnamefont
  {{Lemson}}}, \bibinfo {author} {\bibfnamefont {Y.}~\bibnamefont {{Sigad}}},
  \bibinfo {author} {\bibfnamefont {A.}~\bibnamefont {{Dekel}}}, \bibinfo
  {author} {\bibfnamefont {G.}~\bibnamefont {{Kauffmann}}}, \ and\ \bibinfo
  {author} {\bibfnamefont {S.~D.~M.}\ \bibnamefont {{White}}},\ }\bibfield
  {title} {\enquote {\bibinfo {title} {{Non-linear stochastic galaxy biasing in
  cosmological simulations}},}\ }\href {\doibase
  10.1046/j.1365-8711.2001.03894.x} {\bibfield  {journal} {\bibinfo  {journal}
  {\mnras}\ }\textbf {\bibinfo {volume} {320}},\ \bibinfo {pages} {289--306}
  (\bibinfo {year} {2001})},\ \Eprint
  {http://arxiv.org/abs/arXiv:astro-ph/9912073} {arXiv:astro-ph/9912073}
  \BibitemShut {NoStop}%
\bibitem [{\citenamefont {{Bond}}\ \emph {et~al.}(1991)\citenamefont {{Bond}},
  \citenamefont {{Cole}}, \citenamefont {{Efstathiou}},\ and\ \citenamefont
  {{Kaiser}}}]{1991ApJ...379..440B}%
  \BibitemOpen
  \bibfield  {author} {\bibinfo {author} {\bibfnamefont {J.~R.}\ \bibnamefont
  {{Bond}}}, \bibinfo {author} {\bibfnamefont {S.}~\bibnamefont {{Cole}}},
  \bibinfo {author} {\bibfnamefont {G.}~\bibnamefont {{Efstathiou}}}, \ and\
  \bibinfo {author} {\bibfnamefont {N.}~\bibnamefont {{Kaiser}}},\ }\bibfield
  {title} {\enquote {\bibinfo {title} {{Excursion set mass functions for
  hierarchical Gaussian fluctuations}},}\ }\href {\doibase 10.1086/170520}
  {\bibfield  {journal} {\bibinfo  {journal} {\apj}\ }\textbf {\bibinfo
  {volume} {379}},\ \bibinfo {pages} {440--460} (\bibinfo {year}
  {1991})}\BibitemShut {NoStop}%
\bibitem [{\citenamefont {{Lacey}}\ and\ \citenamefont
  {{Cole}}(1993)}]{1993MNRAS.262..627L}%
  \BibitemOpen
  \bibfield  {author} {\bibinfo {author} {\bibfnamefont {C.}~\bibnamefont
  {{Lacey}}}\ and\ \bibinfo {author} {\bibfnamefont {S.}~\bibnamefont
  {{Cole}}},\ }\bibfield  {title} {\enquote {\bibinfo {title} {{Merger rates in
  hierarchical models of galaxy formation}},}\ }\href@noop {} {\bibfield
  {journal} {\bibinfo  {journal} {\mnras}\ }\textbf {\bibinfo {volume} {262}},\
  \bibinfo {pages} {627--649} (\bibinfo {year} {1993})}\BibitemShut {NoStop}%
\bibitem [{\citenamefont {{Sheth}}(1998)}]{She9811}%
  \BibitemOpen
  \bibfield  {author} {\bibinfo {author} {\bibfnamefont {R.~K.}\ \bibnamefont
  {{Sheth}}},\ }\bibfield  {title} {\enquote {\bibinfo {title} {{An excursion
  set model for the distribution of dark matter and dark matter haloes}},}\
  }\href {\doibase 10.1046/j.1365-8711.1998.01976.x} {\bibfield  {journal}
  {\bibinfo  {journal} {\mnras}\ }\textbf {\bibinfo {volume} {300}},\ \bibinfo
  {pages} {1057--1070} (\bibinfo {year} {1998})},\ \Eprint
  {http://arxiv.org/abs/arXiv:astro-ph/9805319} {arXiv:astro-ph/9805319}
  \BibitemShut {NoStop}%
\bibitem [{\citenamefont {{Bond}}\ and\ \citenamefont
  {{Myers}}(1996)}]{BonMye9603}%
  \BibitemOpen
  \bibfield  {author} {\bibinfo {author} {\bibfnamefont {J.~R.}\ \bibnamefont
  {{Bond}}}\ and\ \bibinfo {author} {\bibfnamefont {S.~T.}\ \bibnamefont
  {{Myers}}},\ }\bibfield  {title} {\enquote {\bibinfo {title} {{The Peak-Patch
  Picture of Cosmic Catalogs. I. Algorithms}},}\ }\href {\doibase
  10.1086/192267} {\bibfield  {journal} {\bibinfo  {journal} {\apjs}\ }\textbf
  {\bibinfo {volume} {103}},\ \bibinfo {pages} {1--+} (\bibinfo {year}
  {1996})}\BibitemShut {NoStop}%
\bibitem [{\citenamefont {{Sheth}}\ \emph {et~al.}(2001)\citenamefont
  {{Sheth}}, \citenamefont {{Mo}},\ and\ \citenamefont
  {{Tormen}}}]{SheMoTor01}%
  \BibitemOpen
  \bibfield  {author} {\bibinfo {author} {\bibfnamefont {R.{}K.}\ \bibnamefont
  {{Sheth}}}, \bibinfo {author} {\bibfnamefont {H.{}J.}\ \bibnamefont {{Mo}}},
  \ and\ \bibinfo {author} {\bibfnamefont {G.}~\bibnamefont {{Tormen}}},\
  }\bibfield  {title} {\enquote {\bibinfo {title} {{Ellipsoidal collapse and an
  improved model for the number and spatial distribution of dark matter
  haloes}},}\ }\href@noop {} {\bibfield  {journal} {\bibinfo  {journal}
  {\mnras}\ }\textbf {\bibinfo {volume} {323}},\ \bibinfo {pages} {1--12}
  (\bibinfo {year} {2001})}\BibitemShut {NoStop}%
\bibitem [{\citenamefont {{Matsubara}}(1995)}]{Mat9511}%
  \BibitemOpen
  \bibfield  {author} {\bibinfo {author} {\bibfnamefont {T.}~\bibnamefont
  {{Matsubara}}},\ }\bibfield  {title} {\enquote {\bibinfo {title}
  {{Diagrammatic Methods in Statistics and Biasing in the Large-Scale Structure
  of the Universe}},}\ }\href {\doibase 10.1086/192231} {\bibfield  {journal}
  {\bibinfo  {journal} {\apjs}\ }\textbf {\bibinfo {volume} {101}},\ \bibinfo
  {pages} {1} (\bibinfo {year} {1995})},\ \Eprint
  {http://arxiv.org/abs/arXiv:astro-ph/9501056} {arXiv:astro-ph/9501056}
  \BibitemShut {NoStop}%
\bibitem [{\citenamefont {{McDonald}}\ and\ \citenamefont
  {{Roy}}(2009)}]{McDRoy0908}%
  \BibitemOpen
  \bibfield  {author} {\bibinfo {author} {\bibfnamefont {P.}~\bibnamefont
  {{McDonald}}}\ and\ \bibinfo {author} {\bibfnamefont {A.}~\bibnamefont
  {{Roy}}},\ }\bibfield  {title} {\enquote {\bibinfo {title} {{Clustering of
  dark matter tracers: generalizing bias for the coming era of precision
  LSS}},}\ }\href {\doibase 10.1088/1475-7516/2009/08/020} {\bibfield
  {journal} {\bibinfo  {journal} {\jcap}\ }\textbf {\bibinfo {volume} {8}},\
  \bibinfo {pages} {20--+} (\bibinfo {year} {2009})},\ \Eprint
  {http://arxiv.org/abs/0902.0991} {arXiv:0902.0991 [astro-ph.CO]} \BibitemShut
  {NoStop}%
\bibitem [{\citenamefont {{Fry}}(1996)}]{Fry9604}%
  \BibitemOpen
  \bibfield  {author} {\bibinfo {author} {\bibfnamefont {J.~N.}\ \bibnamefont
  {{Fry}}},\ }\bibfield  {title} {\enquote {\bibinfo {title} {{The Evolution of
  Bias}},}\ }\href {\doibase 10.1086/310006} {\bibfield  {journal} {\bibinfo
  {journal} {\apjl}\ }\textbf {\bibinfo {volume} {461}},\ \bibinfo {pages}
  {L65+} (\bibinfo {year} {1996})}\BibitemShut {NoStop}%
\bibitem [{\citenamefont {{Pen}}(1998)}]{Pen98}%
  \BibitemOpen
  \bibfield  {author} {\bibinfo {author} {\bibfnamefont {U.-L.}\ \bibnamefont
  {{Pen}}},\ }\bibfield  {title} {\enquote {\bibinfo {title} {{Reconstructing
  Nonlinear Stochastic Bias from Velocity Space Distortions}},}\ }\href
  {\doibase 10.1086/306098} {\bibfield  {journal} {\bibinfo  {journal} {\apj}\
  }\textbf {\bibinfo {volume} {504}},\ \bibinfo {pages} {601--+} (\bibinfo
  {year} {1998})}\BibitemShut {NoStop}%
\bibitem [{\citenamefont {{Tegmark}}\ and\ \citenamefont
  {{Peebles}}(1998)}]{TegPee9806}%
  \BibitemOpen
  \bibfield  {author} {\bibinfo {author} {\bibfnamefont {M.}~\bibnamefont
  {{Tegmark}}}\ and\ \bibinfo {author} {\bibfnamefont {P.~J.~E.}\ \bibnamefont
  {{Peebles}}},\ }\bibfield  {title} {\enquote {\bibinfo {title} {{The Time
  Evolution of Bias}},}\ }\href {\doibase 10.1086/311426} {\bibfield  {journal}
  {\bibinfo  {journal} {\apjl}\ }\textbf {\bibinfo {volume} {500}},\ \bibinfo
  {pages} {L79} (\bibinfo {year} {1998})},\ \Eprint
  {http://arxiv.org/abs/arXiv:astro-ph/9804067} {arXiv:astro-ph/9804067}
  \BibitemShut {NoStop}%
\bibitem [{\citenamefont {{Mo}}\ \emph {et~al.}(1997)\citenamefont {{Mo}},
  \citenamefont {{Jing}},\ and\ \citenamefont {{White}}}]{MoJinWhi97}%
  \BibitemOpen
  \bibfield  {author} {\bibinfo {author} {\bibfnamefont {H.~J.}\ \bibnamefont
  {{Mo}}}, \bibinfo {author} {\bibfnamefont {Y.~P.}\ \bibnamefont {{Jing}}}, \
  and\ \bibinfo {author} {\bibfnamefont {S.~D.~M.}\ \bibnamefont {{White}}},\
  }\bibfield  {title} {\enquote {\bibinfo {title} {{High-order correlations of
  peaks and haloes: a step towards understanding galaxy biasing}},}\
  }\href@noop {} {\bibfield  {journal} {\bibinfo  {journal} {\mnras}\ }\textbf
  {\bibinfo {volume} {284}},\ \bibinfo {pages} {189--201} (\bibinfo {year}
  {1997})}\BibitemShut {NoStop}%
\bibitem [{\citenamefont {{Scoccimarro}}\ \emph
  {et~al.}(2001{\natexlab{b}})\citenamefont {{Scoccimarro}}, \citenamefont
  {{Sheth}}, \citenamefont {{Hui}},\ and\ \citenamefont
  {{Jain}}}]{ScoSheHui01}%
  \BibitemOpen
  \bibfield  {author} {\bibinfo {author} {\bibfnamefont {R.}~\bibnamefont
  {{Scoccimarro}}}, \bibinfo {author} {\bibfnamefont {R.{}K.}\ \bibnamefont
  {{Sheth}}}, \bibinfo {author} {\bibfnamefont {L.}~\bibnamefont {{Hui}}}, \
  and\ \bibinfo {author} {\bibfnamefont {B.}~\bibnamefont {{Jain}}},\
  }\bibfield  {title} {\enquote {\bibinfo {title} {{How Many Galaxies Fit in a
  Halo? Constraints on Galaxy Formation Efficiency from Spatial Clustering}},}\
  }\href@noop {} {\bibfield  {journal} {\bibinfo  {journal} {\apj}\ }\textbf
  {\bibinfo {volume} {546}},\ \bibinfo {pages} {20--34} (\bibinfo {year}
  {2001}{\natexlab{b}})}\BibitemShut {NoStop}%
\bibitem [{\citenamefont {{Catelan}}\ \emph {et~al.}(1998)\citenamefont
  {{Catelan}}, \citenamefont {{Lucchin}}, \citenamefont {{Matarrese}},\ and\
  \citenamefont {{Porciani}}}]{CatLucMat9807}%
  \BibitemOpen
  \bibfield  {author} {\bibinfo {author} {\bibfnamefont {P.}~\bibnamefont
  {{Catelan}}}, \bibinfo {author} {\bibfnamefont {F.}~\bibnamefont
  {{Lucchin}}}, \bibinfo {author} {\bibfnamefont {S.}~\bibnamefont
  {{Matarrese}}}, \ and\ \bibinfo {author} {\bibfnamefont {C.}~\bibnamefont
  {{Porciani}}},\ }\bibfield  {title} {\enquote {\bibinfo {title} {{The bias
  field of dark matter haloes}},}\ }\href {\doibase
  10.1046/j.1365-8711.1998.01455.x} {\bibfield  {journal} {\bibinfo  {journal}
  {\mnras}\ }\textbf {\bibinfo {volume} {297}},\ \bibinfo {pages} {692--712}
  (\bibinfo {year} {1998})},\ \Eprint
  {http://arxiv.org/abs/arXiv:astro-ph/9708067} {arXiv:astro-ph/9708067}
  \BibitemShut {NoStop}%
\bibitem [{\citenamefont {{Catelan}}\ \emph {et~al.}(2000)\citenamefont
  {{Catelan}}, \citenamefont {{Porciani}},\ and\ \citenamefont
  {{Kamionkowski}}}]{CatPorKam0011}%
  \BibitemOpen
  \bibfield  {author} {\bibinfo {author} {\bibfnamefont {P.}~\bibnamefont
  {{Catelan}}}, \bibinfo {author} {\bibfnamefont {C.}~\bibnamefont
  {{Porciani}}}, \ and\ \bibinfo {author} {\bibfnamefont {M.}~\bibnamefont
  {{Kamionkowski}}},\ }\bibfield  {title} {\enquote {\bibinfo {title} {{Two
  ways of biasing galaxy formation}},}\ }\href {\doibase
  10.1046/j.1365-8711.2000.04023.x} {\bibfield  {journal} {\bibinfo  {journal}
  {\mnras}\ }\textbf {\bibinfo {volume} {318}},\ \bibinfo {pages} {L39--L44}
  (\bibinfo {year} {2000})},\ \Eprint
  {http://arxiv.org/abs/arXiv:astro-ph/0005544} {arXiv:astro-ph/0005544}
  \BibitemShut {NoStop}%
\bibitem [{\citenamefont {{Roth}}\ and\ \citenamefont
  {{Porciani}}(2011)}]{RotPor1105}%
  \BibitemOpen
  \bibfield  {author} {\bibinfo {author} {\bibfnamefont {N.}~\bibnamefont
  {{Roth}}}\ and\ \bibinfo {author} {\bibfnamefont {C.}~\bibnamefont
  {{Porciani}}},\ }\bibfield  {title} {\enquote {\bibinfo {title} {{Testing
  standard perturbation theory and the Eulerian local biasing scheme against
  N-body simulations}},}\ }\href {\doibase 10.1111/j.1365-2966.2011.18768.x}
  {\bibfield  {journal} {\bibinfo  {journal} {\mnras}\ ,\ \bibinfo {pages}
  {678--+}} (\bibinfo {year} {2011})},\ \Eprint
  {http://arxiv.org/abs/1101.1520} {arXiv:1101.1520 [astro-ph.CO]} \BibitemShut
  {NoStop}%
\bibitem [{\citenamefont {{Manera}}\ and\ \citenamefont
  {{Gazta{\~n}aga}}(2011)}]{ManGaz1107}%
  \BibitemOpen
  \bibfield  {author} {\bibinfo {author} {\bibfnamefont {M.}~\bibnamefont
  {{Manera}}}\ and\ \bibinfo {author} {\bibfnamefont {E.}~\bibnamefont
  {{Gazta{\~n}aga}}},\ }\bibfield  {title} {\enquote {\bibinfo {title} {{The
  local bias model in the large-scale halo distribution}},}\ }\href {\doibase
  10.1111/j.1365-2966.2011.18705.x} {\bibfield  {journal} {\bibinfo  {journal}
  {\mnras}\ }\textbf {\bibinfo {volume} {415}},\ \bibinfo {pages} {383--398}
  (\bibinfo {year} {2011})},\ \Eprint {http://arxiv.org/abs/0912.0446}
  {arXiv:0912.0446 [astro-ph.CO]} \BibitemShut {NoStop}%
\bibitem [{\citenamefont {{Fukugita}}\ and\ \citenamefont
  {{Peebles}}(2004)}]{FukPee0412}%
  \BibitemOpen
  \bibfield  {author} {\bibinfo {author} {\bibfnamefont {M.}~\bibnamefont
  {{Fukugita}}}\ and\ \bibinfo {author} {\bibfnamefont {P.~J.~E.}\ \bibnamefont
  {{Peebles}}},\ }\bibfield  {title} {\enquote {\bibinfo {title} {{The Cosmic
  Energy Inventory}},}\ }\href {\doibase 10.1086/425155} {\bibfield  {journal}
  {\bibinfo  {journal} {\apj}\ }\textbf {\bibinfo {volume} {616}},\ \bibinfo
  {pages} {643--668} (\bibinfo {year} {2004})},\ \Eprint
  {http://arxiv.org/abs/arXiv:astro-ph/0406095} {arXiv:astro-ph/0406095}
  \BibitemShut {NoStop}%
\bibitem [{\citenamefont {{Somogyi}}\ and\ \citenamefont
  {{Smith}}(2010)}]{SomSmi1001}%
  \BibitemOpen
  \bibfield  {author} {\bibinfo {author} {\bibfnamefont {G.}~\bibnamefont
  {{Somogyi}}}\ and\ \bibinfo {author} {\bibfnamefont {R.~E.}\ \bibnamefont
  {{Smith}}},\ }\bibfield  {title} {\enquote {\bibinfo {title} {{Cosmological
  perturbation theory for baryons and dark matter: One-loop corrections in the
  renormalized perturbation theory framework}},}\ }\href {\doibase
  10.1103/PhysRevD.81.023524} {\bibfield  {journal} {\bibinfo  {journal}
  {\prd}\ }\textbf {\bibinfo {volume} {81}},\ \bibinfo {pages} {023524--+}
  (\bibinfo {year} {2010})},\ \Eprint {http://arxiv.org/abs/0910.5220}
  {arXiv:0910.5220 [astro-ph.CO]} \BibitemShut {NoStop}%
\bibitem [{\citenamefont {{Elia}}\ \emph {et~al.}(2010)\citenamefont {{Elia}},
  \citenamefont {{Kulkarni}}, \citenamefont {{Porciani}}, \citenamefont
  {{Pietroni}},\ and\ \citenamefont {{Matarrese}}}]{EliKulPor1012}%
  \BibitemOpen
  \bibfield  {author} {\bibinfo {author} {\bibfnamefont {A.}~\bibnamefont
  {{Elia}}}, \bibinfo {author} {\bibfnamefont {S.}~\bibnamefont {{Kulkarni}}},
  \bibinfo {author} {\bibfnamefont {C.}~\bibnamefont {{Porciani}}}, \bibinfo
  {author} {\bibfnamefont {M.}~\bibnamefont {{Pietroni}}}, \ and\ \bibinfo
  {author} {\bibfnamefont {S.}~\bibnamefont {{Matarrese}}},\ }\bibfield
  {title} {\enquote {\bibinfo {title} {{Modeling the clustering of dark-matter
  haloes in resummed perturbation theories}},}\ }\href@noop {} {\bibfield
  {journal} {\bibinfo  {journal} {ArXiv e-prints}\ } (\bibinfo {year}
  {2010})},\ \Eprint {http://arxiv.org/abs/1012.4833} {arXiv:1012.4833
  [astro-ph.CO]} \BibitemShut {NoStop}%
\bibitem [{\citenamefont {{Bernardeau}}\ \emph {et~al.}(2011)\citenamefont
  {{Bernardeau}}, \citenamefont {{Van de Rijt}},\ and\ \citenamefont
  {{Vernizzi}}}]{BerVanVer1109}%
  \BibitemOpen
  \bibfield  {author} {\bibinfo {author} {\bibfnamefont {F.}~\bibnamefont
  {{Bernardeau}}}, \bibinfo {author} {\bibfnamefont {N.}~\bibnamefont {{Van de
  Rijt}}}, \ and\ \bibinfo {author} {\bibfnamefont {F.}~\bibnamefont
  {{Vernizzi}}},\ }\bibfield  {title} {\enquote {\bibinfo {title} {{Resummed
  propagators in multi-component cosmic fluids with the eikonal
  approximation}},}\ }\href@noop {} {\bibfield  {journal} {\bibinfo  {journal}
  {ArXiv e-prints}\ } (\bibinfo {year} {2011})},\ \Eprint
  {http://arxiv.org/abs/1109.3400} {arXiv:1109.3400 [astro-ph.CO]} \BibitemShut
  {NoStop}%
\bibitem [{\citenamefont {{Bernardeau}}\ \emph {et~al.}(2002)\citenamefont
  {{Bernardeau}}, \citenamefont {{Colombi}}, \citenamefont {{Gaztanaga}},\ and\
  \citenamefont {{Scoccimarro}}}]{BerColGaz02}%
  \BibitemOpen
  \bibfield  {author} {\bibinfo {author} {\bibfnamefont {F.}~\bibnamefont
  {{Bernardeau}}}, \bibinfo {author} {\bibfnamefont {S.}~\bibnamefont
  {{Colombi}}}, \bibinfo {author} {\bibfnamefont {E.}~\bibnamefont
  {{Gaztanaga}}}, \ and\ \bibinfo {author} {\bibfnamefont {R.}~\bibnamefont
  {{Scoccimarro}}},\ }\bibfield  {title} {\enquote {\bibinfo {title}
  {{Large-scale structure of the Universe and cosmological perturbation
  theory.}}}\ }\href@noop {} {\bibfield  {journal} {\bibinfo  {journal}
  {\physrep}\ }\textbf {\bibinfo {volume} {367}},\ \bibinfo {pages} {1--128}
  (\bibinfo {year} {2002})}\BibitemShut {NoStop}%
\bibitem [{\citenamefont {{Scoccimarro}}(1998)}]{Sco98}%
  \BibitemOpen
  \bibfield  {author} {\bibinfo {author} {\bibfnamefont {R.}~\bibnamefont
  {{Scoccimarro}}},\ }\bibfield  {title} {\enquote {\bibinfo {title}
  {{Transients from initial conditions: a perturbative analysis}},}\
  }\href@noop {} {\bibfield  {journal} {\bibinfo  {journal} {\mnras}\ }\textbf
  {\bibinfo {volume} {299}},\ \bibinfo {pages} {1097--1118} (\bibinfo {year}
  {1998})}\BibitemShut {NoStop}%
\bibitem [{\citenamefont {{Scoccimarro}}(2001)}]{Sco01}%
  \BibitemOpen
  \bibfield  {author} {\bibinfo {author} {\bibfnamefont {R.}~\bibnamefont
  {{Scoccimarro}}},\ }\bibfield  {title} {\enquote {\bibinfo {title} {{A New
  Angle on Gravitational Clustering}},}\ }\href@noop {} {\bibfield  {journal}
  {\bibinfo  {journal} {ArXiv:astro-ph/0008277, Annals New York Academy
  Sciences}\ }\textbf {\bibinfo {volume} {927}},\ \bibinfo {pages} {13--23}
  (\bibinfo {year} {2001})}\BibitemShut {NoStop}%
\bibitem [{\citenamefont {{Scoccimarro}}\ \emph {et~al.}(1998)\citenamefont
  {{Scoccimarro}}, \citenamefont {{Colombi}}, \citenamefont {{Fry}},
  \citenamefont {{Frieman}}, \citenamefont {{Hivon}},\ and\ \citenamefont
  {{Melott}}}]{ScoColFry9803}%
  \BibitemOpen
  \bibfield  {author} {\bibinfo {author} {\bibfnamefont {R.}~\bibnamefont
  {{Scoccimarro}}}, \bibinfo {author} {\bibfnamefont {S.}~\bibnamefont
  {{Colombi}}}, \bibinfo {author} {\bibfnamefont {J.~N.}\ \bibnamefont
  {{Fry}}}, \bibinfo {author} {\bibfnamefont {J.~A.}\ \bibnamefont
  {{Frieman}}}, \bibinfo {author} {\bibfnamefont {E.}~\bibnamefont {{Hivon}}},
  \ and\ \bibinfo {author} {\bibfnamefont {A.}~\bibnamefont {{Melott}}},\
  }\bibfield  {title} {\enquote {\bibinfo {title} {{Nonlinear Evolution of the
  Bispectrum of Cosmological Perturbations}},}\ }\href {\doibase
  10.1086/305399} {\bibfield  {journal} {\bibinfo  {journal} {\apj}\ }\textbf
  {\bibinfo {volume} {496}},\ \bibinfo {pages} {586} (\bibinfo {year}
  {1998})},\ \Eprint {http://arxiv.org/abs/arXiv:astro-ph/9704075}
  {arXiv:astro-ph/9704075} \BibitemShut {NoStop}%
\bibitem [{\citenamefont {{Col{\'{\i}}n}}\ \emph {et~al.}(2000)\citenamefont
  {{Col{\'{\i}}n}}, \citenamefont {{Klypin}},\ and\ \citenamefont
  {{Kravtsov}}}]{ColKlyKra0008}%
  \BibitemOpen
  \bibfield  {author} {\bibinfo {author} {\bibfnamefont {P.}~\bibnamefont
  {{Col{\'{\i}}n}}}, \bibinfo {author} {\bibfnamefont {A.~A.}\ \bibnamefont
  {{Klypin}}}, \ and\ \bibinfo {author} {\bibfnamefont {A.~V.}\ \bibnamefont
  {{Kravtsov}}},\ }\bibfield  {title} {\enquote {\bibinfo {title} {{Velocity
  Bias in a {$\Lambda$} Cold Dark Matter Model}},}\ }\href {\doibase
  10.1086/309248} {\bibfield  {journal} {\bibinfo  {journal} {\apj}\ }\textbf
  {\bibinfo {volume} {539}},\ \bibinfo {pages} {561--569} (\bibinfo {year}
  {2000})},\ \Eprint {http://arxiv.org/abs/arXiv:astro-ph/9907337}
  {arXiv:astro-ph/9907337} \BibitemShut {NoStop}%
\bibitem [{\citenamefont {{Berlind}}\ \emph {et~al.}(2003)\citenamefont
  {{Berlind}}, \citenamefont {{Weinberg}}, \citenamefont {{Benson}},
  \citenamefont {{Baugh}}, \citenamefont {{Cole}}, \citenamefont {{Dav{\'e}}},
  \citenamefont {{Frenk}}, \citenamefont {{Jenkins}}, \citenamefont {{Katz}},\
  and\ \citenamefont {{Lacey}}}]{BerWeiBen0308}%
  \BibitemOpen
  \bibfield  {author} {\bibinfo {author} {\bibfnamefont {A.~A.}\ \bibnamefont
  {{Berlind}}}, \bibinfo {author} {\bibfnamefont {D.~H.}\ \bibnamefont
  {{Weinberg}}}, \bibinfo {author} {\bibfnamefont {A.~J.}\ \bibnamefont
  {{Benson}}}, \bibinfo {author} {\bibfnamefont {C.~M.}\ \bibnamefont
  {{Baugh}}}, \bibinfo {author} {\bibfnamefont {S.}~\bibnamefont {{Cole}}},
  \bibinfo {author} {\bibfnamefont {R.}~\bibnamefont {{Dav{\'e}}}}, \bibinfo
  {author} {\bibfnamefont {C.~S.}\ \bibnamefont {{Frenk}}}, \bibinfo {author}
  {\bibfnamefont {A.}~\bibnamefont {{Jenkins}}}, \bibinfo {author}
  {\bibfnamefont {N.}~\bibnamefont {{Katz}}}, \ and\ \bibinfo {author}
  {\bibfnamefont {C.~G.}\ \bibnamefont {{Lacey}}},\ }\bibfield  {title}
  {\enquote {\bibinfo {title} {{The Halo Occupation Distribution and the
  Physics of Galaxy Formation}},}\ }\href {\doibase 10.1086/376517} {\bibfield
  {journal} {\bibinfo  {journal} {\apj}\ }\textbf {\bibinfo {volume} {593}},\
  \bibinfo {pages} {1--25} (\bibinfo {year} {2003})},\ \Eprint
  {http://arxiv.org/abs/arXiv:astro-ph/0212357} {arXiv:astro-ph/0212357}
  \BibitemShut {NoStop}%
\bibitem [{\citenamefont {{Diemand}}\ \emph {et~al.}(2004)\citenamefont
  {{Diemand}}, \citenamefont {{Moore}},\ and\ \citenamefont
  {{Stadel}}}]{DieMooSta0408}%
  \BibitemOpen
  \bibfield  {author} {\bibinfo {author} {\bibfnamefont {J.}~\bibnamefont
  {{Diemand}}}, \bibinfo {author} {\bibfnamefont {B.}~\bibnamefont {{Moore}}},
  \ and\ \bibinfo {author} {\bibfnamefont {J.}~\bibnamefont {{Stadel}}},\
  }\bibfield  {title} {\enquote {\bibinfo {title} {{Velocity and spatial biases
  in cold dark matter subhalo distributions}},}\ }\href {\doibase
  10.1111/j.1365-2966.2004.07940.x} {\bibfield  {journal} {\bibinfo  {journal}
  {\mnras}\ }\textbf {\bibinfo {volume} {352}},\ \bibinfo {pages} {535--546}
  (\bibinfo {year} {2004})},\ \Eprint
  {http://arxiv.org/abs/arXiv:astro-ph/0402160} {arXiv:astro-ph/0402160}
  \BibitemShut {NoStop}%
\bibitem [{\citenamefont {{Desjacques}}\ and\ \citenamefont
  {{Sheth}}(2010)}]{DesShe1001}%
  \BibitemOpen
  \bibfield  {author} {\bibinfo {author} {\bibfnamefont {V.}~\bibnamefont
  {{Desjacques}}}\ and\ \bibinfo {author} {\bibfnamefont {R.~K.}\ \bibnamefont
  {{Sheth}}},\ }\bibfield  {title} {\enquote {\bibinfo {title} {{Redshift space
  correlations and scale-dependent stochastic biasing of density peaks}},}\
  }\href {\doibase 10.1103/PhysRevD.81.023526} {\bibfield  {journal} {\bibinfo
  {journal} {\prd}\ }\textbf {\bibinfo {volume} {81}},\ \bibinfo {pages}
  {023526--+} (\bibinfo {year} {2010})},\ \Eprint
  {http://arxiv.org/abs/0909.4544} {arXiv:0909.4544 [astro-ph.CO]} \BibitemShut
  {NoStop}%
\bibitem [{\citenamefont {{Desjacques}}\ \emph {et~al.}(2010)\citenamefont
  {{Desjacques}}, \citenamefont {{Crocce}}, \citenamefont {{Scoccimarro}},\
  and\ \citenamefont {{Sheth}}}]{DesCroSco1011}%
  \BibitemOpen
  \bibfield  {author} {\bibinfo {author} {\bibfnamefont {V.}~\bibnamefont
  {{Desjacques}}}, \bibinfo {author} {\bibfnamefont {M.}~\bibnamefont
  {{Crocce}}}, \bibinfo {author} {\bibfnamefont {R.}~\bibnamefont
  {{Scoccimarro}}}, \ and\ \bibinfo {author} {\bibfnamefont {R.~K.}\
  \bibnamefont {{Sheth}}},\ }\bibfield  {title} {\enquote {\bibinfo {title}
  {{Modeling scale-dependent bias on the baryonic acoustic scale with the
  statistics of peaks of Gaussian random fields}},}\ }\href {\doibase
  10.1103/PhysRevD.82.103529} {\bibfield  {journal} {\bibinfo  {journal}
  {\prd}\ }\textbf {\bibinfo {volume} {82}},\ \bibinfo {pages} {103529--+}
  (\bibinfo {year} {2010})},\ \Eprint {http://arxiv.org/abs/1009.3449}
  {arXiv:1009.3449 [astro-ph.CO]} \BibitemShut {NoStop}%
\bibitem [{\citenamefont {{Tseliakhovich}}\ and\ \citenamefont
  {{Hirata}}(2010)}]{TseHir1010}%
  \BibitemOpen
  \bibfield  {author} {\bibinfo {author} {\bibfnamefont {D.}~\bibnamefont
  {{Tseliakhovich}}}\ and\ \bibinfo {author} {\bibfnamefont {C.}~\bibnamefont
  {{Hirata}}},\ }\bibfield  {title} {\enquote {\bibinfo {title} {{Relative
  velocity of dark matter and baryonic fluids and the formation of the first
  structures}},}\ }\href {\doibase 10.1103/PhysRevD.82.083520} {\bibfield
  {journal} {\bibinfo  {journal} {\prd}\ }\textbf {\bibinfo {volume} {82}},\
  \bibinfo {eid} {083520} (\bibinfo {year} {2010})},\ \Eprint
  {http://arxiv.org/abs/1005.2416} {arXiv:1005.2416 [astro-ph.CO]} \BibitemShut
  {NoStop}%
\bibitem [{\citenamefont {{Dalal}}\ \emph {et~al.}(2010)\citenamefont
  {{Dalal}}, \citenamefont {{Pen}},\ and\ \citenamefont
  {{Seljak}}}]{DalPenSel1011}%
  \BibitemOpen
  \bibfield  {author} {\bibinfo {author} {\bibfnamefont {N.}~\bibnamefont
  {{Dalal}}}, \bibinfo {author} {\bibfnamefont {U.-L.}\ \bibnamefont {{Pen}}},
  \ and\ \bibinfo {author} {\bibfnamefont {U.}~\bibnamefont {{Seljak}}},\
  }\bibfield  {title} {\enquote {\bibinfo {title} {{Large-scale BAO signatures
  of the smallest galaxies}},}\ }\href {\doibase 10.1088/1475-7516/2010/11/007}
  {\bibfield  {journal} {\bibinfo  {journal} {\jcap}\ }\textbf {\bibinfo
  {volume} {11}},\ \bibinfo {pages} {7} (\bibinfo {year} {2010})},\ \Eprint
  {http://arxiv.org/abs/1009.4704} {arXiv:1009.4704 [astro-ph.CO]} \BibitemShut
  {NoStop}%
\bibitem [{\citenamefont {{Grin}}\ \emph {et~al.}(2011)\citenamefont {{Grin}},
  \citenamefont {{Dor{\'e}}},\ and\ \citenamefont
  {{Kamionkowski}}}]{GriDorKam1112}%
  \BibitemOpen
  \bibfield  {author} {\bibinfo {author} {\bibfnamefont {D.}~\bibnamefont
  {{Grin}}}, \bibinfo {author} {\bibfnamefont {O.}~\bibnamefont {{Dor{\'e}}}},
  \ and\ \bibinfo {author} {\bibfnamefont {M.}~\bibnamefont {{Kamionkowski}}},\
  }\bibfield  {title} {\enquote {\bibinfo {title} {{Do Baryons Trace Dark
  Matter in the Early Universe?}}}\ }\href {\doibase
  10.1103/PhysRevLett.107.261301} {\bibfield  {journal} {\bibinfo  {journal}
  {Physical Review Letters}\ }\textbf {\bibinfo {volume} {107}},\ \bibinfo
  {eid} {261301} (\bibinfo {year} {2011})},\ \Eprint
  {http://arxiv.org/abs/1107.1716} {arXiv:1107.1716 [astro-ph.CO]} \BibitemShut
  {NoStop}%
\bibitem [{\citenamefont {{Crocce}}\ and\ \citenamefont
  {{Scoccimarro}}(2008)}]{CroSco0801}%
  \BibitemOpen
  \bibfield  {author} {\bibinfo {author} {\bibfnamefont {M.}~\bibnamefont
  {{Crocce}}}\ and\ \bibinfo {author} {\bibfnamefont {R.}~\bibnamefont
  {{Scoccimarro}}},\ }\bibfield  {title} {\enquote {\bibinfo {title}
  {{Nonlinear evolution of baryon acoustic oscillations}},}\ }\href {\doibase
  10.1103/PhysRevD.77.023533} {\bibfield  {journal} {\bibinfo  {journal}
  {\prd}\ }\textbf {\bibinfo {volume} {77}},\ \bibinfo {pages} {023533--+}
  (\bibinfo {year} {2008})},\ \Eprint {http://arxiv.org/abs/0704.2783}
  {arXiv:0704.2783} \BibitemShut {NoStop}%
\bibitem [{\citenamefont {{Bartolo}}\ \emph {et~al.}(2010)\citenamefont
  {{Bartolo}}, \citenamefont {{Matarrese}},\ and\ \citenamefont
  {{Riotto}}}]{BarMatRio1011}%
  \BibitemOpen
  \bibfield  {author} {\bibinfo {author} {\bibfnamefont {N.}~\bibnamefont
  {{Bartolo}}}, \bibinfo {author} {\bibfnamefont {S.}~\bibnamefont
  {{Matarrese}}}, \ and\ \bibinfo {author} {\bibfnamefont {A.}~\bibnamefont
  {{Riotto}}},\ }\bibfield  {title} {\enquote {\bibinfo {title} {{The
  Gauge-Invariant Bias of Dark Matter Haloes with Primordial
  non-Gaussianity}},}\ }\href@noop {} {\bibfield  {journal} {\bibinfo
  {journal} {ArXiv e-prints}\ } (\bibinfo {year} {2010})},\ \Eprint
  {http://arxiv.org/abs/1011.4374} {arXiv:1011.4374 [astro-ph.CO]} \BibitemShut
  {NoStop}%
\bibitem [{\citenamefont {{Zel'Dovich}}(1970)}]{Zel7003}%
  \BibitemOpen
  \bibfield  {author} {\bibinfo {author} {\bibfnamefont {Y.~B.}\ \bibnamefont
  {{Zel'Dovich}}},\ }\bibfield  {title} {\enquote {\bibinfo {title}
  {{Gravitational instability: An approximate theory for large density
  perturbations.}}}\ }\href@noop {} {\bibfield  {journal} {\bibinfo  {journal}
  {\aap}\ }\textbf {\bibinfo {volume} {5}},\ \bibinfo {pages} {84--89}
  (\bibinfo {year} {1970})}\BibitemShut {NoStop}%
\bibitem [{\citenamefont {{Matarrese}}\ \emph {et~al.}(1997)\citenamefont
  {{Matarrese}}, \citenamefont {{Coles}}, \citenamefont {{Lucchin}},\ and\
  \citenamefont {{Moscardini}}}]{MatColLuc9703}%
  \BibitemOpen
  \bibfield  {author} {\bibinfo {author} {\bibfnamefont {S.}~\bibnamefont
  {{Matarrese}}}, \bibinfo {author} {\bibfnamefont {P.}~\bibnamefont
  {{Coles}}}, \bibinfo {author} {\bibfnamefont {F.}~\bibnamefont {{Lucchin}}},
  \ and\ \bibinfo {author} {\bibfnamefont {L.}~\bibnamefont {{Moscardini}}},\
  }\bibfield  {title} {\enquote {\bibinfo {title} {{Redshift evolution of
  clustering}},}\ }\href@noop {} {\bibfield  {journal} {\bibinfo  {journal}
  {\mnras}\ }\textbf {\bibinfo {volume} {286}},\ \bibinfo {pages} {115--132}
  (\bibinfo {year} {1997})},\ \Eprint
  {http://arxiv.org/abs/arXiv:astro-ph/9608004} {arXiv:astro-ph/9608004}
  \BibitemShut {NoStop}%
\bibitem [{\citenamefont {{Moscardini}}\ \emph {et~al.}(1998)\citenamefont
  {{Moscardini}}, \citenamefont {{Coles}}, \citenamefont {{Lucchin}},\ and\
  \citenamefont {{Matarrese}}}]{MosColLuc9808}%
  \BibitemOpen
  \bibfield  {author} {\bibinfo {author} {\bibfnamefont {L.}~\bibnamefont
  {{Moscardini}}}, \bibinfo {author} {\bibfnamefont {P.}~\bibnamefont
  {{Coles}}}, \bibinfo {author} {\bibfnamefont {F.}~\bibnamefont {{Lucchin}}},
  \ and\ \bibinfo {author} {\bibfnamefont {S.}~\bibnamefont {{Matarrese}}},\
  }\bibfield  {title} {\enquote {\bibinfo {title} {{Modelling galaxy clustering
  at high redshift}},}\ }\href {\doibase 10.1046/j.1365-8711.1998.01728.x}
  {\bibfield  {journal} {\bibinfo  {journal} {\mnras}\ }\textbf {\bibinfo
  {volume} {299}},\ \bibinfo {pages} {95--110} (\bibinfo {year} {1998})},\
  \Eprint {http://arxiv.org/abs/arXiv:astro-ph/9712184}
  {arXiv:astro-ph/9712184} \BibitemShut {NoStop}%
\bibitem [{\citenamefont {{Col{\'{\i}}n}}\ \emph {et~al.}(1999)\citenamefont
  {{Col{\'{\i}}n}}, \citenamefont {{Klypin}}, \citenamefont {{Kravtsov}},\ and\
  \citenamefont {{Khokhlov}}}]{ColKlyKra9909}%
  \BibitemOpen
  \bibfield  {author} {\bibinfo {author} {\bibfnamefont {P.}~\bibnamefont
  {{Col{\'{\i}}n}}}, \bibinfo {author} {\bibfnamefont {A.~A.}\ \bibnamefont
  {{Klypin}}}, \bibinfo {author} {\bibfnamefont {A.~V.}\ \bibnamefont
  {{Kravtsov}}}, \ and\ \bibinfo {author} {\bibfnamefont {A.~M.}\ \bibnamefont
  {{Khokhlov}}},\ }\bibfield  {title} {\enquote {\bibinfo {title} {{Evolution
  of Bias in Different Cosmological Models}},}\ }\href {\doibase
  10.1086/307710} {\bibfield  {journal} {\bibinfo  {journal} {\apj}\ }\textbf
  {\bibinfo {volume} {523}},\ \bibinfo {pages} {32--53} (\bibinfo {year}
  {1999})},\ \Eprint {http://arxiv.org/abs/arXiv:astro-ph/9809202}
  {arXiv:astro-ph/9809202} \BibitemShut {NoStop}%
\bibitem [{\citenamefont {{Marinoni}}\ \emph {et~al.}(2005)\citenamefont
  {{Marinoni}}, \citenamefont {{Le F{\`e}vre}}, \citenamefont {{Meneux}},
  \citenamefont {{Iovino}}, \citenamefont {{Pollo}}, \citenamefont {{Ilbert}},
  \citenamefont {{Zamorani}}, \citenamefont {{Guzzo}}, \citenamefont
  {{Mazure}}, \citenamefont {{Scaramella}}, \citenamefont {{Cappi}},
  \citenamefont {{McCracken}}, \citenamefont {{Bottini}}, \citenamefont
  {{Garilli}}, \citenamefont {{Le Brun}}, \citenamefont {{Maccagni}},
  \citenamefont {{Picat}}, \citenamefont {{Scodeggio}}, \citenamefont
  {{Tresse}}, \citenamefont {{Vettolani}}, \citenamefont {{Zanichelli}},
  \citenamefont {{Adami}}, \citenamefont {{Arnouts}}, \citenamefont
  {{Bardelli}}, \citenamefont {{Blaizot}}, \citenamefont {{Bolzonella}},
  \citenamefont {{Charlot}}, \citenamefont {{Ciliegi}}, \citenamefont
  {{Contini}}, \citenamefont {{Foucaud}}, \citenamefont {{Franzetti}},
  \citenamefont {{Gavignaud}}, \citenamefont {{Marano}}, \citenamefont
  {{Mathez}}, \citenamefont {{Merighi}}, \citenamefont {{Paltani}},
  \citenamefont {{Pell{\`o}}}, \citenamefont {{Pozzetti}}, \citenamefont
  {{Radovich}}, \citenamefont {{Zucca}}, \citenamefont {{Bondi}}, \citenamefont
  {{Bongiorno}}, \citenamefont {{Busarello}}, \citenamefont {{Colombi}},
  \citenamefont {{Cucciati}}, \citenamefont {{Lamareille}}, \citenamefont
  {{Mellier}}, \citenamefont {{Merluzzi}}, \citenamefont {{Ripepi}},\ and\
  \citenamefont {{Rizzo}}}]{MarLe-Men0511}%
  \BibitemOpen
  \bibfield  {author} {\bibinfo {author} {\bibfnamefont {C.}~\bibnamefont
  {{Marinoni}}}, \bibinfo {author} {\bibfnamefont {O.}~\bibnamefont {{Le
  F{\`e}vre}}}, \bibinfo {author} {\bibfnamefont {B.}~\bibnamefont {{Meneux}}},
  \bibinfo {author} {\bibfnamefont {A.}~\bibnamefont {{Iovino}}}, \bibinfo
  {author} {\bibfnamefont {A.}~\bibnamefont {{Pollo}}}, \bibinfo {author}
  {\bibfnamefont {O.}~\bibnamefont {{Ilbert}}}, \bibinfo {author}
  {\bibfnamefont {G.}~\bibnamefont {{Zamorani}}}, \bibinfo {author}
  {\bibfnamefont {L.}~\bibnamefont {{Guzzo}}}, \bibinfo {author} {\bibfnamefont
  {A.}~\bibnamefont {{Mazure}}}, \bibinfo {author} {\bibfnamefont
  {R.}~\bibnamefont {{Scaramella}}}, \bibinfo {author} {\bibfnamefont
  {A.}~\bibnamefont {{Cappi}}}, \bibinfo {author} {\bibfnamefont {H.~J.}\
  \bibnamefont {{McCracken}}}, \bibinfo {author} {\bibfnamefont
  {D.}~\bibnamefont {{Bottini}}}, \bibinfo {author} {\bibfnamefont
  {B.}~\bibnamefont {{Garilli}}}, \bibinfo {author} {\bibfnamefont
  {V.}~\bibnamefont {{Le Brun}}}, \bibinfo {author} {\bibfnamefont
  {D.}~\bibnamefont {{Maccagni}}}, \bibinfo {author} {\bibfnamefont {J.~P.}\
  \bibnamefont {{Picat}}}, \bibinfo {author} {\bibfnamefont {M.}~\bibnamefont
  {{Scodeggio}}}, \bibinfo {author} {\bibfnamefont {L.}~\bibnamefont
  {{Tresse}}}, \bibinfo {author} {\bibfnamefont {G.}~\bibnamefont
  {{Vettolani}}}, \bibinfo {author} {\bibfnamefont {A.}~\bibnamefont
  {{Zanichelli}}}, \bibinfo {author} {\bibfnamefont {C.}~\bibnamefont
  {{Adami}}}, \bibinfo {author} {\bibfnamefont {S.}~\bibnamefont {{Arnouts}}},
  \bibinfo {author} {\bibfnamefont {S.}~\bibnamefont {{Bardelli}}}, \bibinfo
  {author} {\bibfnamefont {J.}~\bibnamefont {{Blaizot}}}, \bibinfo {author}
  {\bibfnamefont {M.}~\bibnamefont {{Bolzonella}}}, \bibinfo {author}
  {\bibfnamefont {S.}~\bibnamefont {{Charlot}}}, \bibinfo {author}
  {\bibfnamefont {P.}~\bibnamefont {{Ciliegi}}}, \bibinfo {author}
  {\bibfnamefont {T.}~\bibnamefont {{Contini}}}, \bibinfo {author}
  {\bibfnamefont {S.}~\bibnamefont {{Foucaud}}}, \bibinfo {author}
  {\bibfnamefont {P.}~\bibnamefont {{Franzetti}}}, \bibinfo {author}
  {\bibfnamefont {I.}~\bibnamefont {{Gavignaud}}}, \bibinfo {author}
  {\bibfnamefont {B.}~\bibnamefont {{Marano}}}, \bibinfo {author}
  {\bibfnamefont {G.}~\bibnamefont {{Mathez}}}, \bibinfo {author}
  {\bibfnamefont {R.}~\bibnamefont {{Merighi}}}, \bibinfo {author}
  {\bibfnamefont {S.}~\bibnamefont {{Paltani}}}, \bibinfo {author}
  {\bibfnamefont {R.}~\bibnamefont {{Pell{\`o}}}}, \bibinfo {author}
  {\bibfnamefont {L.}~\bibnamefont {{Pozzetti}}}, \bibinfo {author}
  {\bibfnamefont {M.}~\bibnamefont {{Radovich}}}, \bibinfo {author}
  {\bibfnamefont {E.}~\bibnamefont {{Zucca}}}, \bibinfo {author} {\bibfnamefont
  {M.}~\bibnamefont {{Bondi}}}, \bibinfo {author} {\bibfnamefont
  {A.}~\bibnamefont {{Bongiorno}}}, \bibinfo {author} {\bibfnamefont
  {G.}~\bibnamefont {{Busarello}}}, \bibinfo {author} {\bibfnamefont
  {S.}~\bibnamefont {{Colombi}}}, \bibinfo {author} {\bibfnamefont
  {O.}~\bibnamefont {{Cucciati}}}, \bibinfo {author} {\bibfnamefont
  {F.}~\bibnamefont {{Lamareille}}}, \bibinfo {author} {\bibfnamefont
  {Y.}~\bibnamefont {{Mellier}}}, \bibinfo {author} {\bibfnamefont
  {P.}~\bibnamefont {{Merluzzi}}}, \bibinfo {author} {\bibfnamefont
  {V.}~\bibnamefont {{Ripepi}}}, \ and\ \bibinfo {author} {\bibfnamefont
  {D.}~\bibnamefont {{Rizzo}}},\ }\bibfield  {title} {\enquote {\bibinfo
  {title} {{The VIMOS VLT Deep Survey. Evolution of the non-linear galaxy bias
  up to z = 1.5}},}\ }\href {\doibase 10.1051/0004-6361:20052966} {\bibfield
  {journal} {\bibinfo  {journal} {\aap}\ }\textbf {\bibinfo {volume} {442}},\
  \bibinfo {pages} {801--825} (\bibinfo {year} {2005})},\ \Eprint
  {http://arxiv.org/abs/arXiv:astro-ph/0506561} {arXiv:astro-ph/0506561}
  \BibitemShut {NoStop}%
\bibitem [{\citenamefont {{White}}\ \emph {et~al.}(2007)\citenamefont
  {{White}}, \citenamefont {{Zheng}}, \citenamefont {{Brown}}, \citenamefont
  {{Dey}},\ and\ \citenamefont {{Jannuzi}}}]{WhiZheBro0702}%
  \BibitemOpen
  \bibfield  {author} {\bibinfo {author} {\bibfnamefont {M.}~\bibnamefont
  {{White}}}, \bibinfo {author} {\bibfnamefont {Z.}~\bibnamefont {{Zheng}}},
  \bibinfo {author} {\bibfnamefont {M.~J.~I.}\ \bibnamefont {{Brown}}},
  \bibinfo {author} {\bibfnamefont {A.}~\bibnamefont {{Dey}}}, \ and\ \bibinfo
  {author} {\bibfnamefont {B.~T.}\ \bibnamefont {{Jannuzi}}},\ }\bibfield
  {title} {\enquote {\bibinfo {title} {{Evidence for Merging or Disruption of
  Red Galaxies from the Evolution of Their Clustering}},}\ }\href {\doibase
  10.1086/512015} {\bibfield  {journal} {\bibinfo  {journal} {\apjl}\ }\textbf
  {\bibinfo {volume} {655}},\ \bibinfo {pages} {L69--L72} (\bibinfo {year}
  {2007})},\ \Eprint {http://arxiv.org/abs/arXiv:astro-ph/0611901}
  {arXiv:astro-ph/0611901} \BibitemShut {NoStop}%
\bibitem [{\citenamefont {{Brown}}\ \emph {et~al.}(2008)\citenamefont
  {{Brown}}, \citenamefont {{Zheng}}, \citenamefont {{White}}, \citenamefont
  {{Dey}}, \citenamefont {{Jannuzi}}, \citenamefont {{Benson}}, \citenamefont
  {{Brand}}, \citenamefont {{Brodwin}},\ and\ \citenamefont
  {{Croton}}}]{BroZheWhi0808}%
  \BibitemOpen
  \bibfield  {author} {\bibinfo {author} {\bibfnamefont {M.~J.~I.}\
  \bibnamefont {{Brown}}}, \bibinfo {author} {\bibfnamefont {Z.}~\bibnamefont
  {{Zheng}}}, \bibinfo {author} {\bibfnamefont {M.}~\bibnamefont {{White}}},
  \bibinfo {author} {\bibfnamefont {A.}~\bibnamefont {{Dey}}}, \bibinfo
  {author} {\bibfnamefont {B.~T.}\ \bibnamefont {{Jannuzi}}}, \bibinfo {author}
  {\bibfnamefont {A.~J.}\ \bibnamefont {{Benson}}}, \bibinfo {author}
  {\bibfnamefont {K.}~\bibnamefont {{Brand}}}, \bibinfo {author} {\bibfnamefont
  {M.}~\bibnamefont {{Brodwin}}}, \ and\ \bibinfo {author} {\bibfnamefont
  {D.~J.}\ \bibnamefont {{Croton}}},\ }\bibfield  {title} {\enquote {\bibinfo
  {title} {{Red Galaxy Growth and the Halo Occupation Distribution}},}\ }\href
  {\doibase 10.1086/589538} {\bibfield  {journal} {\bibinfo  {journal} {\apj}\
  }\textbf {\bibinfo {volume} {682}},\ \bibinfo {pages} {937--963} (\bibinfo
  {year} {2008})},\ \Eprint {http://arxiv.org/abs/0804.2293} {arXiv:0804.2293}
  \BibitemShut {NoStop}%
\bibitem [{\citenamefont {{Wake}}\ \emph {et~al.}(2008)\citenamefont {{Wake}},
  \citenamefont {{Croom}}, \citenamefont {{Sadler}},\ and\ \citenamefont
  {{Johnston}}}]{WakCroSad0812}%
  \BibitemOpen
  \bibfield  {author} {\bibinfo {author} {\bibfnamefont {D.~A.}\ \bibnamefont
  {{Wake}}}, \bibinfo {author} {\bibfnamefont {S.~M.}\ \bibnamefont {{Croom}}},
  \bibinfo {author} {\bibfnamefont {E.~M.}\ \bibnamefont {{Sadler}}}, \ and\
  \bibinfo {author} {\bibfnamefont {H.~M.}\ \bibnamefont {{Johnston}}},\
  }\bibfield  {title} {\enquote {\bibinfo {title} {{The clustering of radio
  galaxies at z \~{}= 0.55 from the 2SLAQ LRG survey}},}\ }\href {\doibase
  10.1111/j.1365-2966.2008.14039.x} {\bibfield  {journal} {\bibinfo  {journal}
  {\mnras}\ }\textbf {\bibinfo {volume} {391}},\ \bibinfo {pages} {1674--1684}
  (\bibinfo {year} {2008})},\ \Eprint {http://arxiv.org/abs/0810.1050}
  {arXiv:0810.1050} \BibitemShut {NoStop}%
\bibitem [{\citenamefont {{Tojeiro}}\ and\ \citenamefont
  {{Percival}}(2010)}]{TojPer1007}%
  \BibitemOpen
  \bibfield  {author} {\bibinfo {author} {\bibfnamefont {R.}~\bibnamefont
  {{Tojeiro}}}\ and\ \bibinfo {author} {\bibfnamefont {W.~J.}\ \bibnamefont
  {{Percival}}},\ }\bibfield  {title} {\enquote {\bibinfo {title} {{The
  evolution of luminous red galaxies in the Sloan Digital Sky Survey 7th data
  release}},}\ }\href {\doibase 10.1111/j.1365-2966.2010.16630.x} {\bibfield
  {journal} {\bibinfo  {journal} {\mnras}\ }\textbf {\bibinfo {volume} {405}},\
  \bibinfo {pages} {2534--2548} (\bibinfo {year} {2010})},\ \Eprint
  {http://arxiv.org/abs/1001.2015} {arXiv:1001.2015 [astro-ph.CO]} \BibitemShut
  {NoStop}%
\bibitem [{\citenamefont {{Kovac}}\ \emph {et~al.}(2011)\citenamefont
  {{Kovac}}, \citenamefont {{Porciani}}, \citenamefont {{Lilly}}, \citenamefont
  {{Marinoni}}, \citenamefont {{Guzzo}}, \citenamefont {{Cucciati}},
  \citenamefont {{Zamorani}}, \citenamefont {{Iovino}}, \citenamefont
  {{Oesch}}, \citenamefont {{Bolzonella}}, \citenamefont {{Peng}},
  \citenamefont {{Meneux}}, \citenamefont {{Zucca}}, \citenamefont
  {{Bardelli}}, \citenamefont {{Carollo}}, \citenamefont {{Contini}},
  \citenamefont {{Kneib}}, \citenamefont {{Le F{\`e}vre}}, \citenamefont
  {{Mainieri}}, \citenamefont {{Renzini}}, \citenamefont {{Scodeggio}},
  \citenamefont {{Bongiorno}}, \citenamefont {{Caputi}}, \citenamefont
  {{Coppa}}, \citenamefont {{de la Torre}}, \citenamefont {{de Ravel}},
  \citenamefont {{Finoguenov}}, \citenamefont {{Franzetti}}, \citenamefont
  {{Garilli}}, \citenamefont {{Kampczyk}}, \citenamefont {{Knobel}},
  \citenamefont {{Lamareille}}, \citenamefont {{Le Borgne}}, \citenamefont {{Le
  Brun}}, \citenamefont {{Maier}}, \citenamefont {{Mignoli}}, \citenamefont
  {{Pello}}, \citenamefont {{Perez-Montero}}, \citenamefont {{Pozzetti}},
  \citenamefont {{Ricciardelli}}, \citenamefont {{Silverman}}, \citenamefont
  {{Tanaka}}, \citenamefont {{Tasca}}, \citenamefont {{Tresse}}, \citenamefont
  {{Vergani}}, \citenamefont {{Abbas}}, \citenamefont {{Bottini}},
  \citenamefont {{Cappi}}, \citenamefont {{Cassata}}, \citenamefont
  {{Cimatti}}, \citenamefont {{Fumana}}, \citenamefont {{Koekemoer}},
  \citenamefont {{Leauthaud}}, \citenamefont {{Maccagni}}, \citenamefont
  {{McCracken}}, \citenamefont {{Memeo}}, \citenamefont {{Scaramella}},\ and\
  \citenamefont {{Scoville}}}]{KovPorLil1104}%
  \BibitemOpen
  \bibfield  {author} {\bibinfo {author} {\bibfnamefont {K.}~\bibnamefont
  {{Kovac}}}, \bibinfo {author} {\bibfnamefont {C.}~\bibnamefont {{Porciani}}},
  \bibinfo {author} {\bibfnamefont {S.~J.}\ \bibnamefont {{Lilly}}}, \bibinfo
  {author} {\bibfnamefont {C.}~\bibnamefont {{Marinoni}}}, \bibinfo {author}
  {\bibfnamefont {L.}~\bibnamefont {{Guzzo}}}, \bibinfo {author} {\bibfnamefont
  {O.}~\bibnamefont {{Cucciati}}}, \bibinfo {author} {\bibfnamefont
  {G.}~\bibnamefont {{Zamorani}}}, \bibinfo {author} {\bibfnamefont
  {A.}~\bibnamefont {{Iovino}}}, \bibinfo {author} {\bibfnamefont
  {P.}~\bibnamefont {{Oesch}}}, \bibinfo {author} {\bibfnamefont
  {M.}~\bibnamefont {{Bolzonella}}}, \bibinfo {author} {\bibfnamefont
  {Y.}~\bibnamefont {{Peng}}}, \bibinfo {author} {\bibfnamefont
  {B.}~\bibnamefont {{Meneux}}}, \bibinfo {author} {\bibfnamefont
  {E.}~\bibnamefont {{Zucca}}}, \bibinfo {author} {\bibfnamefont
  {S.}~\bibnamefont {{Bardelli}}}, \bibinfo {author} {\bibfnamefont {C.~M.}\
  \bibnamefont {{Carollo}}}, \bibinfo {author} {\bibfnamefont {T.}~\bibnamefont
  {{Contini}}}, \bibinfo {author} {\bibfnamefont {J.-P.}\ \bibnamefont
  {{Kneib}}}, \bibinfo {author} {\bibfnamefont {O.}~\bibnamefont {{Le
  F{\`e}vre}}}, \bibinfo {author} {\bibfnamefont {V.}~\bibnamefont
  {{Mainieri}}}, \bibinfo {author} {\bibfnamefont {A.}~\bibnamefont
  {{Renzini}}}, \bibinfo {author} {\bibfnamefont {M.}~\bibnamefont
  {{Scodeggio}}}, \bibinfo {author} {\bibfnamefont {A.}~\bibnamefont
  {{Bongiorno}}}, \bibinfo {author} {\bibfnamefont {K.}~\bibnamefont
  {{Caputi}}}, \bibinfo {author} {\bibfnamefont {G.}~\bibnamefont {{Coppa}}},
  \bibinfo {author} {\bibfnamefont {S.}~\bibnamefont {{de la Torre}}}, \bibinfo
  {author} {\bibfnamefont {L.}~\bibnamefont {{de Ravel}}}, \bibinfo {author}
  {\bibfnamefont {A.}~\bibnamefont {{Finoguenov}}}, \bibinfo {author}
  {\bibfnamefont {P.}~\bibnamefont {{Franzetti}}}, \bibinfo {author}
  {\bibfnamefont {B.}~\bibnamefont {{Garilli}}}, \bibinfo {author}
  {\bibfnamefont {P.}~\bibnamefont {{Kampczyk}}}, \bibinfo {author}
  {\bibfnamefont {C.}~\bibnamefont {{Knobel}}}, \bibinfo {author}
  {\bibfnamefont {F.}~\bibnamefont {{Lamareille}}}, \bibinfo {author}
  {\bibfnamefont {J.-F.}\ \bibnamefont {{Le Borgne}}}, \bibinfo {author}
  {\bibfnamefont {V.}~\bibnamefont {{Le Brun}}}, \bibinfo {author}
  {\bibfnamefont {C.}~\bibnamefont {{Maier}}}, \bibinfo {author} {\bibfnamefont
  {M.}~\bibnamefont {{Mignoli}}}, \bibinfo {author} {\bibfnamefont
  {R.}~\bibnamefont {{Pello}}}, \bibinfo {author} {\bibfnamefont
  {E.}~\bibnamefont {{Perez-Montero}}}, \bibinfo {author} {\bibfnamefont
  {L.}~\bibnamefont {{Pozzetti}}}, \bibinfo {author} {\bibfnamefont
  {E.}~\bibnamefont {{Ricciardelli}}}, \bibinfo {author} {\bibfnamefont
  {J.~D.}\ \bibnamefont {{Silverman}}}, \bibinfo {author} {\bibfnamefont
  {M.}~\bibnamefont {{Tanaka}}}, \bibinfo {author} {\bibfnamefont {L.~A.~M.}\
  \bibnamefont {{Tasca}}}, \bibinfo {author} {\bibfnamefont {L.}~\bibnamefont
  {{Tresse}}}, \bibinfo {author} {\bibfnamefont {D.}~\bibnamefont {{Vergani}}},
  \bibinfo {author} {\bibfnamefont {U.}~\bibnamefont {{Abbas}}}, \bibinfo
  {author} {\bibfnamefont {D.}~\bibnamefont {{Bottini}}}, \bibinfo {author}
  {\bibfnamefont {A.}~\bibnamefont {{Cappi}}}, \bibinfo {author} {\bibfnamefont
  {P.}~\bibnamefont {{Cassata}}}, \bibinfo {author} {\bibfnamefont
  {A.}~\bibnamefont {{Cimatti}}}, \bibinfo {author} {\bibfnamefont
  {M.}~\bibnamefont {{Fumana}}}, \bibinfo {author} {\bibfnamefont {A.~M.}\
  \bibnamefont {{Koekemoer}}}, \bibinfo {author} {\bibfnamefont
  {A.}~\bibnamefont {{Leauthaud}}}, \bibinfo {author} {\bibfnamefont
  {D.}~\bibnamefont {{Maccagni}}}, \bibinfo {author} {\bibfnamefont {H.~J.}\
  \bibnamefont {{McCracken}}}, \bibinfo {author} {\bibfnamefont
  {P.}~\bibnamefont {{Memeo}}}, \bibinfo {author} {\bibfnamefont
  {R.}~\bibnamefont {{Scaramella}}}, \ and\ \bibinfo {author} {\bibfnamefont
  {N.~Z.}\ \bibnamefont {{Scoville}}},\ }\bibfield  {title} {\enquote {\bibinfo
  {title} {{The Nonlinear Biasing of the zCOSMOS Galaxies up to z \~{} 1 from
  the 10k Sample}},}\ }\href {\doibase 10.1088/0004-637X/731/2/102} {\bibfield
  {journal} {\bibinfo  {journal} {\apj}\ }\textbf {\bibinfo {volume} {731}},\
  \bibinfo {eid} {102} (\bibinfo {year} {2011})},\ \Eprint
  {http://arxiv.org/abs/0910.0004} {arXiv:0910.0004 [astro-ph.CO]} \BibitemShut
  {NoStop}%
\bibitem [{\citenamefont {{Nicolis}}\ \emph {et~al.}(2009)\citenamefont
  {{Nicolis}}, \citenamefont {{Rattazzi}},\ and\ \citenamefont
  {{Trincherini}}}]{NicRatTri0903}%
  \BibitemOpen
  \bibfield  {author} {\bibinfo {author} {\bibfnamefont {A.}~\bibnamefont
  {{Nicolis}}}, \bibinfo {author} {\bibfnamefont {R.}~\bibnamefont
  {{Rattazzi}}}, \ and\ \bibinfo {author} {\bibfnamefont {E.}~\bibnamefont
  {{Trincherini}}},\ }\bibfield  {title} {\enquote {\bibinfo {title} {{Galileon
  as a local modification of gravity}},}\ }\href {\doibase
  10.1103/PhysRevD.79.064036} {\bibfield  {journal} {\bibinfo  {journal}
  {\prd}\ }\textbf {\bibinfo {volume} {79}},\ \bibinfo {eid} {064036} (\bibinfo
  {year} {2009})},\ \Eprint {http://arxiv.org/abs/0811.2197} {arXiv:0811.2197
  [hep-th]} \BibitemShut {NoStop}%
\bibitem [{\citenamefont {{Scoccimarro}}\ and\ \citenamefont
  {{Frieman}}(1996)}]{ScoFri9607}%
  \BibitemOpen
  \bibfield  {author} {\bibinfo {author} {\bibfnamefont {R.}~\bibnamefont
  {{Scoccimarro}}}\ and\ \bibinfo {author} {\bibfnamefont {J.}~\bibnamefont
  {{Frieman}}},\ }\bibfield  {title} {\enquote {\bibinfo {title} {{Loop
  Corrections in Nonlinear Cosmological Perturbation Theory}},}\ }\href
  {\doibase 10.1086/192306} {\bibfield  {journal} {\bibinfo  {journal} {\apjs}\
  }\textbf {\bibinfo {volume} {105}},\ \bibinfo {pages} {37} (\bibinfo {year}
  {1996})},\ \Eprint {http://arxiv.org/abs/arXiv:astro-ph/9509047}
  {arXiv:astro-ph/9509047} \BibitemShut {NoStop}%
\bibitem [{\citenamefont {{Munshi}}\ and\ \citenamefont
  {{Starobinsky}}(1994)}]{MunSta9406}%
  \BibitemOpen
  \bibfield  {author} {\bibinfo {author} {\bibfnamefont {D.}~\bibnamefont
  {{Munshi}}}\ and\ \bibinfo {author} {\bibfnamefont {A.~A.}\ \bibnamefont
  {{Starobinsky}}},\ }\bibfield  {title} {\enquote {\bibinfo {title}
  {{Nonlinear approximations to gravitational instability: A comparison in
  second-order perturbation theory}},}\ }\href {\doibase 10.1086/174255}
  {\bibfield  {journal} {\bibinfo  {journal} {\apj}\ }\textbf {\bibinfo
  {volume} {428}},\ \bibinfo {pages} {433--438} (\bibinfo {year} {1994})},\
  \Eprint {http://arxiv.org/abs/arXiv:astro-ph/9311056}
  {arXiv:astro-ph/9311056} \BibitemShut {NoStop}%
\bibitem [{\citenamefont {{Hui}}\ and\ \citenamefont
  {{Bertschinger}}(1996)}]{HuiBer9611}%
  \BibitemOpen
  \bibfield  {author} {\bibinfo {author} {\bibfnamefont {L.}~\bibnamefont
  {{Hui}}}\ and\ \bibinfo {author} {\bibfnamefont {E.}~\bibnamefont
  {{Bertschinger}}},\ }\bibfield  {title} {\enquote {\bibinfo {title} {{Local
  Approximations to the Gravitational Collapse of Cold Matter}},}\ }\href
  {\doibase 10.1086/177948} {\bibfield  {journal} {\bibinfo  {journal} {\apj}\
  }\textbf {\bibinfo {volume} {471}},\ \bibinfo {pages} {1} (\bibinfo {year}
  {1996})},\ \Eprint {http://arxiv.org/abs/arXiv:astro-ph/9508114}
  {arXiv:astro-ph/9508114} \BibitemShut {NoStop}%
\bibitem [{\citenamefont {{Bouchet}}\ \emph {et~al.}(1995)\citenamefont
  {{Bouchet}}, \citenamefont {{Colombi}}, \citenamefont {{Hivon}},\ and\
  \citenamefont {{Juszkiewicz}}}]{BouColHiv9504}%
  \BibitemOpen
  \bibfield  {author} {\bibinfo {author} {\bibfnamefont {F.~R.}\ \bibnamefont
  {{Bouchet}}}, \bibinfo {author} {\bibfnamefont {S.}~\bibnamefont
  {{Colombi}}}, \bibinfo {author} {\bibfnamefont {E.}~\bibnamefont {{Hivon}}},
  \ and\ \bibinfo {author} {\bibfnamefont {R.}~\bibnamefont {{Juszkiewicz}}},\
  }\bibfield  {title} {\enquote {\bibinfo {title} {{Perturbative Lagrangian
  approach to gravitational instability.}}}\ }\href@noop {} {\bibfield
  {journal} {\bibinfo  {journal} {\aap}\ }\textbf {\bibinfo {volume} {296}},\
  \bibinfo {pages} {575} (\bibinfo {year} {1995})},\ \Eprint
  {http://arxiv.org/abs/arXiv:astro-ph/9406013} {arXiv:astro-ph/9406013}
  \BibitemShut {NoStop}%
\bibitem [{\citenamefont {{Springel}}(2005)}]{2005MNRAS.364.1105S}%
  \BibitemOpen
  \bibfield  {author} {\bibinfo {author} {\bibfnamefont {V.}~\bibnamefont
  {{Springel}}},\ }\bibfield  {title} {\enquote {\bibinfo {title} {{The
  cosmological simulation code GADGET-2}},}\ }\href {\doibase
  10.1111/j.1365-2966.2005.09655.x} {\bibfield  {journal} {\bibinfo  {journal}
  {\mnras}\ }\textbf {\bibinfo {volume} {364}},\ \bibinfo {pages} {1105--1134}
  (\bibinfo {year} {2005})},\ \Eprint
  {http://arxiv.org/abs/arXiv:astro-ph/0505010} {arXiv:astro-ph/0505010}
  \BibitemShut {NoStop}%
\bibitem [{\citenamefont {{Crocce}}\ \emph {et~al.}(2006)\citenamefont
  {{Crocce}}, \citenamefont {{Pueblas}},\ and\ \citenamefont
  {{Scoccimarro}}}]{2006MNRAS.373..369C}%
  \BibitemOpen
  \bibfield  {author} {\bibinfo {author} {\bibfnamefont {M.}~\bibnamefont
  {{Crocce}}}, \bibinfo {author} {\bibfnamefont {S.}~\bibnamefont {{Pueblas}}},
  \ and\ \bibinfo {author} {\bibfnamefont {R.}~\bibnamefont {{Scoccimarro}}},\
  }\bibfield  {title} {\enquote {\bibinfo {title} {{Transients from initial
  conditions in cosmological simulations}},}\ }\href {\doibase
  10.1111/j.1365-2966.2006.11040.x} {\bibfield  {journal} {\bibinfo  {journal}
  {\mnras}\ }\textbf {\bibinfo {volume} {373}},\ \bibinfo {pages} {369--381}
  (\bibinfo {year} {2006})},\ \Eprint
  {http://arxiv.org/abs/arXiv:astro-ph/0606505} {arXiv:astro-ph/0606505}
  \BibitemShut {NoStop}%
\bibitem [{\citenamefont {{Pueblas}}\ and\ \citenamefont
  {{Scoccimarro}}(2009)}]{PueSco0908}%
  \BibitemOpen
  \bibfield  {author} {\bibinfo {author} {\bibfnamefont {S.}~\bibnamefont
  {{Pueblas}}}\ and\ \bibinfo {author} {\bibfnamefont {R.}~\bibnamefont
  {{Scoccimarro}}},\ }\bibfield  {title} {\enquote {\bibinfo {title}
  {{Generation of vorticity and velocity dispersion by orbit crossing}},}\
  }\href {\doibase 10.1103/PhysRevD.80.043504} {\bibfield  {journal} {\bibinfo
  {journal} {\prd}\ }\textbf {\bibinfo {volume} {80}},\ \bibinfo {pages}
  {043504--+} (\bibinfo {year} {2009})},\ \Eprint
  {http://arxiv.org/abs/0809.4606} {arXiv:0809.4606} \BibitemShut {NoStop}%
\bibitem [{\citenamefont {{Scoccimarro}}(2004)}]{2004PhRvD..70h3007S}%
  \BibitemOpen
  \bibfield  {author} {\bibinfo {author} {\bibfnamefont {R.}~\bibnamefont
  {{Scoccimarro}}},\ }\bibfield  {title} {\enquote {\bibinfo {title}
  {{Redshift-space distortions, pairwise velocities, and nonlinearities}},}\
  }\href {\doibase 10.1103/PhysRevD.70.083007} {\bibfield  {journal} {\bibinfo
  {journal} {\prd}\ }\textbf {\bibinfo {volume} {70}},\ \bibinfo {pages}
  {083007} (\bibinfo {year} {2004})}\BibitemShut {NoStop}%
\bibitem [{\citenamefont {{Smith}}\ \emph {et~al.}(2007)\citenamefont
  {{Smith}}, \citenamefont {{Scoccimarro}},\ and\ \citenamefont
  {{Sheth}}}]{2007PhRvD..75f3512S}%
  \BibitemOpen
  \bibfield  {author} {\bibinfo {author} {\bibfnamefont {R.~E.}\ \bibnamefont
  {{Smith}}}, \bibinfo {author} {\bibfnamefont {R.}~\bibnamefont
  {{Scoccimarro}}}, \ and\ \bibinfo {author} {\bibfnamefont {R.~K.}\
  \bibnamefont {{Sheth}}},\ }\bibfield  {title} {\enquote {\bibinfo {title}
  {{Scale dependence of halo and galaxy bias: Effects in real space}},}\ }\href
  {\doibase 10.1103/PhysRevD.75.063512} {\bibfield  {journal} {\bibinfo
  {journal} {\prd}\ }\textbf {\bibinfo {volume} {75}},\ \bibinfo {pages}
  {063512--+} (\bibinfo {year} {2007})},\ \Eprint
  {http://arxiv.org/abs/arXiv:astro-ph/0609547} {arXiv:astro-ph/0609547}
  \BibitemShut {NoStop}%
\bibitem [{\citenamefont {{Manera}}\ and\ \citenamefont
  {{Gaztanaga}}(2011)}]{2009arXiv0912.0446M}%
  \BibitemOpen
  \bibfield  {author} {\bibinfo {author} {\bibfnamefont {M.}~\bibnamefont
  {{Manera}}}\ and\ \bibinfo {author} {\bibfnamefont {E.}~\bibnamefont
  {{Gaztanaga}}},\ }\bibfield  {title} {\enquote {\bibinfo {title} {{The Local
  Bias Model in the Large Scale Halo Distribution}},}\ }\href@noop {}
  {\bibfield  {journal} {\bibinfo  {journal} {\mnras}\ }\textbf {\bibinfo
  {volume} {415}},\ \bibinfo {pages} {383--398} (\bibinfo {year} {2011})},\
  \Eprint {http://arxiv.org/abs/0912.0446} {arXiv:0912.0446 [astro-ph.CO]}
  \BibitemShut {NoStop}%
\bibitem [{\citenamefont {{Pollack}}\ \emph {et~al.}(2011)\citenamefont
  {{Pollack}}, \citenamefont {{Smith}},\ and\ \citenamefont
  {{Porciani}}}]{PolSmiPor1109}%
  \BibitemOpen
  \bibfield  {author} {\bibinfo {author} {\bibfnamefont {J.~E.}\ \bibnamefont
  {{Pollack}}}, \bibinfo {author} {\bibfnamefont {R.~E.}\ \bibnamefont
  {{Smith}}}, \ and\ \bibinfo {author} {\bibfnamefont {C.}~\bibnamefont
  {{Porciani}}},\ }\bibfield  {title} {\enquote {\bibinfo {title} {{Modelling
  large-scale halo bias using the bispectrum}},}\ }\href@noop {} {\bibfield
  {journal} {\bibinfo  {journal} {ArXiv e-prints}\ } (\bibinfo {year}
  {2011})},\ \Eprint {http://arxiv.org/abs/1109.3458} {arXiv:1109.3458
  [astro-ph.CO]} \BibitemShut {NoStop}%
\bibitem [{\citenamefont {{Sefusatti}}\ \emph {et~al.}(2011)\citenamefont
  {{Sefusatti}}, \citenamefont {{Crocce}},\ and\ \citenamefont
  {{Desjacques}}}]{SefCroDes1111}%
  \BibitemOpen
  \bibfield  {author} {\bibinfo {author} {\bibfnamefont {E.}~\bibnamefont
  {{Sefusatti}}}, \bibinfo {author} {\bibfnamefont {M.}~\bibnamefont
  {{Crocce}}}, \ and\ \bibinfo {author} {\bibfnamefont {V.}~\bibnamefont
  {{Desjacques}}},\ }\bibfield  {title} {\enquote {\bibinfo {title} {{The Halo
  Bispectrum in N-body Simulations with non-Gaussian Initial Conditions}},}\
  }\href@noop {} {\bibfield  {journal} {\bibinfo  {journal} {ArXiv e-prints}\ }
  (\bibinfo {year} {2011})},\ \Eprint {http://arxiv.org/abs/1111.6966}
  {arXiv:1111.6966 [astro-ph.CO]} \BibitemShut {NoStop}%
\bibitem [{\citenamefont {{Sefusatti}}\ \emph {et~al.}(2006)\citenamefont
  {{Sefusatti}}, \citenamefont {{Crocce}}, \citenamefont {{Pueblas}},\ and\
  \citenamefont {{Scoccimarro}}}]{2006PhRvD..74b3522S}%
  \BibitemOpen
  \bibfield  {author} {\bibinfo {author} {\bibfnamefont {E.}~\bibnamefont
  {{Sefusatti}}}, \bibinfo {author} {\bibfnamefont {M.}~\bibnamefont
  {{Crocce}}}, \bibinfo {author} {\bibfnamefont {S.}~\bibnamefont {{Pueblas}}},
  \ and\ \bibinfo {author} {\bibfnamefont {R.}~\bibnamefont {{Scoccimarro}}},\
  }\bibfield  {title} {\enquote {\bibinfo {title} {{Cosmology and the
  bispectrum}},}\ }\href {\doibase 10.1103/PhysRevD.74.023522} {\bibfield
  {journal} {\bibinfo  {journal} {\prd}\ }\textbf {\bibinfo {volume} {74}},\
  \bibinfo {pages} {023522--+} (\bibinfo {year} {2006})},\ \Eprint
  {http://arxiv.org/abs/arXiv:astro-ph/0604505} {arXiv:astro-ph/0604505}
  \BibitemShut {NoStop}%
\bibitem [{\citenamefont {{Chan}}\ and\ \citenamefont
  {{Scoccimarro}}(2009)}]{2009PhRvD..80j4005C}%
  \BibitemOpen
  \bibfield  {author} {\bibinfo {author} {\bibfnamefont {K.~C.}\ \bibnamefont
  {{Chan}}}\ and\ \bibinfo {author} {\bibfnamefont {R.}~\bibnamefont
  {{Scoccimarro}}},\ }\bibfield  {title} {\enquote {\bibinfo {title}
  {{Large-scale structure in brane-induced gravity. II. Numerical
  simulations}},}\ }\href {\doibase 10.1103/PhysRevD.80.104005} {\bibfield
  {journal} {\bibinfo  {journal} {\prd}\ }\textbf {\bibinfo {volume} {80}},\
  \bibinfo {pages} {104005--+} (\bibinfo {year} {2009})},\ \Eprint
  {http://arxiv.org/abs/0906.4548} {arXiv:0906.4548 [astro-ph.CO]} \BibitemShut
  {NoStop}%
\bibitem [{\citenamefont {{Scoccimarro}}(2009)}]{2009PhRvD..80j4006S}%
  \BibitemOpen
  \bibfield  {author} {\bibinfo {author} {\bibfnamefont {R.}~\bibnamefont
  {{Scoccimarro}}},\ }\bibfield  {title} {\enquote {\bibinfo {title}
  {{Large-scale structure in brane-induced gravity. I. Perturbation theory}},}\
  }\href {\doibase 10.1103/PhysRevD.80.104006} {\bibfield  {journal} {\bibinfo
  {journal} {\prd}\ }\textbf {\bibinfo {volume} {80}},\ \bibinfo {pages}
  {104006--+} (\bibinfo {year} {2009})},\ \Eprint
  {http://arxiv.org/abs/0906.4545} {arXiv:0906.4545 [astro-ph.CO]} \BibitemShut
  {NoStop}%
\bibitem [{\citenamefont {{Scoccimarro}}\ \emph {et~al.}(2011)\citenamefont
  {{Scoccimarro}}, \citenamefont {{Hui}}, \citenamefont {{Manera}},\ and\
  \citenamefont {{Chan}}}]{ScoHuiMan1108}%
  \BibitemOpen
  \bibfield  {author} {\bibinfo {author} {\bibfnamefont {R.}~\bibnamefont
  {{Scoccimarro}}}, \bibinfo {author} {\bibfnamefont {L.}~\bibnamefont
  {{Hui}}}, \bibinfo {author} {\bibfnamefont {M.}~\bibnamefont {{Manera}}}, \
  and\ \bibinfo {author} {\bibfnamefont {K.~C.}\ \bibnamefont {{Chan}}},\
  }\bibfield  {title} {\enquote {\bibinfo {title} {{Large-scale Bias and
  Efficient Generation of Initial Conditions for Non-Local Primordial
  Non-Gaussianity}},}\ }\href@noop {} {\bibfield  {journal} {\bibinfo
  {journal} {ArXiv e-prints}\ } (\bibinfo {year} {2011})},\ \Eprint
  {http://arxiv.org/abs/1108.5512} {arXiv:1108.5512 [astro-ph.CO]} \BibitemShut
  {NoStop}%
\bibitem [{\citenamefont {{Ohta}}\ \emph {et~al.}(2004)\citenamefont {{Ohta}},
  \citenamefont {{Kayo}},\ and\ \citenamefont {{Taruya}}}]{OhtKayTar0406}%
  \BibitemOpen
  \bibfield  {author} {\bibinfo {author} {\bibfnamefont {Y.}~\bibnamefont
  {{Ohta}}}, \bibinfo {author} {\bibfnamefont {I.}~\bibnamefont {{Kayo}}}, \
  and\ \bibinfo {author} {\bibfnamefont {A.}~\bibnamefont {{Taruya}}},\
  }\bibfield  {title} {\enquote {\bibinfo {title} {{Cosmological Density
  Distribution Function from the Ellipsoidal Collapse Model in Real Space}},}\
  }\href {\doibase 10.1086/420762} {\bibfield  {journal} {\bibinfo  {journal}
  {\apj}\ }\textbf {\bibinfo {volume} {608}},\ \bibinfo {pages} {647--662}
  (\bibinfo {year} {2004})},\ \Eprint
  {http://arxiv.org/abs/arXiv:astro-ph/0402618} {arXiv:astro-ph/0402618}
  \BibitemShut {NoStop}%
\end{thebibliography}%

\end{document}